\documentclass[a4paper,11pt]{article}

\usepackage{jheppub}

\usepackage{graphicx}

\title{New analyses of event shape observables in electron-positron annihilation and the determination of $\alpha_s$ running behavior in perturbative domain}

\author{Sheng-Quan Wang$^{1}$,} \emailAdd{sqwang@cqu.edu.cn}
\author{Chao-Qin Luo$^{1}$,}
\author{Xing-Gang Wu$^{2}$,} \emailAdd{wuxg@cqu.edu.cn}
\author{Jian-Ming Shen$^{3}$,} \emailAdd{shenjm@hnu.edu.cn}
\author{Leonardo Di Giustino$^{4}$} \emailAdd{ldigiustino@uninsubria.it}

\affiliation{ $^{1}$Department of Physics, Guizhou Minzu University, Guiyang 550025, P.R. China}
\affiliation{ $^{2}$Department of Physics, Chongqing University, Chongqing 401331, P.R. China}
\affiliation{ $^{3}$School of Physics and Electronics, Hunan University, Changsha 410082, P.R. China}
\affiliation{ $^{4}$Department of Science and High Technology, University of Insubria, via valleggio 11, I-22100, Como, Italy}

\abstract{
In this paper, we give comprehensive analyses for event shape observables in electron-positron annihilation by using the Principle of Maximum Conformality (PMC) which is a rigorous scale-setting method to eliminate the renormalization scheme and scale ambiguities in perturbative QCD predictions. Conventionally the renormalization scale and theoretical uncertainties in event shape observables are often evaluated by setting the scale to the center-of-mass energy $\sqrt{s}$. The event shape distributions using this conventional scale setting are plagued by the large renormalization scale uncertainty and underestimate the experimental data. Moreover, since the renormalization scale is simply fixed to the center-of-mass energy $\sqrt{s}$, only one value of the coupling $\alpha_s$ at the single scale $\sqrt{s}$ can be extracted. In contrast, the PMC renormalization scales are determined by absorbing the nonconformal $\beta$ contributions that govern the behavior of the running coupling via the Renormalization Group Equation (RGE). The resulting PMC scales change with event shape kinematics, reflecting the virtuality of the underlying quark and gluon subprocess. The PMC scales thus yield the correct physical behavior of the scale and the PMC predictions agree with precise event shape distributions measured at the LEP experiment. More importantly, we can precisely determine the running of the QCD coupling constant $\alpha_s(Q^2)$ over a wide range of $Q^2$ in perturbative domain from event shape distributions measured at a single center-of-mass energy $\sqrt{s}$.}


\begin{document}

\maketitle

\flushbottom

\section{Introduction}
\label{sec:1}

Event shape observables provide an ideal platform for testing the QCD. Given that the strong interaction occurs at the lowest order in perturbative QCD and leading term is proportional to the QCD coupling $\alpha_s$, a reliable value $\alpha_s$ can be extracted from the comparison of theoretical predictions with experimental data. Thus event shape observables have been extensively studied experimentally and theoretically. The experiments at LEP and at SLAC have measured event shape distributions with a high precision, especially at $Z^0$ peak~\cite{Heister:2003aj, Abdallah:2003xz, Abbiendi:2004qz, Achard:2004sv, Abe:1994mf}. The precision of these measured event shape distributions calls at least for equally precise theoretical predictions.

Significant efforts have been made on improving the precision of theoretical calculations for event shapes. The pQCD corrections have been calculated up to the next-to-next-to-leading order (NNLO) level~\cite{Ellis:1980wv, Kunszt:1980vt, Vermaseren:1980qz, Fabricius:1981sx, Giele:1991vf, Catani:1996jh, Gehrmann-DeRidder:2007nzq, GehrmannDeRidder:2007hr, Ridder:2014wza, Weinzierl:2008iv, Weinzierl:2009ms, DelDuca:2016csb, DelDuca:2016ily}. The resummation of logarithmically enhanced contributions has also been studied in the literature (see e.g.,~\cite{Catani:1992ua, Banfi:2004yd, Banfi:2014sua, Chien:2010kc, Becher:2012qc, Abbate:2010xh, Hoang:2014wka, Chiu:2011qc, Chiu:2012ir}). The main purpose of improving the precision of theoretical calculations is to obtain reliable values of $\alpha_s$, which has been extensively studied (see e.g.,~\cite{PDG:2020} for a summary from Particle Data Group).

Comparing the different sources of error in the extraction of the QCD coupling $\alpha_s$, one finds that the main obstacle for achieving a highly precise $\alpha_s$ is the uncertainties of theoretical calculations especially from the renormalization scale ambiguity. According to the conventional practice, the renormalization scale $\mu_r$ is often set to the typical scale $Q$ of the process, and theoretical uncertainties are estimated by varying the scale over a range of $\mu_r\in[Q/2, 2Q]$ for the pQCD predictions. For the calculation of event shape observables, one also often set the renormalization scale to the center-of-mass energy $\sqrt{s}$ to eliminate the logarithmic term $\ln(\mu_r/\sqrt{s})$, and vary the scale $\mu_r\in[\sqrt{s}/2, 2\sqrt{s}]$ to estimate the theoretical uncertainty. In order to match theoretical predictions to experimental results, in the past it was customary to take the renormalization scale at a give energy as an additional fit parameter. By using the conventional scale-setting method of setting $\mu_r\in[\sqrt{s}/2, 2\sqrt{s}]$, the event shape distributions are plagued by the large renormalization scale uncertainty, and even up to NNLO pQCD corrections underestimate the experimental data. Moreover, if the renormalization scale is always fixed to $\sqrt{s}$, only one value of $\alpha_s$ at the scale $\sqrt{s}$ can be determined. The large renormalization scale uncertainty for conventional event shape predictions affects also the value of $\alpha_s$. As a matter of fact the variation of the scale $\mu_r\in[\sqrt{s}/2, 2\sqrt{s}]$ can give only a mere indication about the theoretical errors; we actually do not know what is the correct range of variation of the renormalization scale in order to have reliable quantitative predictions for the theoretical uncertainties. The variation of the renormalization scale is only sensitive to the non-conformal $\beta$ terms but not to the conformal terms. Thus, it is important to find a correct method to achieve a reliable and accurate prediction for event shape observables.

Event shape observables in electron-positron annihilation is an ideal testing ground for the validity of a scale-setting method. The Principle of Maximum Conformality (PMC)~\cite{Brodsky:2011ta, Brodsky:2012rj, Brodsky:2011ig, Mojaza:2012mf, Brodsky:2013vpa} has been proposed for eliminating the renormalization scale and scheme ambiguities in pQCD predictions. We have successfully applied the PMC to eliminate the conventional scale uncertainty for two event shapes: the thrust ($T$) and C-parameter ($C$) measured at $\sqrt{s}=91.2$ GeV~\cite{Wang:2019ljl, Wang:2019isi}. Different from the conventional method of simply fixing the renormalization scale $\mu_r=\sqrt{s}$, the PMC scales are determined by absorbing the $\beta$ terms that govern the behavior of the running $\alpha_s$ via the Renormalization Group Equation (RGE). Since the $\beta$ terms do not appear, there is no renormalon divergence in the pQCD series and a more convergent perturbative series can be achieved. The PMC method extends the Brodsky-Lepage-Mackenzie (BLM) scale-setting method~\cite{Brodsky:1982gc} to all orders, and it reduces in the Abelian limit to the Gell-Mann-Low method~\cite{GellMann:1954fq}. PMC predictions do not depend on the renormalization scheme, satisfying the principles of Renormalization Group invariance (RGI)~\cite{Brodsky:2012ms, Wu:2014iba, Wu:2019mky}. After applying the PMC method to the thrust ($T$) and C-parameter ($C$), we observed that the PMC scales depend dynamically on the virtuality of the underlying quark and gluon subprocess and thus the specific kinematics of each event. The PMC predictions agree with the precise experimental data with high precision. We provided a novel method for precise determination of the running of the QCD coupling constant $\alpha_s(Q^2)$ over a wide range of $Q^2$ in perturbative domain from the thrust ($T$) and C-parameter ($C$) distributions measured at a single center-of-mass energy $\sqrt{s}=91.2$ GeV.

In this paper, as a step forward of our previous calculations for the thrust ($T$) and C-parameter ($C$) distributions at $\sqrt{s}=91.2$ GeV~\cite{Wang:2019ljl, Wang:2019isi}, we will give comprehensive analyses for event shape observables and then determine the $\alpha_s$ running behavior in perturbative domain by applying the PMC method. As will be show in Section~\ref{sec:2}, the PMC scale itself is a perturbative expansion series in $\alpha_s$, due to unknown higher-order contributions, this leads to residual scale dependence for the PMC scale (first kind of residual scale dependence); in addition, the last terms of the pQCD approximant are unfixed because of its PMC scale cannot be determined (second kind of residual scale dependence)~\cite{Zheng:2013uja}. However, these residual scale dependencies are distinct from the conventional scale ambiguities. As will be show in the following, the first kind of residual scale dependence is highly suppressed. The PMC single-scale method~\cite{Shen:2017pdu} exactly removes the second kind of residual scale dependence, which has been adopted for our analyses in present paper.
A reliable and accurate prediction will be achieved by using the PMC to calculate event shape observables.

The remaining sections of this paper are organized as follows: In Sec.~\ref{sec:2}, we present the detailed technology for applying the PMC to event shape observables in electron-positron annihilation. In Sec.~\ref{sec:3}, we present theoretical predictions and discussions for event shape distributions using the conventional and PMC scale-setting methods. In Sec.~\ref{sec:4}, we extract the QCD running coupling $\alpha_s(Q^2)$ over a large range of $Q^2$ in perturbative domain by comparing the PMC predictions with the experimental data. Sec.~\ref{sec:5} is reserved for a summary.

\section{PMC scale-setting for event shape observables}
\label{sec:2}

The pQCD calculation for the distribution of an event shape observable $y$ at the center-of-mass energy $\sqrt{s}$ can be written as
\begin{eqnarray}
\frac{1}{\sigma_0}\frac{d\sigma}{dy}&=&A(y)\,a_s(\sqrt{s})+B(y)\,a^2_s(\sqrt{s}) + {\cal O}(a^3_s),
\end{eqnarray}
where $a_s(\sqrt{s})=\alpha_s(\sqrt{s})/(2\pi)$. The born cross-section $\sigma_0$ is for $e^+e^-\rightarrow$ hadrons. The $A(y)$, $B(y)$, ... are perturbative coefficients, which are computed at a renormalization scale fixed to the center-of-mass energy $\sqrt{s}$, and thus depend only on the value of the event shape observable $y$. The measured event shape distribution is normalized to total cross-section $\sigma_{tot}$ of the $e^+e^-\rightarrow$ hadrons,
\begin{eqnarray}
\frac{1}{\sigma_{tot}}\frac{d\sigma}{dy}&=&\bar{A}(y)\,a_s(\sqrt{s})+\bar{B}(y)\,a^2_s(\sqrt{s})+{\cal O}(a^3_s),
\label{eventshapet}
\end{eqnarray}
where, the perturbative coefficients $\bar{A}(y)$ and $\bar{B}(y)$ are related to the coefficients $A(y)$ and $B(y)$ by
\begin{eqnarray}
\bar{A}(y)&=&A(y), \nonumber\\
\bar{B}(y)&=&B(y)-\frac{3}{2}\,C_F\,A(y),
\end{eqnarray}
and the total cross-section $\sigma_{tot}$ is
\begin{eqnarray}
\sigma_{tot}=\sigma_0\left(1+\frac{3}{2}\,C_F\,a_s(\sqrt{s})+{\cal O}(a^2_s)\right).
\end{eqnarray}

The perturbative coefficients can be expressed by the $n_f$-term, e.g., the NLO coefficient is
\begin{eqnarray}
\bar{B}(y)=\bar{B}(y)_{\rm in}+\bar{B}(y)_{n_f}\cdot n_f,
\label{NLOefficnf}
\end{eqnarray}
where the coefficients $\bar{B}(y)_{\rm in}$ is independent of $n_f$-term and $\bar{B}(y)_{n_f}$ is the coefficient of $n_f$-term. The arbitrary renormalization scale $\mu_r$ dependence for the perturbative coefficients $\bar{A}(y)$, $\bar{B}(y)$, ... can be restored from the RGE, i.e.,
\begin{eqnarray}
\bar{A}(y,\mu_r)&=&\bar{A}(y), \nonumber\\
\bar{B}(y,\mu_r)&=&\bar{B}(y)+\frac{1}{2}\beta_0\ln\left(\frac{\mu^2_r}{s}\right)\bar{A}(y),
\end{eqnarray}
where, $\beta_0=11\,C_A/3-4/3\,T_R\,n_f$. The renormalization scale $\mu_r$ dependence of the NLO coefficients in Eq.(\ref{NLOefficnf}) is changed to
\begin{eqnarray}
\bar{B}(y,\mu_r)=\bar{B}(y,\mu_r)_{\rm in}+\bar{B}(y,\mu_r)_{n_f}\cdot n_f.
\end{eqnarray}

In order to apply the PMC method to event shape observables, the coefficients need to be divided into conformal terms and non-conformal terms~\cite{Mojaza:2012mf, Brodsky:2013vpa, Wang:2019ljl, Wang:2019isi}. The NLO coefficient changes to
\begin{eqnarray}
\bar{B}(y,\mu_r)=\bar{B}(y,\mu_r)_{\rm con}+\bar{B}(y,\mu_r)_{\beta_0}\cdot \beta_0,
\end{eqnarray}
where the coefficients $\bar{B}(y,\mu_r)_{\rm con}$ and $\bar{B}(y,\mu_r)_{\beta_0}$ are conformal and non-conformal coefficients, respectively. By using the PMC scale-setting method, the event shape distribution in Eq.(\ref{eventshapet}) changes to the following conformal series,
\begin{eqnarray}
\frac{1}{\sigma_{tot}}\frac{d\sigma}{dy}&=&\bar{A}(y)\,a_s(Q_\star)+\bar{B}(y,\mu_r)_{\rm con}\,a^2_s(Q_\star)+{\cal O}(a^3_s),
\label{eventshapePMC}
\end{eqnarray}
where $Q_\star$ stands for the PMC scale which is determined by absorbing all of the non-conformal terms and can be given by
\begin{eqnarray}
\ln\frac{Q^2_\star}{\mu^2_r}=-\frac{2\,\bar{B}(y,\mu_r)_{\beta_0}}{\bar{A}(y,\mu_r)}+{\cal O}(a_s).
\label{evenPMCscale}
\end{eqnarray}
The conformal coefficient is
\begin{eqnarray}
\bar{B}(y,\mu_r)_{\rm con}=\frac{11\,C_A}{4\,T_R}\bar{B}(y,\mu_r)_{n_f}+\bar{B}(y,\mu_r)_{\rm in},
\label{evenPMCcon}
\end{eqnarray}
where $C_A=3$, and $T_R$=1/2. The conventional results are obtained by using Eq.(\ref{eventshapet}), while the PMC results are obtained by using Eq.(\ref{eventshapePMC}). The non-conformal $\beta$ terms disappear, and only the conformal terms are retained and thus there is no renormalon divergence in the pQCD series. It is noted that many studies on renormalons in soft-collinear effective theory (SCET) are given in Refs.\cite{Webber:1994cp,Beneke:1995pq,Gardi:2001ny,Hoang:2007vb,Gracia:2021nut}. Since the PMC scale $Q_\star$ is independent of the choice of renormalization scale $\mu_r$ and the conformal coefficient $\bar{B}(y,\mu_r)_{\rm con}$ is also renormalization scale-independent, the PMC prediction in Eq.(\ref{eventshapePMC}) eliminates the renormalization scale uncertainty.

Currently, pQCD corrections to event shapes have been calculated up to NNLO. However, how to fully distinguish the conformal and non-conformal contributions at NNLO for event shapes is complicated, which is beyond the scope of this paper. We can only provide a NLO PMC analysis and determine one PMC scale for event shapes. At NLO, the $n_f$-dependent terms are non-conformal and the PMC scale $Q_\star$ can be determined unambiguity. It is noted that the non-conformal $\beta^2_0$ terms at NNLO are easy to identify, the PMC scale $Q_\star$ can be determined up to next-to-leading-log (NLL) accuracy.

\section{Theoretical predictions and discussions}
\label{sec:3}

The detailed technology for applying the PMC to the thrust ($T$) distribution at $\sqrt{s}=91.2$ GeV has also been presented in Ref.~\cite{Wang:2019ljl}. We calculate all the event shape observables following a similar procedure. For the evaluation of the QCD running coupling, we adopt the RunDec program\cite{Chetyrkin:2000yt} from world average $\alpha_s(M^2_Z)=0.1179\pm0.0010$~\cite{PDG:2020} in the $\overline{\rm MS}$ scheme.

\subsection{Event shape distributions using the conventional scale-setting method }

In this paper, we calculate the classical event shape observables consisting of the thrust ($T$), heavy jet mass ($\rho=M^2_H/s$), wide jet broadening ($B_W$), total jet broadening ($B_T$) and the C-parameter ($C$). These event shape distributions have been measured at LEP experiment from $91.2$ to $206$ GeV~\cite{Heister:2003aj}. There are other observables such as the jet rates and the jet transition parameters which are logarithmic distributions and associated to the different jet definitions; we do not calculate these observables in the present paper.

\begin{figure*}
\begin{center}
\includegraphics[width=0.45\textwidth]{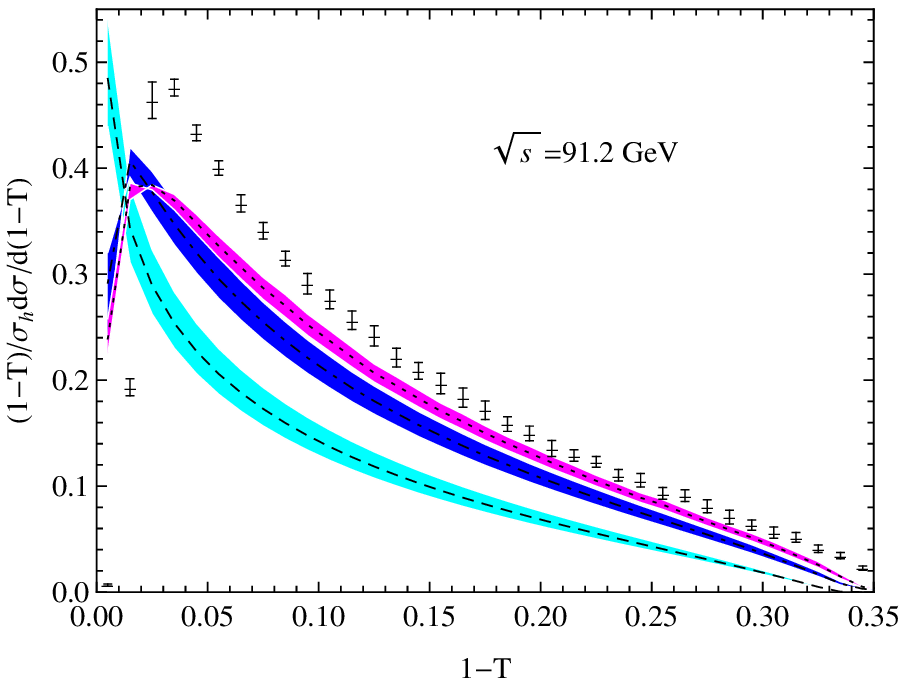} 
\includegraphics[width=0.45\textwidth]{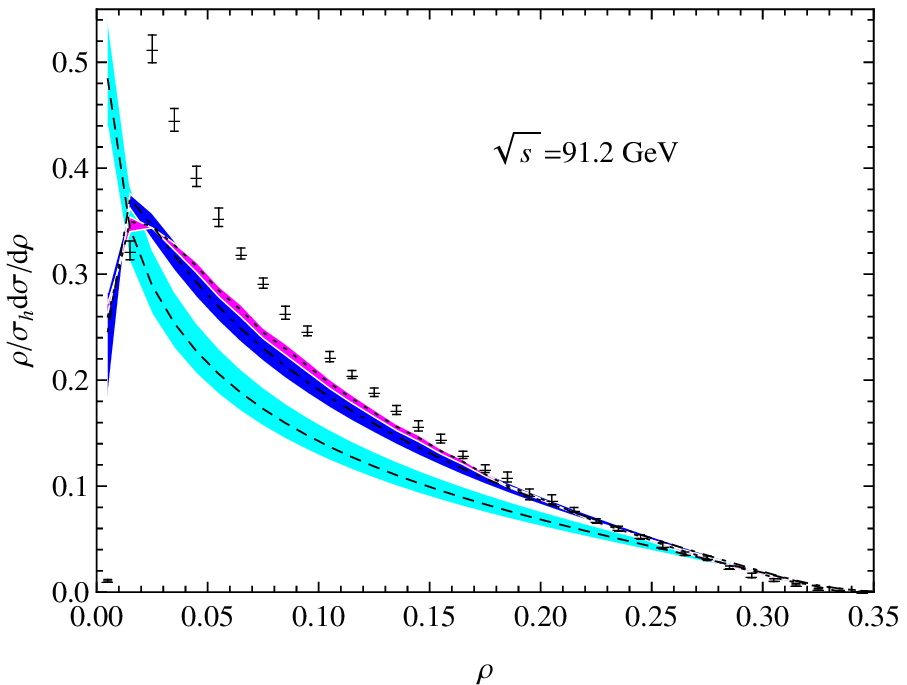} 
\includegraphics[width=0.45\textwidth]{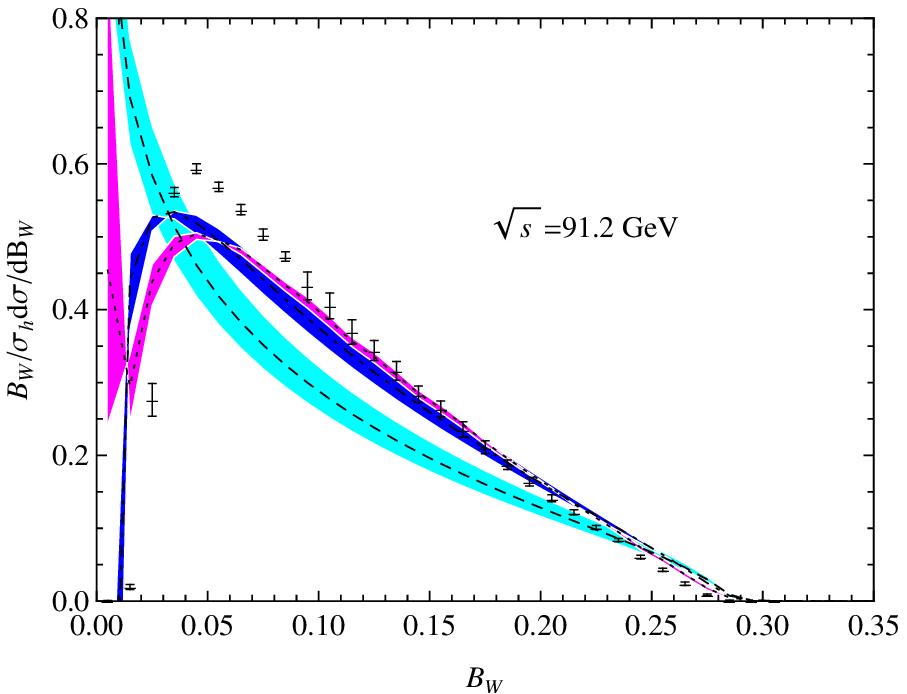} 
\includegraphics[width=0.45\textwidth]{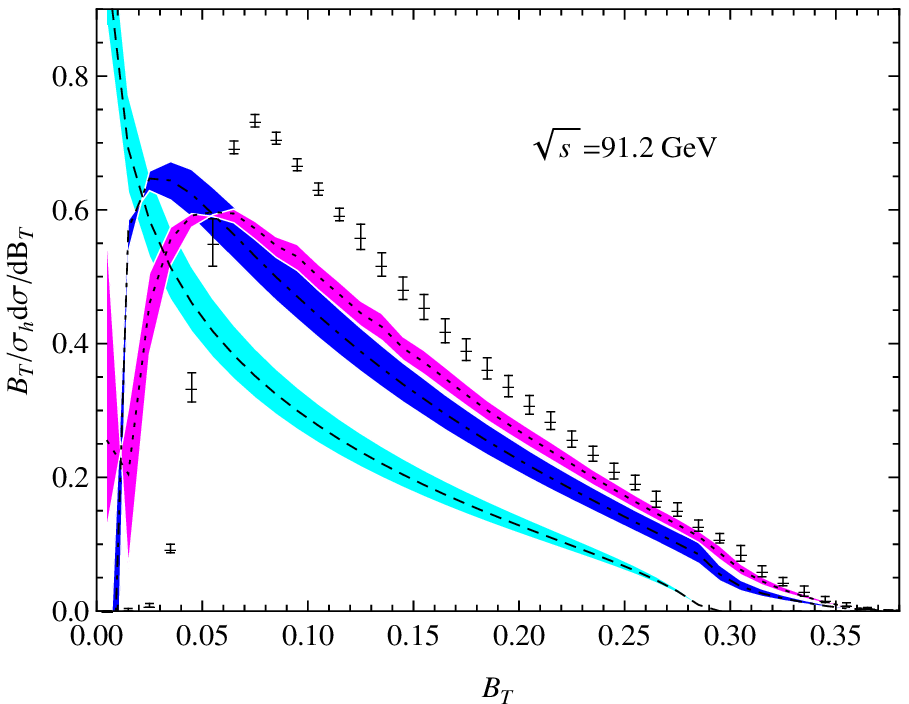} 
\includegraphics[width=0.45\textwidth]{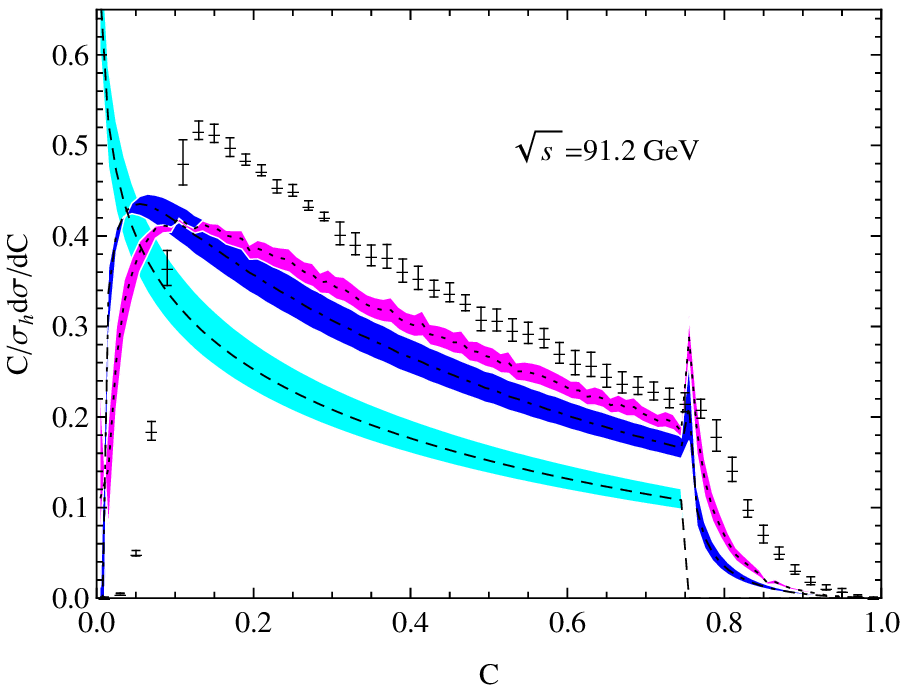} 
\caption{The event shape observables thrust ($1-T$), heavy jet mass ($\rho$), wide jet broadening ($B_W$), total jet broadening ($B_T$) and C-parameter ($C$) using the conventional scale-setting method at $\sqrt{s}=91.2$ GeV. For all the figures, the dashed, dotdashed and dotted lines are the conventional results at LO, NLO and NNLO~\cite{GehrmannDeRidder:2007hr, Weinzierl:2009ms}, respectively. The bands for the theoretical predictions are obtained by varying $\mu_r\in[\sqrt{s}/2,2\sqrt{s}]$. The experimental data are taken from the ALEPH Collaboration~\cite{Heister:2003aj}. }
\label{eventshapeConv}
\end{center}
\end{figure*}

The results for the event shape observables thrust ($T$), heavy jet mass ($\rho$), wide jet broadening ($B_W$), total jet broadening ($B_T$) and C-parameter ($C$) using the conventional scale-setting method are presented in Fig.(\ref{eventshapeConv}). In order to draw definitive conclusions, we only present the event shape distributions at $\sqrt{s}=91.2$ GeV. We observe from Figure (\ref{eventshapeConv}) that even up to NNLO QCD corrections the conventional results are plagued by the large renormalization scale $\mu_r$ uncertainty and underestimate the precise experimental data for all the event shape observables. There are only some matches between the predictions and the experimental data exist in small regions for the heavy jet mass and the wide jet broadening. The pQCD series shows a slow convergence in a wide region of the distributions. Moreover, for all event shape observables, the NLO results do not overlap with LO predictions; for the thrust, total jet broadening and the C-parameter, the NNLO results still do not overlap with NLO predictions, while for the heavy jet mass and wide jet broadening, the NNLO results only have some overlap with NLO predictions in the intermediate region. Thus, the conventional estimation of unknown higher-order QCD corrections by simply varying the renormalization scale $\mu_r\in[\sqrt{s}/2,2\sqrt{s}]$ is unreliable.

In fact, by using the conventional scale-setting method of simply fixing the scale $\mu_r=\sqrt{s}$, the pQCD series is in general factorially divergent, behaving as $n!\beta^n_0\alpha^n_s$--the ``renormalon" problem~\cite{Beneke:1994qe, Beneke:1998ui}. The conventional wisdom that the conventional scale $\mu_r$ dependence shall be suppressed by the inclusion of higher-order QCD corrections has always been discussed; however, one cannot judge whether a poor convergence is the intrinsic property of pQCD series, or is due to improper choice of the renormalization scale. Moreover, the variation of the renormalization scale is only sensitive to the non-conformal $\beta$ terms, but not to the conformal terms. We also do not know how wide a range the renormalization scale should vary in order to achieve reasonable predictions of its error; one often find significant renormalization scale and scheme ambiguities~\cite{Wu:2013ei}. The pQCD series shows a slow convergence by simply fixing the scale $\mu_r=\sqrt{s}$, and the estimation of unknown higher-order terms by varying $\mu_r\in[\sqrt{s}/2,2\sqrt{s}]$ is unreliable for event shape observables. Predictions based on conventional scale setting are also incorrect for Abelian theory--Quantum Electrodynamics (QED), where the renormalization scale of the QED coupling $\alpha$ can be set unambiguously by using the Gell-Mann-Low method~\cite{GellMann:1954fq}.

\subsection{Event shape distributions using the PMC scale-setting method }

\begin{figure*}
\begin{center}
\includegraphics[width=0.45\textwidth]{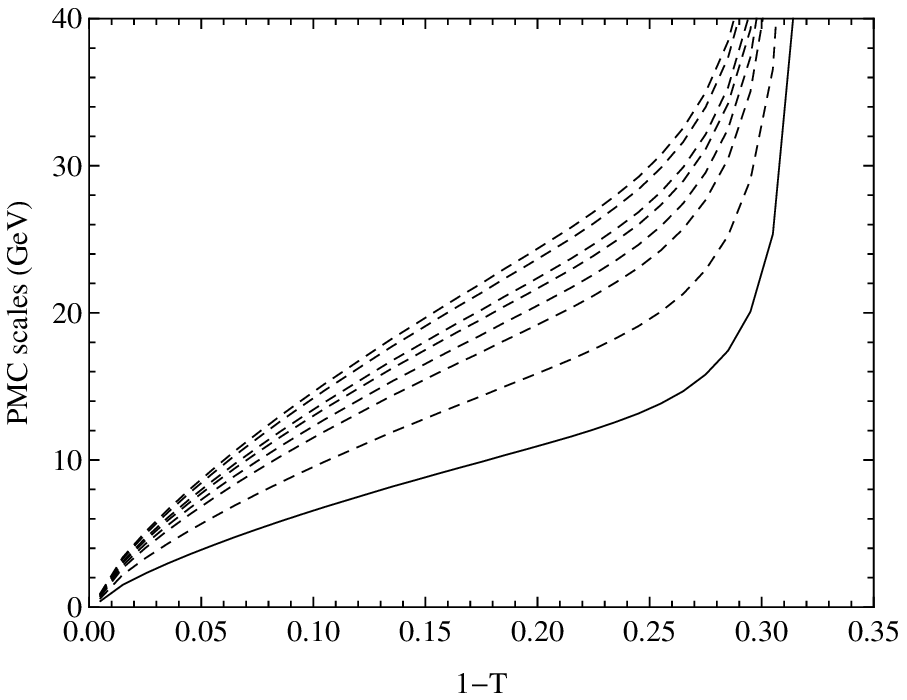} 
\includegraphics[width=0.45\textwidth]{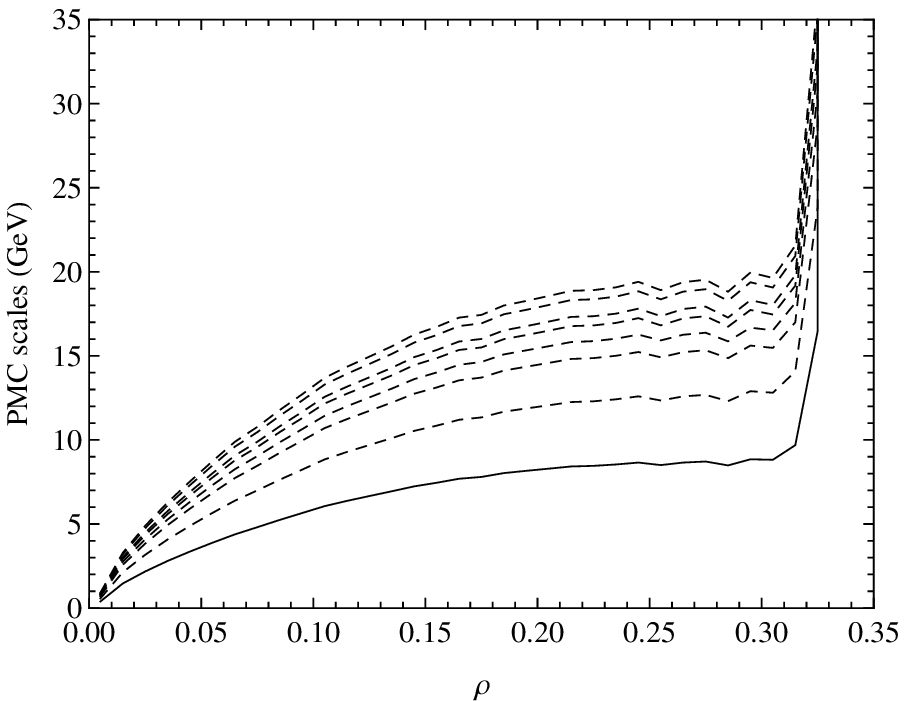} 
\includegraphics[width=0.45\textwidth]{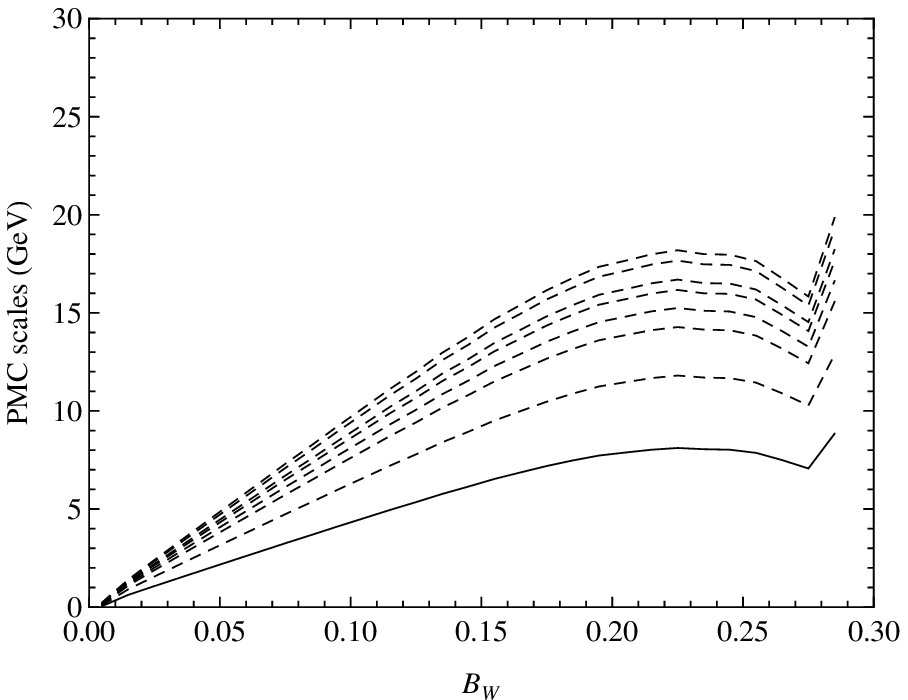} 
\includegraphics[width=0.45\textwidth]{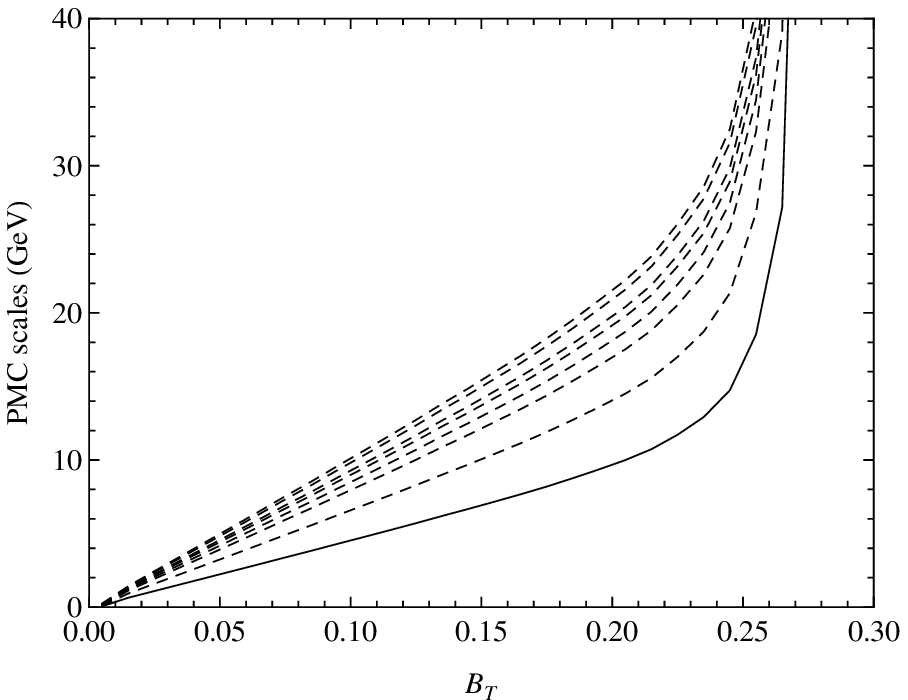} 
\includegraphics[width=0.45\textwidth]{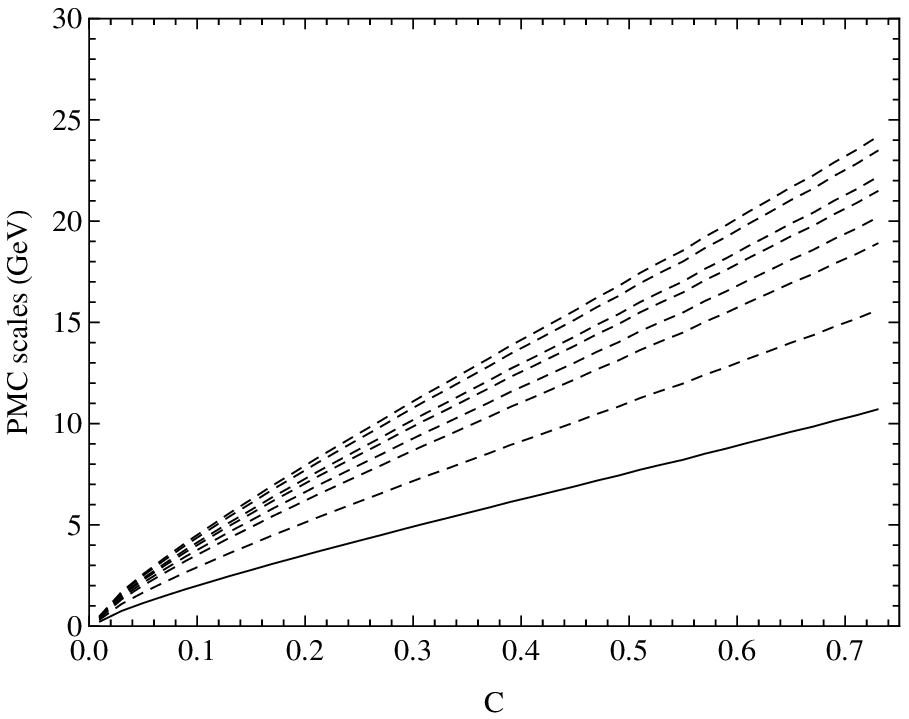} 
\caption{The PMC scales for the event shape observables thrust ($1-T$), heavy jet mass ($\rho$), wide jet broadening ($B_W$), total jet broadening ($B_T$) and C-parameter ($C$). In each figure, there are one solid line and seven dashed lines, which represent the PMC scales for $\sqrt{s}=91.2$ GeV and $\sqrt{s}=133$, $161$, $172$, $183$, $189$, $200$, $206$ GeV, respectively. These figures show clearly that the PMC scales monotonously increase with the center-of-mass energy $\sqrt{s}$. }
\label{eventshapePMCscale}
\end{center}
\end{figure*}

After applying PMC scale setting, the PMC scales in $\alpha_s$ are determined by absorbing the non-conformal $\beta$ terms which depend on the value of event shape observables. We present the PMC scales for the event shape observables thrust ($1-T$), heavy jet mass ($\rho$), wide jet broadening ($B_W$), total jet broadening ($B_T$) and C-parameter ($C$) at $\sqrt{s}=91.2$, $133$, $161$, $172$, $183$, $189$, $200$, $206$ GeV in Fig.(\ref{eventshapePMCscale}). We observe from Figure (\ref{eventshapePMCscale}) that:
\begin{itemize}
\item Remarkably, the PMC scales are not a single value but they change dynamically with event shapes, reflecting the virtuality of the underlying quark and gluon subprocess. Thus the number of active flavors $n_f$ changes with event shape observables according to the PMC scales.
\item It is noted that the quarks and gluons have soft virtuality near the two-jet region. As the argument of the $\alpha_s$ approaches the two-jet region, the PMC scales are very soft and thus the non-perturbative effects must be taken into account, while in the regions away from the two-jet region, the PMC scales are increased, as expected.
\item The PMC scales are very small in the wide kinematic regions compared to the conventional method of simply fixing $\mu_r=\sqrt{s}$.
\item The PMC scales are independent of the choice of $\mu_r$, and they are proportional to the center-of-mass energy $\sqrt{s}$ and thus increased with the $\sqrt{s}$.
\end{itemize}

These features yield the correct physical behavior of the scale. The typical analysis based on SCET also show that the scale is very soft near the two-jet region. Thus, both the PMC method and the SCET analysis get the same physical behavior for the scale in the two-jet region. For the very small scale in the two-jet region, we adopt the light-front holographic QCD~\cite{Brodsky:2014yha} to evaluate the running coupling $\alpha_s$. The PMC is consistent with its QED analog obtained using the GM-L scale-setting approach to the $e^+e^-\rightarrow \gamma^*\rightarrow X$ (QED) where the final-state particles are restricted to leptons and photons. Previous analyses also show that the renormalization scale is not a single value for event shape observables~\cite{Kramer:1990zt, Gehrmann:2014uva, Abbate:2010xh, Hoang:2014wka} and the analysis of event shape observables using variable flavour number scheme have been suggested in the literature (see e.g.,~\cite{Pietrulewicz:2014qza}).

\begin{figure}[htb]
\centering
\includegraphics[width=0.50\textwidth]{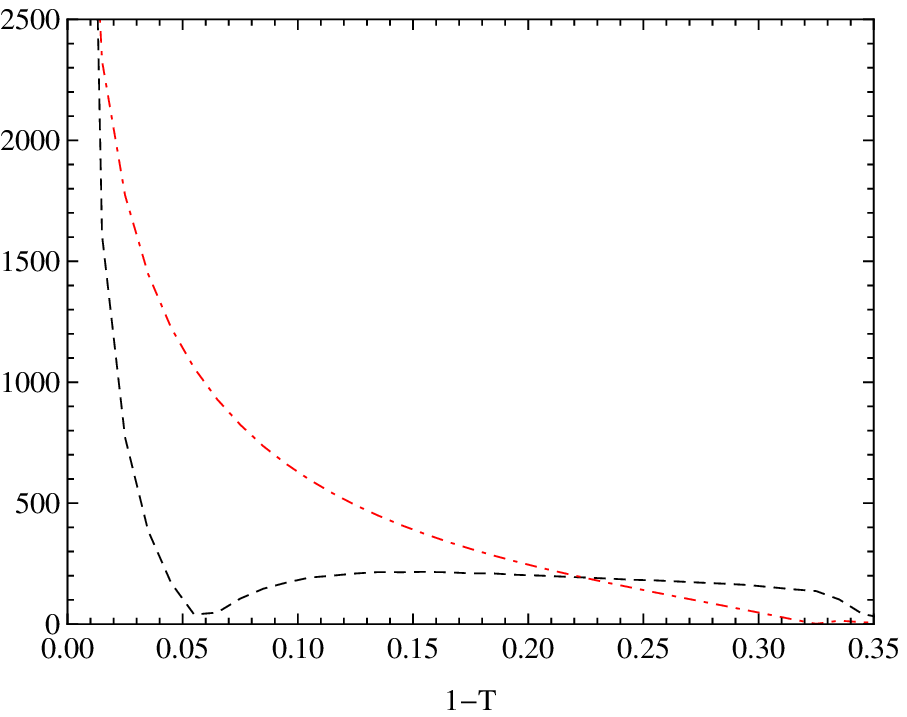}
\caption{The absolute value of the non-conformal term $\bar{B}(y,\mu_r)_{\beta_0}\cdot\beta_0$ and conformal term $\bar{B}(y,\mu_r)_{\rm con}$ for the thrust ($1-T$), where the dashed line stands for the $\bar{B}(y,\mu_r)_{\rm con}$, and the dot-dashed line represents the $\bar{B}(y,\mu_r)_{\beta_0}\cdot\beta_0$. }
\label{connonconT}
\end{figure}

\begin{figure}[htb]
\centering
\includegraphics[width=0.50\textwidth]{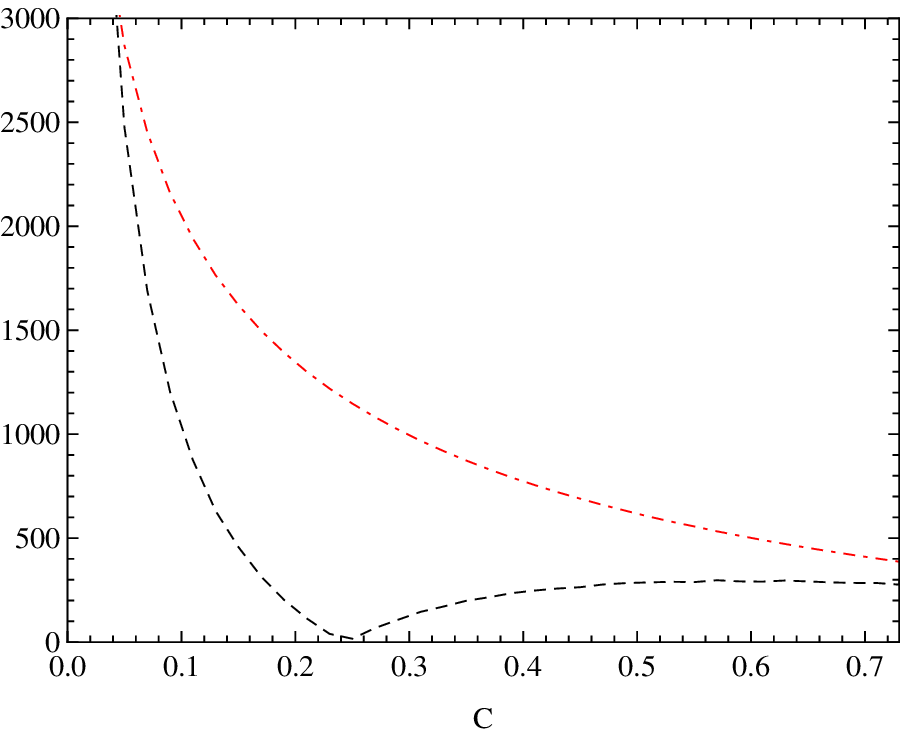}
\caption{The absolute value of the non-conformal term $\bar{B}(y,\mu_r)_{\beta_0}\cdot\beta_0$ and conformal term $\bar{B}(y,\mu_r)_{\rm con}$ for the C-parameter ($C$), where the dashed line stands for the $\bar{B}(y,\mu_r)_{\rm con}$, and the dot-dashed line represents the $\bar{B}(y,\mu_r)_{\beta_0}\cdot\beta_0$. }
\label{connonconC}
\end{figure}

The analysis for event shape observables at $\sqrt{s}=91.2$ GeV using BLM scale-setting method has been given in Ref.\cite{Gehrmann:2014uva}. At NLO, our results for event shape distributions at $\sqrt{s}=91.2$ GeV are identical to those of Ref.\cite{Gehrmann:2014uva}. In this paper, we give comprehensive analyses for event shape observables measured over a wide range of $\sqrt{s}$. More importantly, due to the PMC scales are not a single value but dynamically changing with event shapes, we thus provide a novel method for the precise determination of the running of $\alpha_s(Q^2)$ over a wide range of $Q^2$ in perturbative domain. Even though strictly speaking the $\beta_0$-dependent terms at NLO does not come from vacuum-bubble diagrams, guided by the fact that these terms seem to numerically dominate (see Figs.(\ref{connonconT}) and (\ref{connonconC})) and following Ref.\cite{Gehrmann:2014uva} we take them as non-conformal.

Since the PMC scale $Q_\star$ is a perturbative expansion series in $\alpha_s$, it can be determined up to leading-log (LL) and next-to-leading-log (NLL) accuracy using the non-conformal $\beta_0$ term at NLO and $\beta^2_0$ term at NNLO, respectively. We have found that the PMC scales $Q_\star$ for all event shape observables show a fast pQCD convergence, and the inclusion of the $\beta^2_0$ term at NNLO only slightly changes the PMC scales $Q_\star$ at LL. More explicitly, in the case of $\sqrt{s}=91.2$ GeV, the magnitude of the NLL-terms of the PMC scales $Q_\star$ are only about $2\%-4\%$ for the thrust, $1\%-3\%$ for the heavy jet mass, $1\%-2\%$ for the wide jet broadening, $2\%-4\%$ for the total jet broadening, and $2\%-4\%$ for the C-parameter of the LL-terms in the wide intermediate region. For example, the $2\%-4\%$ correction of NLL-terms for the thrust only yields $0.2-1$ GeV to its PMC scale $Q_\star$ at LL, which was explicitly shown in Ref.\cite{Wang:2019ljl}. This leads us to believe that the unknown higher-order correction to the PMC scale $Q_\star$ and thus the first kind of residual scale dependence are greatly suppressed.

\begin{figure*}
\begin{center}
\includegraphics[width=0.45\textwidth]{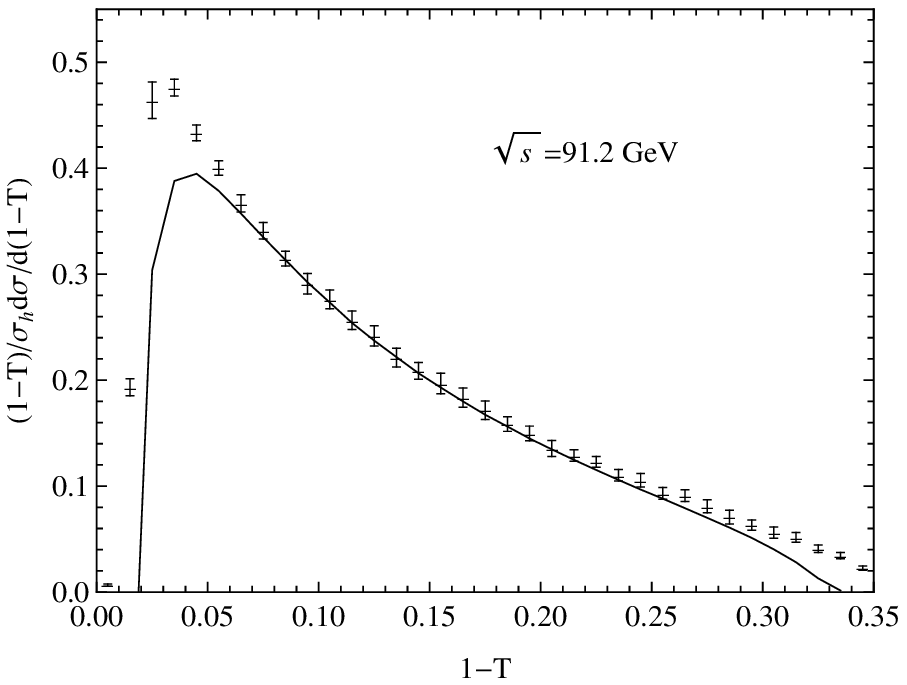}
\includegraphics[width=0.45\textwidth]{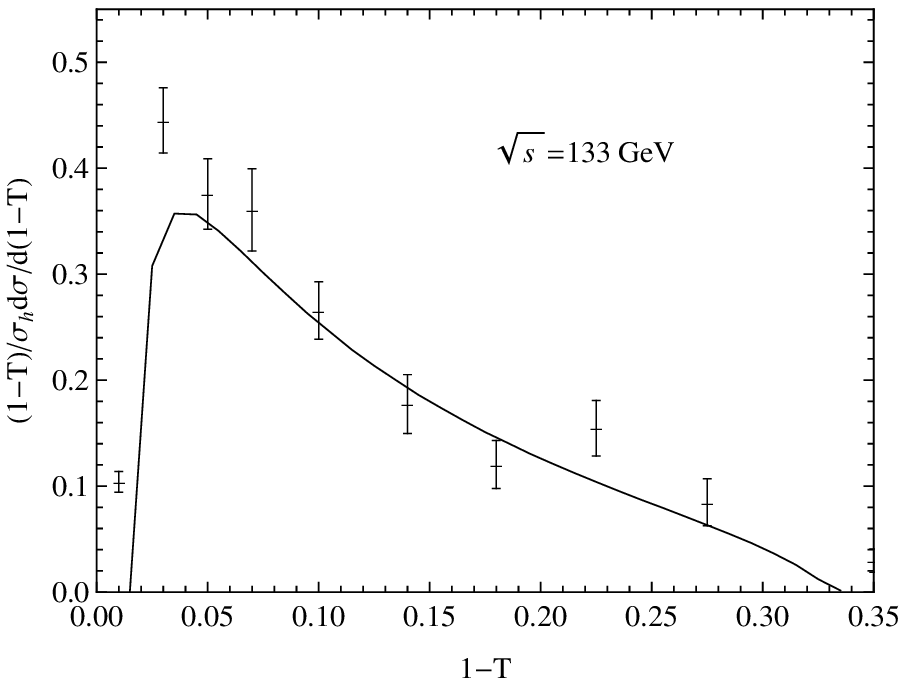}
\includegraphics[width=0.45\textwidth]{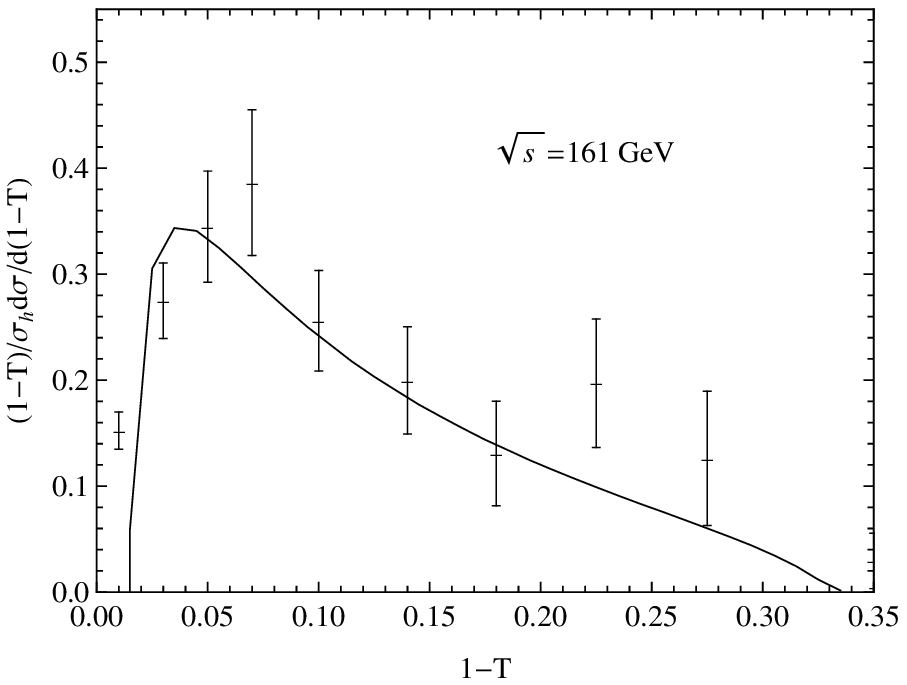}
\includegraphics[width=0.45\textwidth]{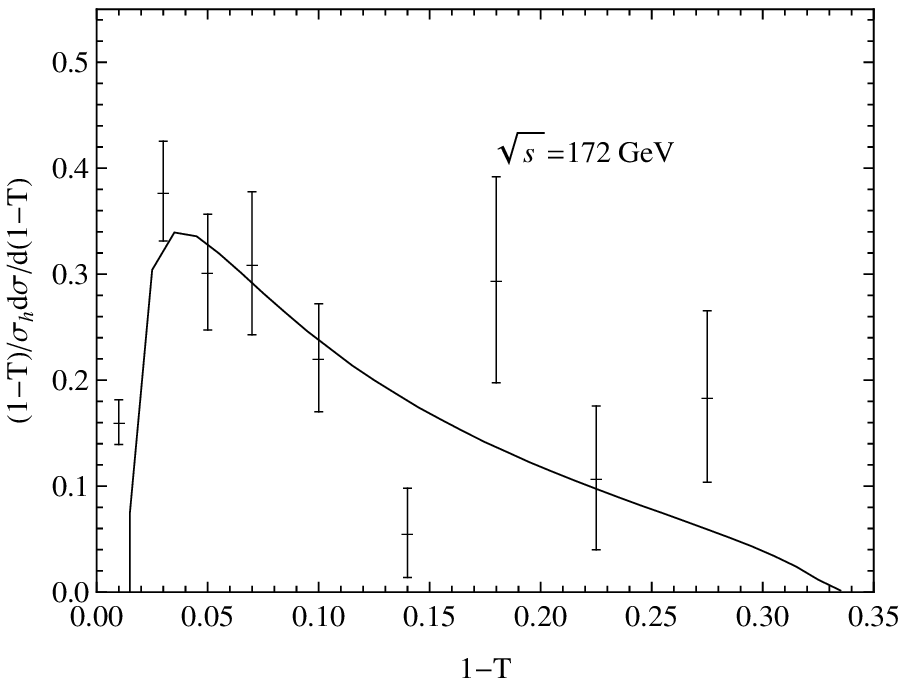}
\includegraphics[width=0.45\textwidth]{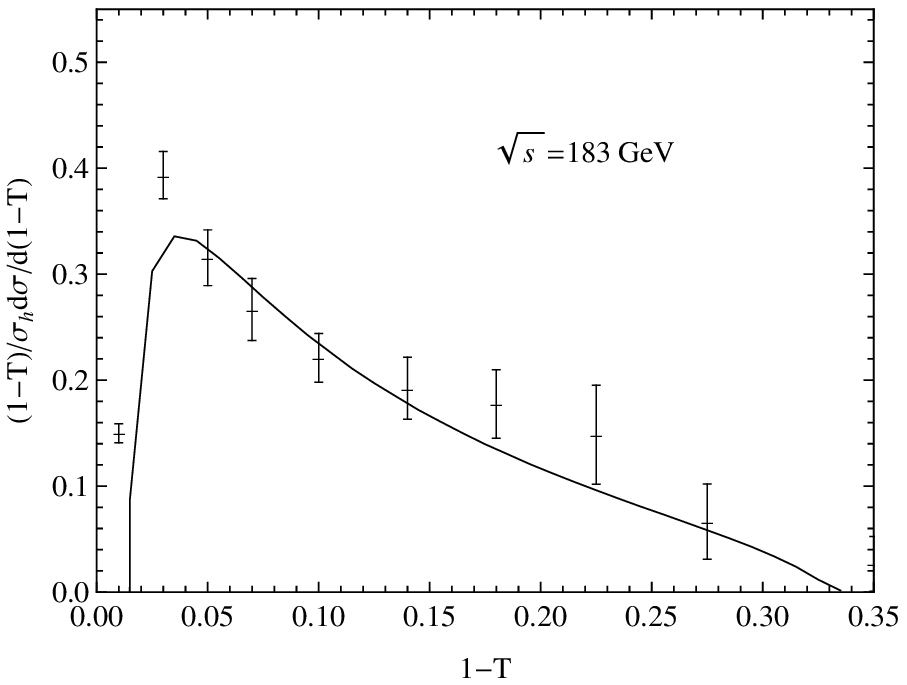}
\includegraphics[width=0.45\textwidth]{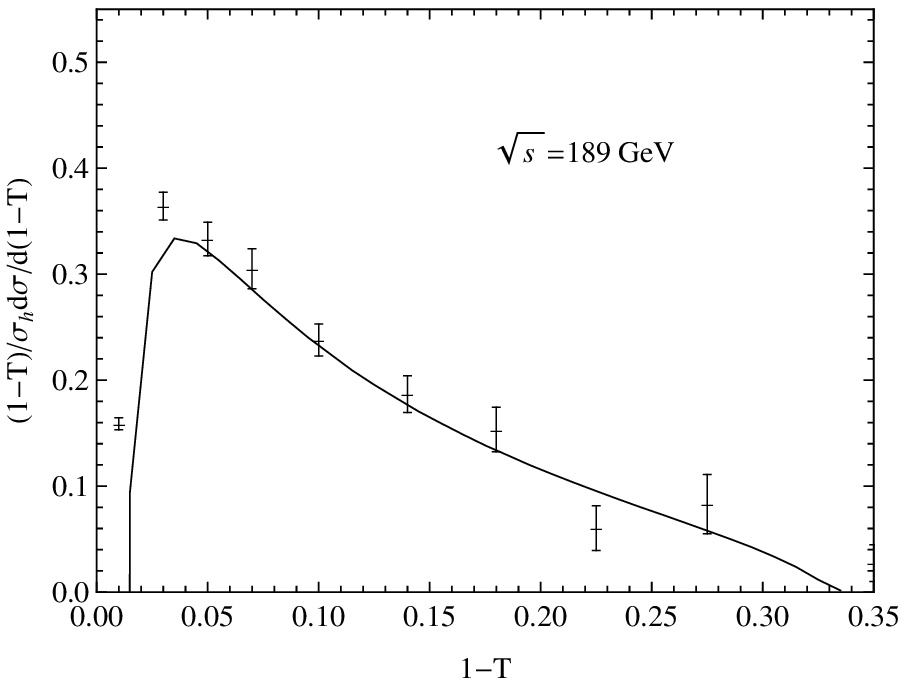}
\includegraphics[width=0.45\textwidth]{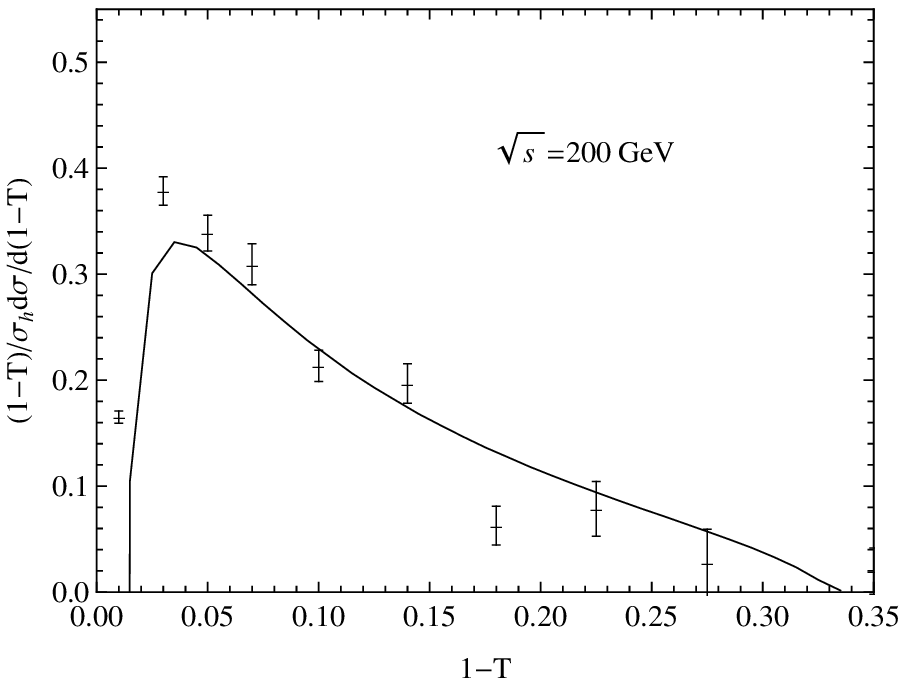}
\includegraphics[width=0.45\textwidth]{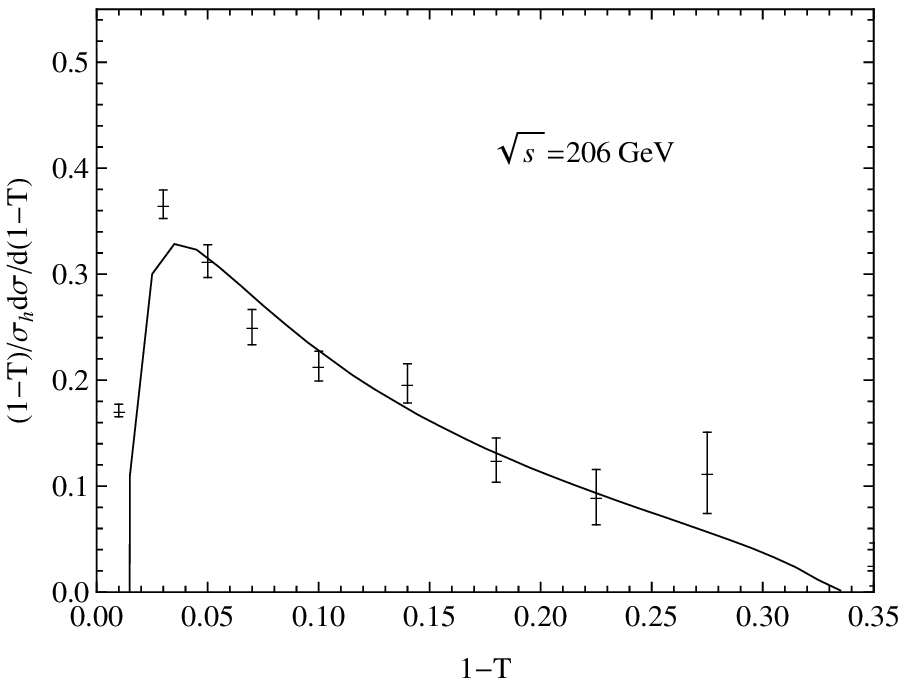}
\caption{The thrust ($1-T$) distributions using PMC scale setting for $\sqrt{s}=91.2$, $133$, $161$, $172$, $183$, $189$, $200$, $206$ GeV. The experimental data are taken from the ALEPH Collaboration~\cite{Heister:2003aj}.}
\label{distriPMCT}
\end{center}
\end{figure*}

\begin{figure*}
\begin{center}
\includegraphics[width=0.45\textwidth]{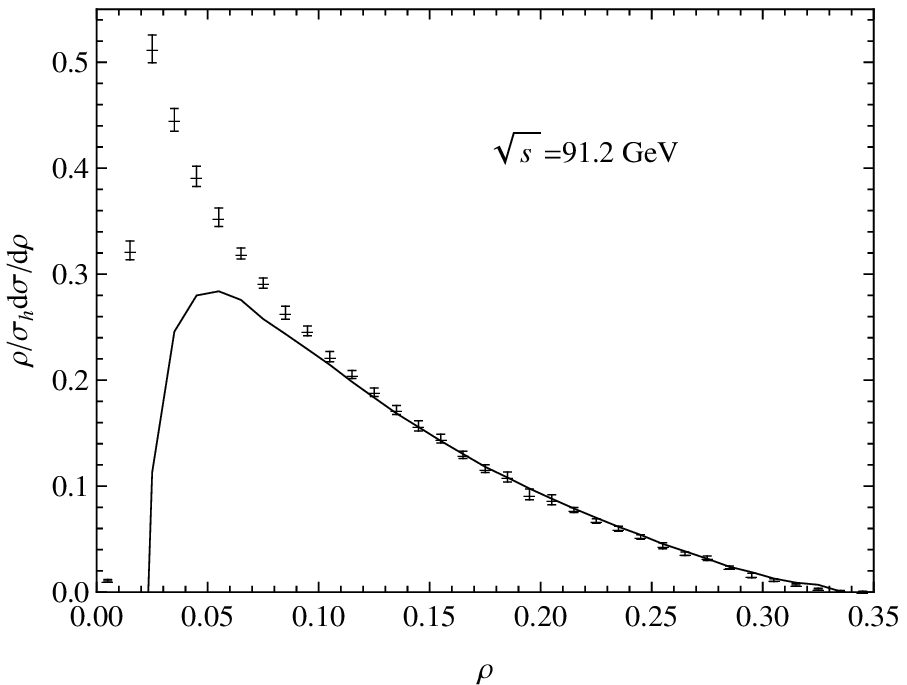}
\includegraphics[width=0.45\textwidth]{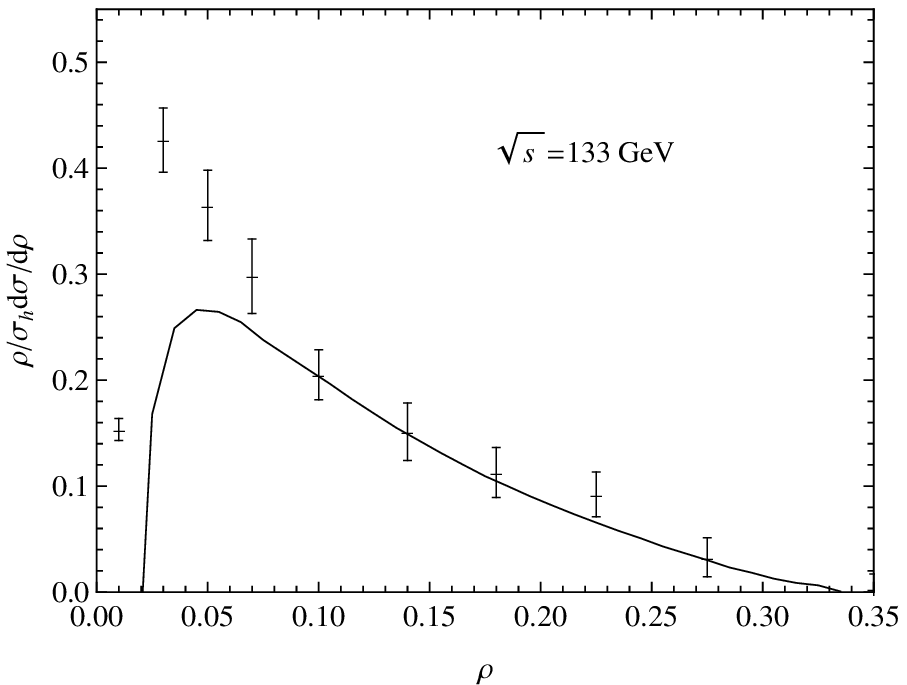}
\includegraphics[width=0.45\textwidth]{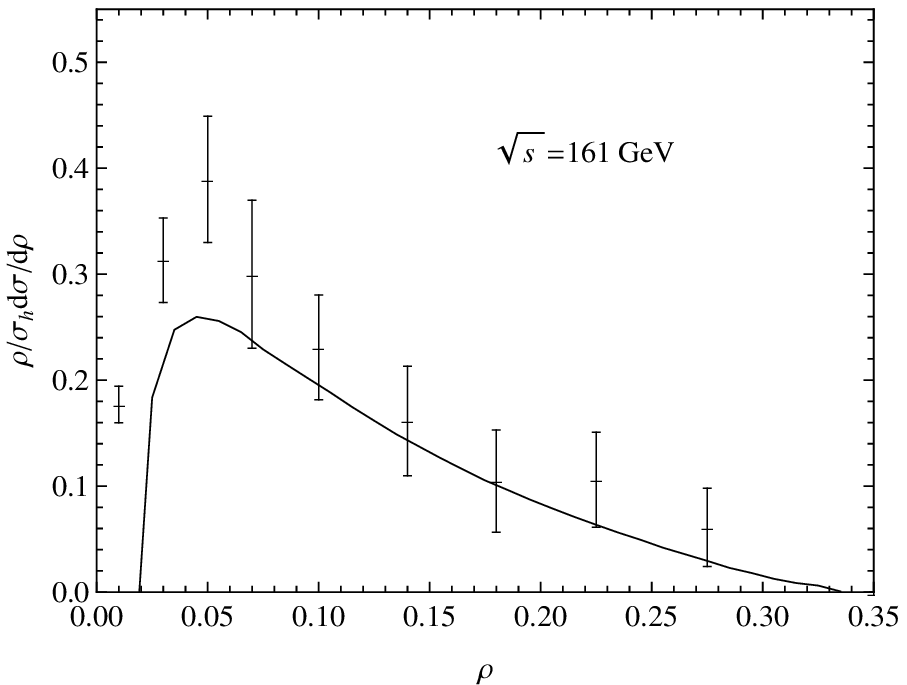}
\includegraphics[width=0.45\textwidth]{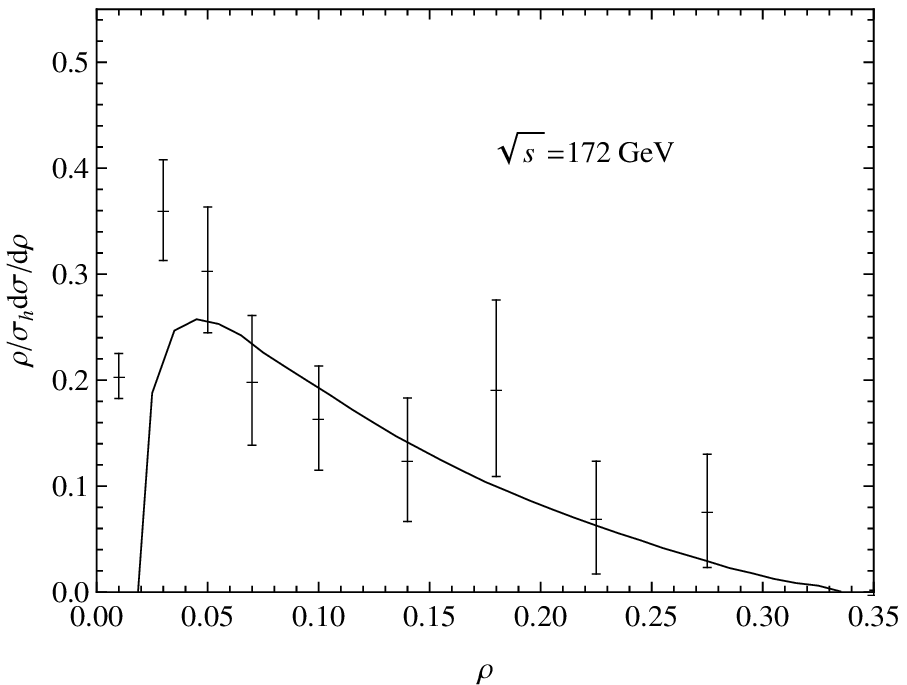}
\includegraphics[width=0.45\textwidth]{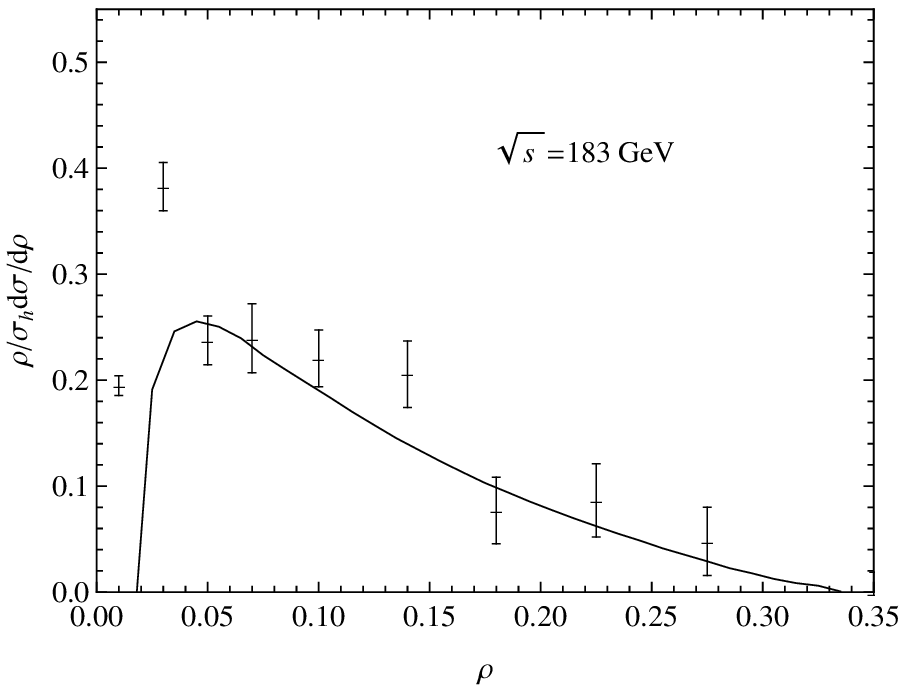}
\includegraphics[width=0.45\textwidth]{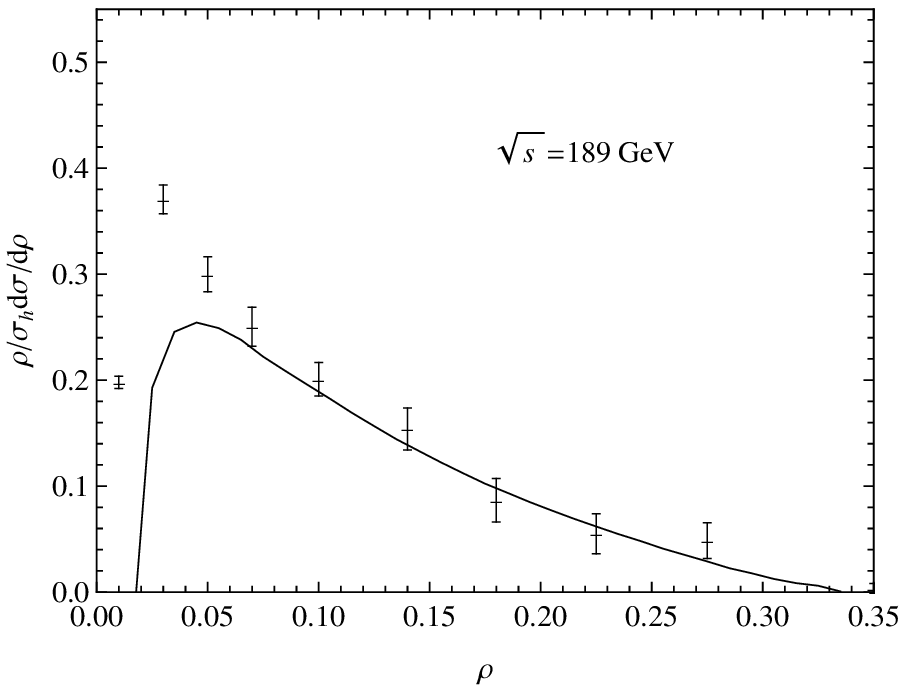}
\includegraphics[width=0.45\textwidth]{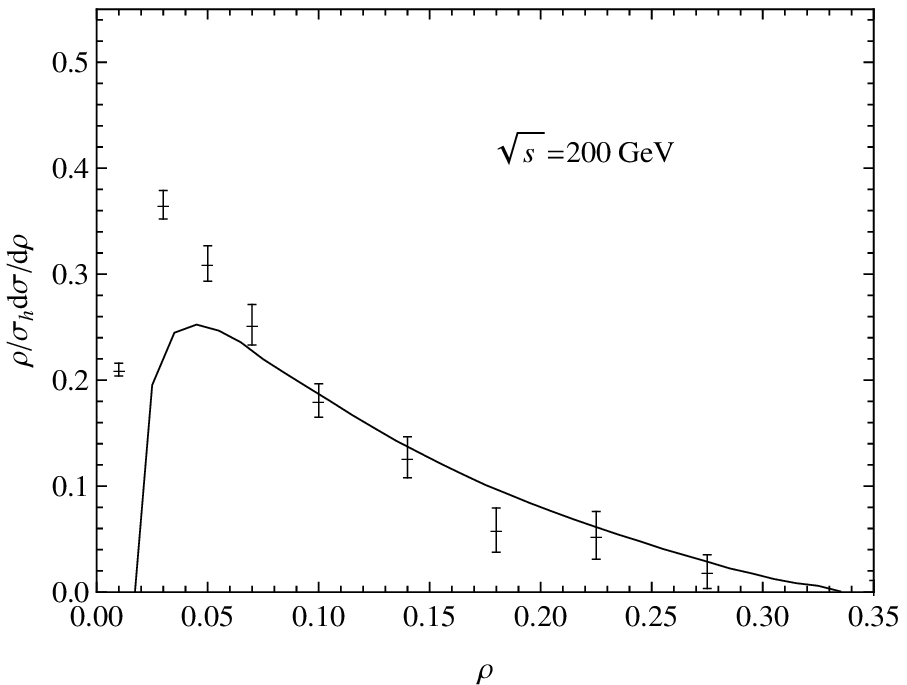}
\includegraphics[width=0.45\textwidth]{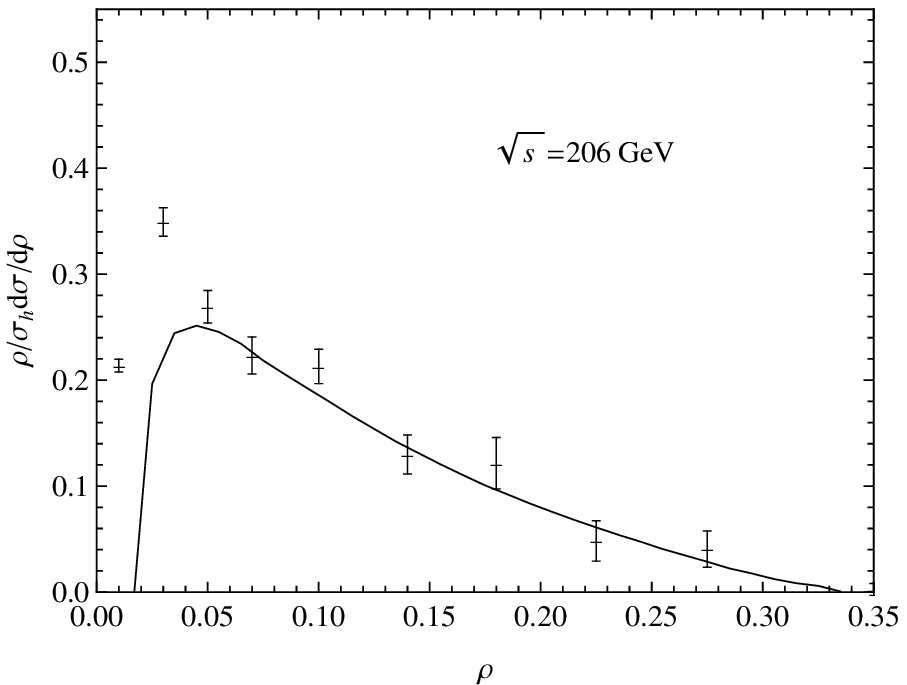}
\caption{The heavy jet mass ($\rho$) distributions using PMC scale setting for $\sqrt{s}=91.2$, $133$, $161$, $172$, $183$, $189$, $200$, $206$ GeV. The experimental data are taken from the ALEPH Collaboration~\cite{Heister:2003aj}.}
\label{distriPMCR}
\end{center}
\end{figure*}

\begin{figure*}
\begin{center}
\includegraphics[width=0.45\textwidth]{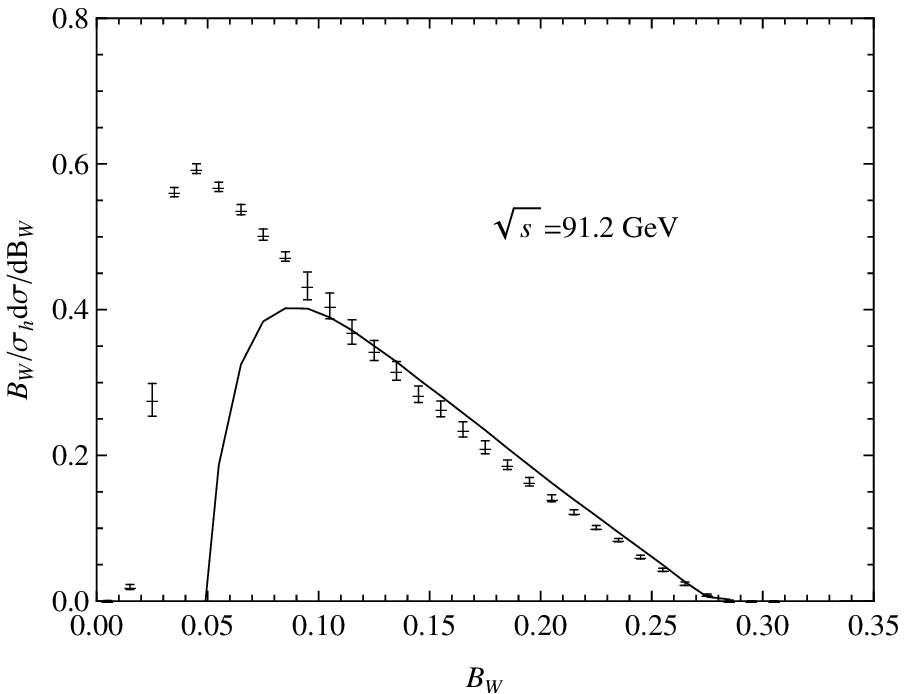}
\includegraphics[width=0.45\textwidth]{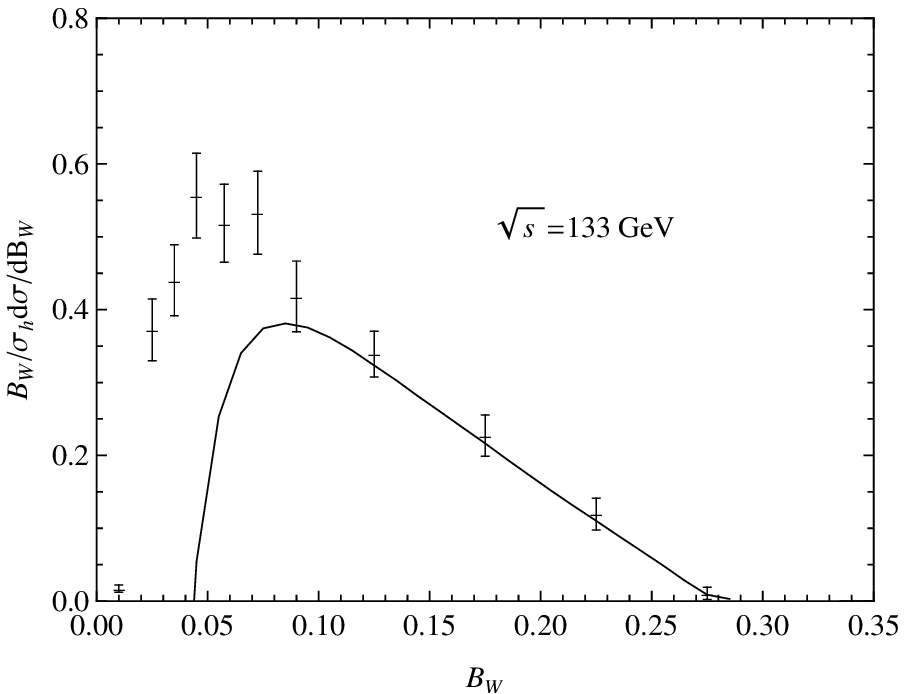}
\includegraphics[width=0.45\textwidth]{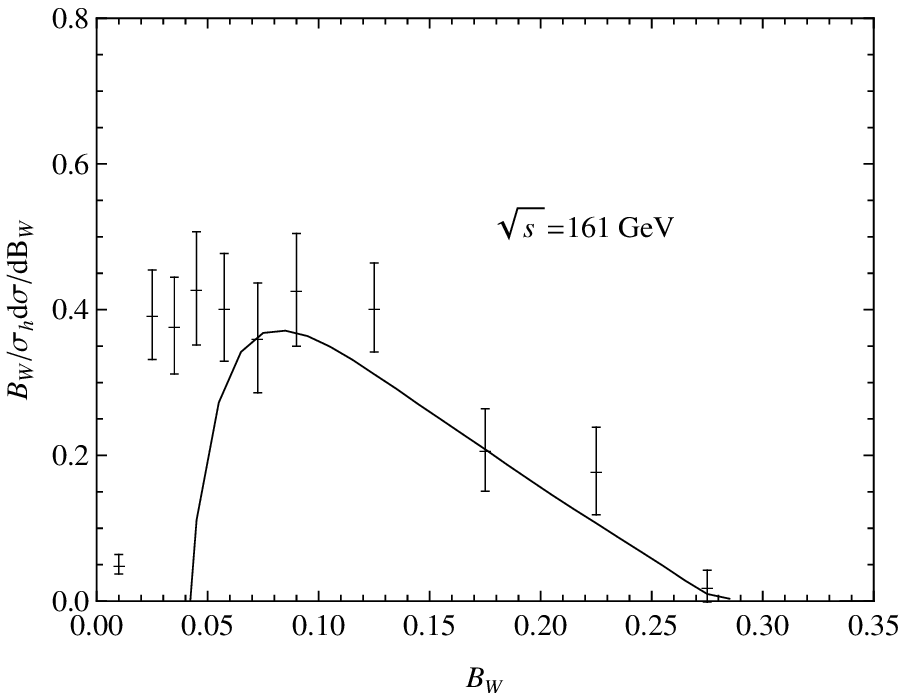}
\includegraphics[width=0.45\textwidth]{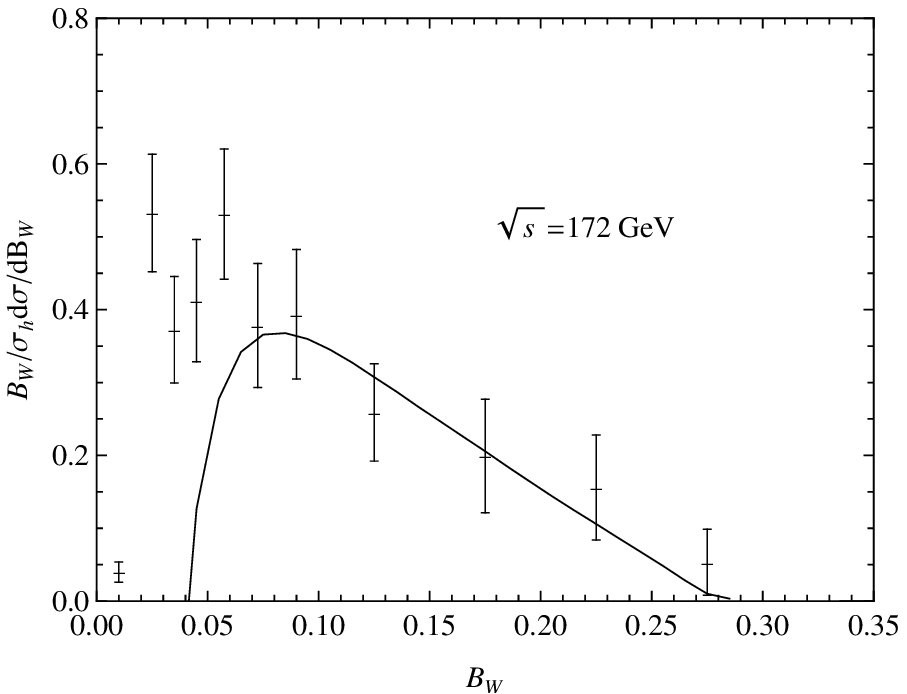}
\includegraphics[width=0.45\textwidth]{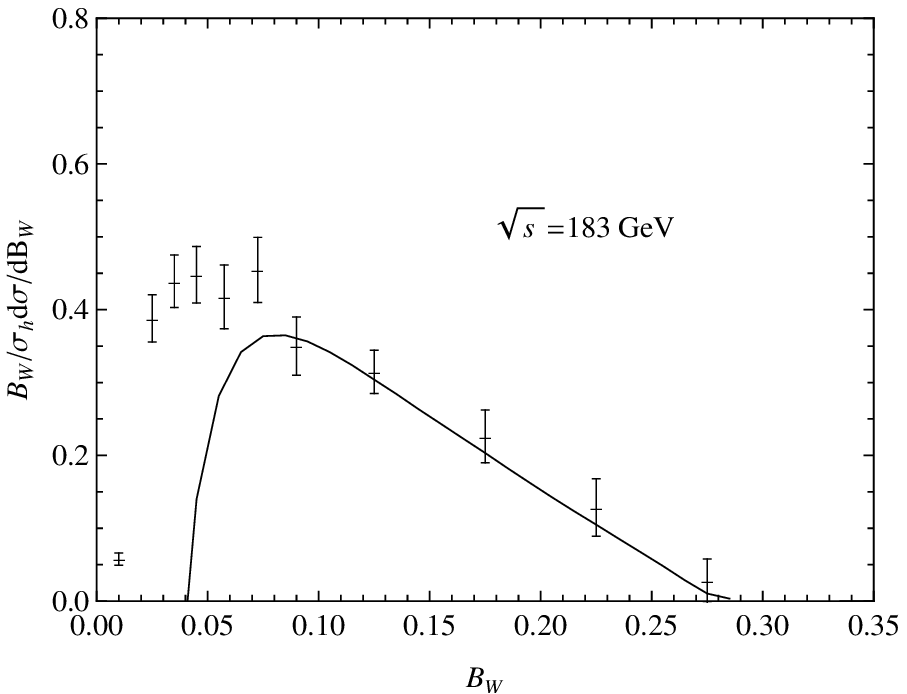}
\includegraphics[width=0.45\textwidth]{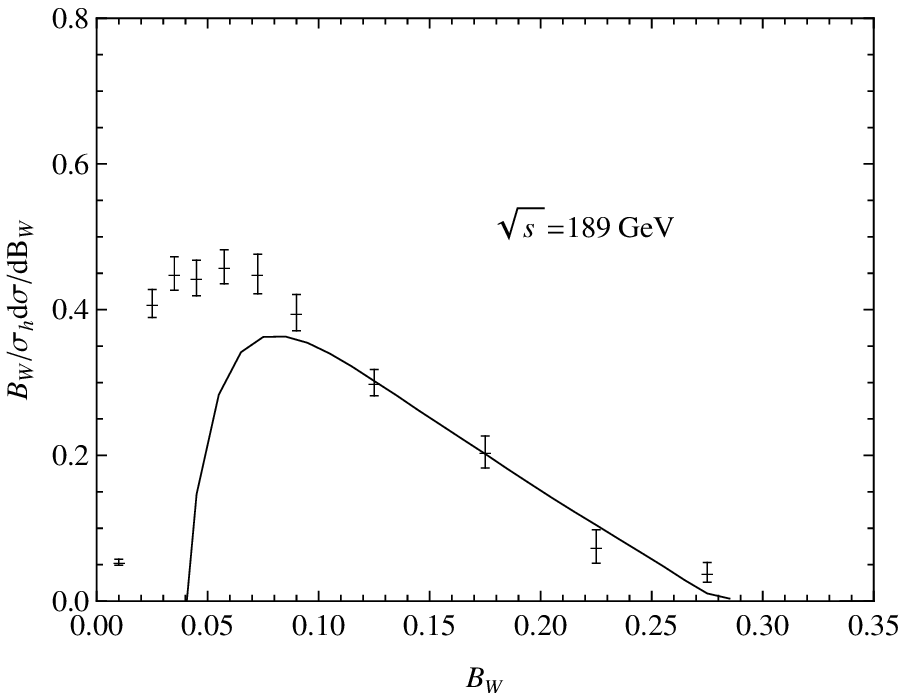}
\includegraphics[width=0.45\textwidth]{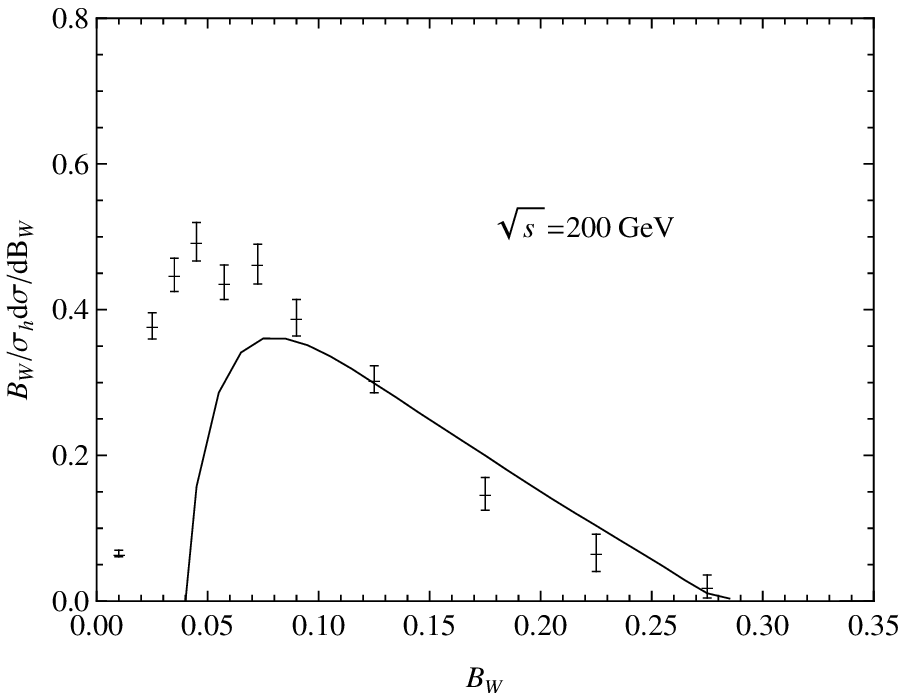}
\includegraphics[width=0.45\textwidth]{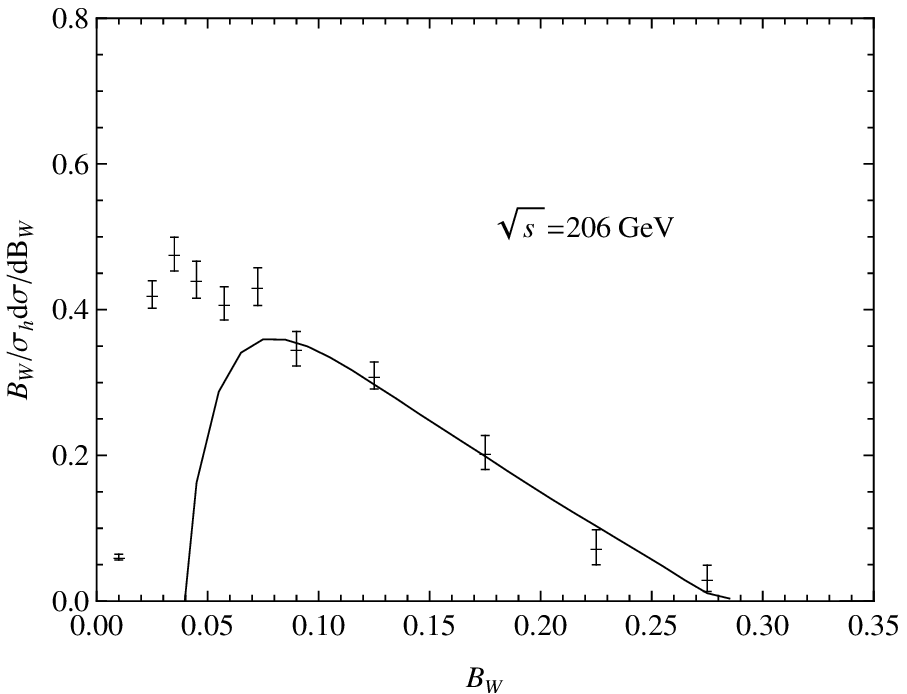}
\caption{The wide jet broadening ($B_W$) distributions using PMC scale setting for $\sqrt{s}=91.2$, $133$, $161$, $172$, $183$, $189$, $200$, $206$ GeV. The experimental data are taken from the ALEPH Collaboration~\cite{Heister:2003aj}.}
\label{distriPMCBW}
\end{center}
\end{figure*}

\begin{figure*}
\begin{center}
\includegraphics[width=0.45\textwidth]{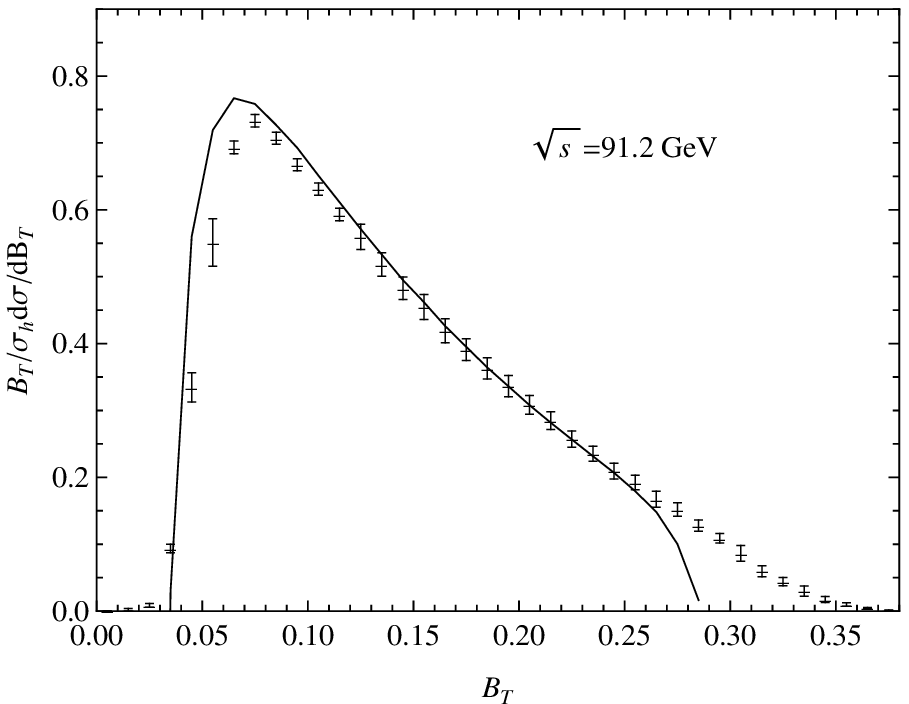}
\includegraphics[width=0.45\textwidth]{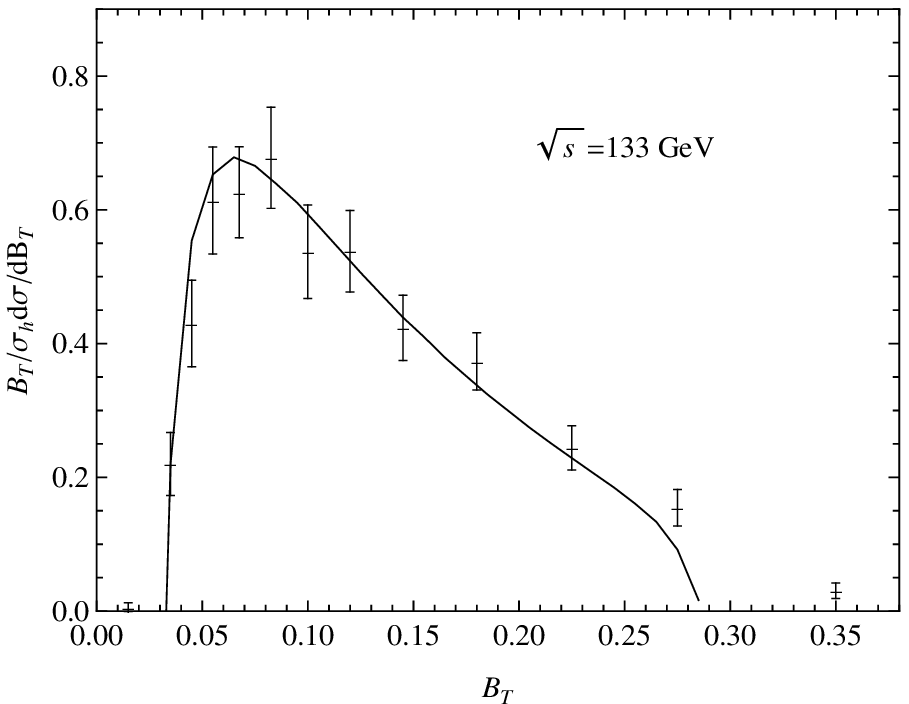}
\includegraphics[width=0.45\textwidth]{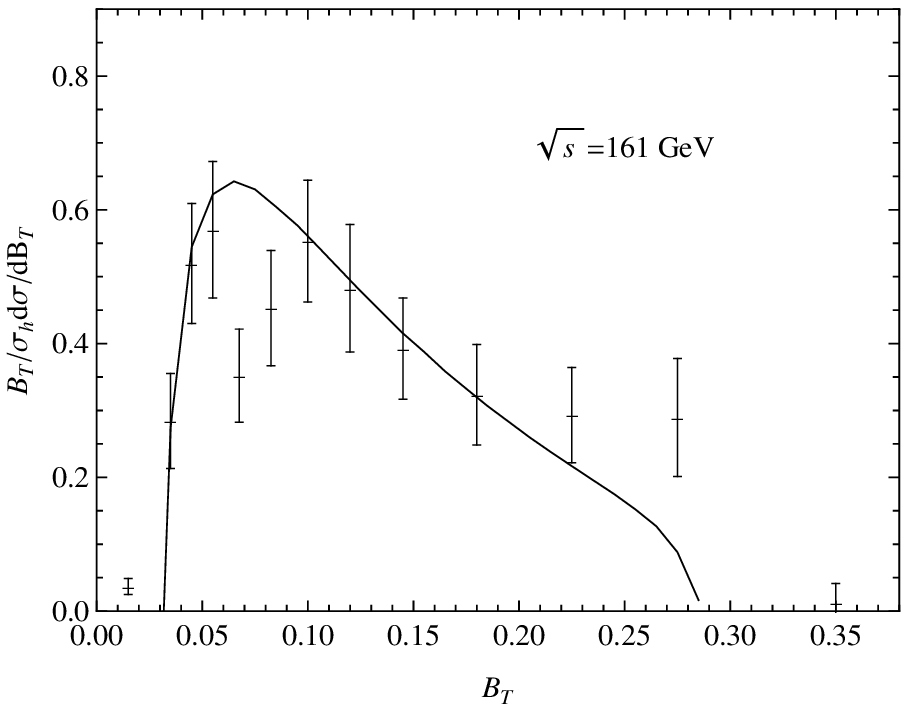}
\includegraphics[width=0.45\textwidth]{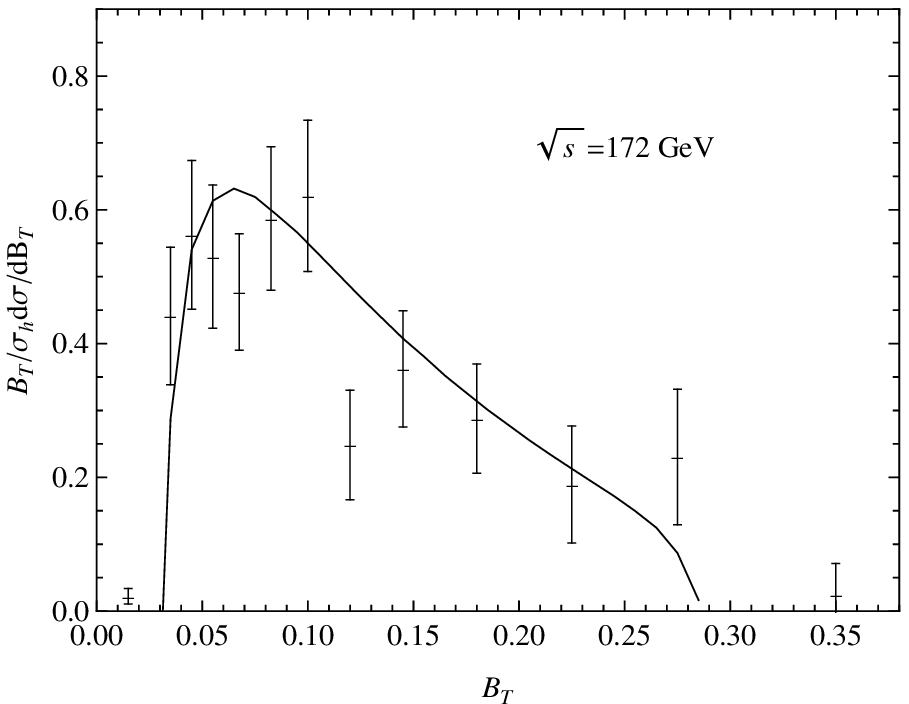}
\includegraphics[width=0.45\textwidth]{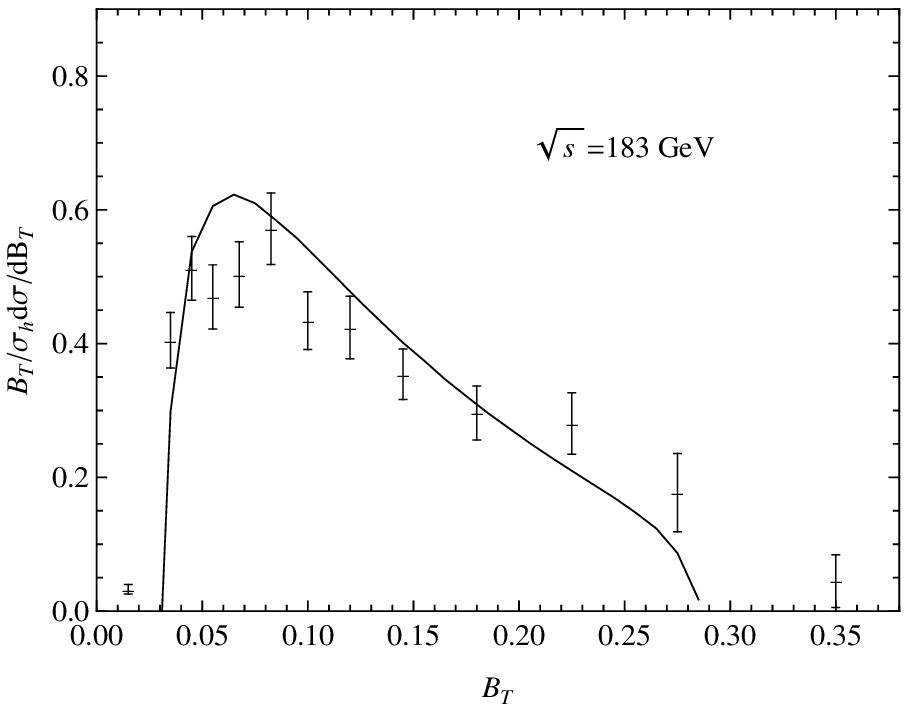}
\includegraphics[width=0.45\textwidth]{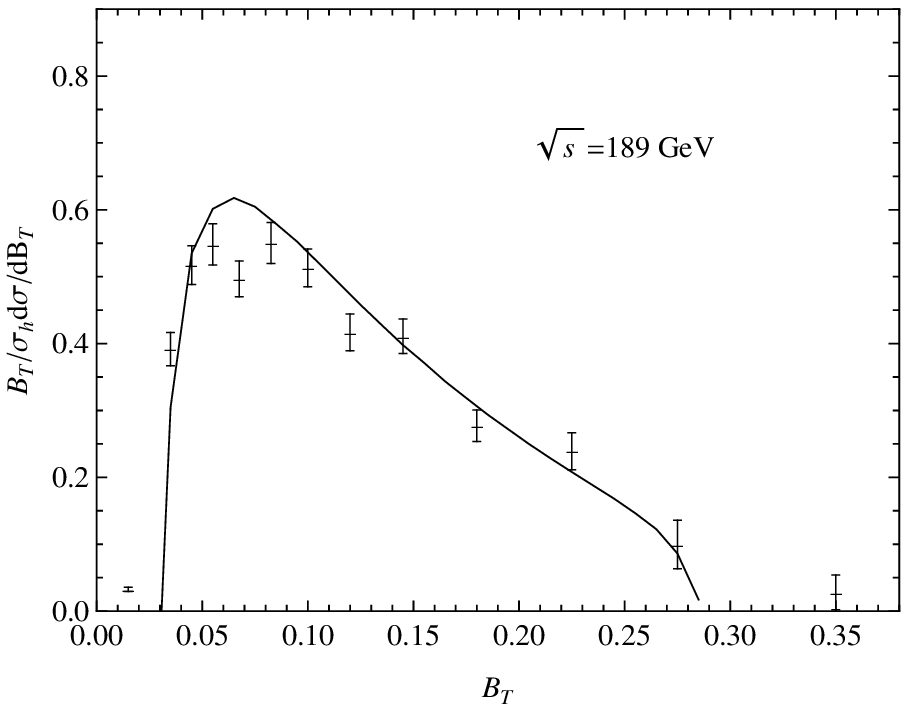}
\includegraphics[width=0.45\textwidth]{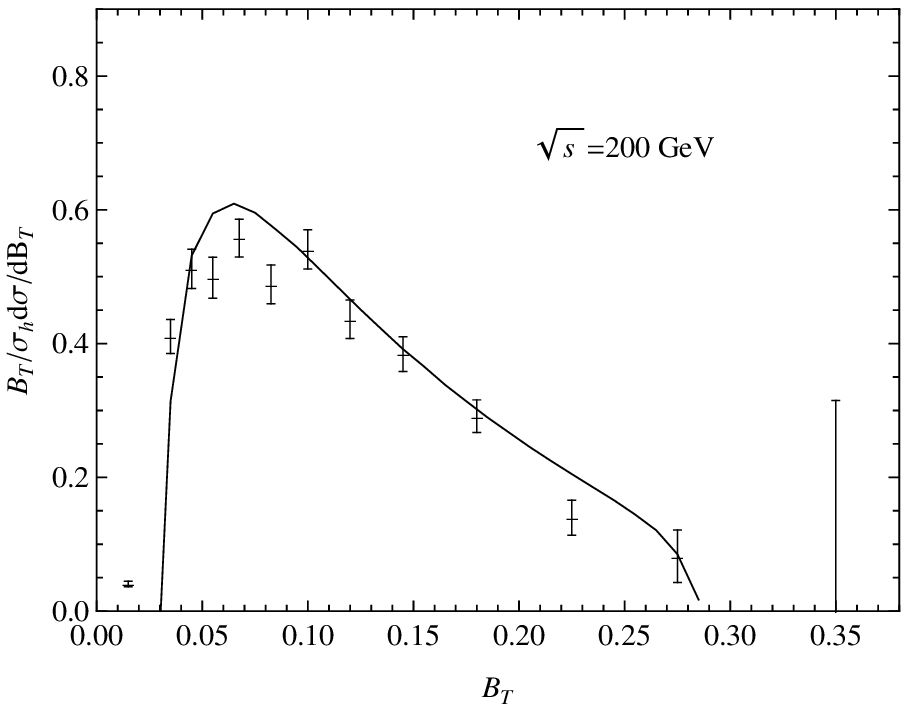}
\includegraphics[width=0.45\textwidth]{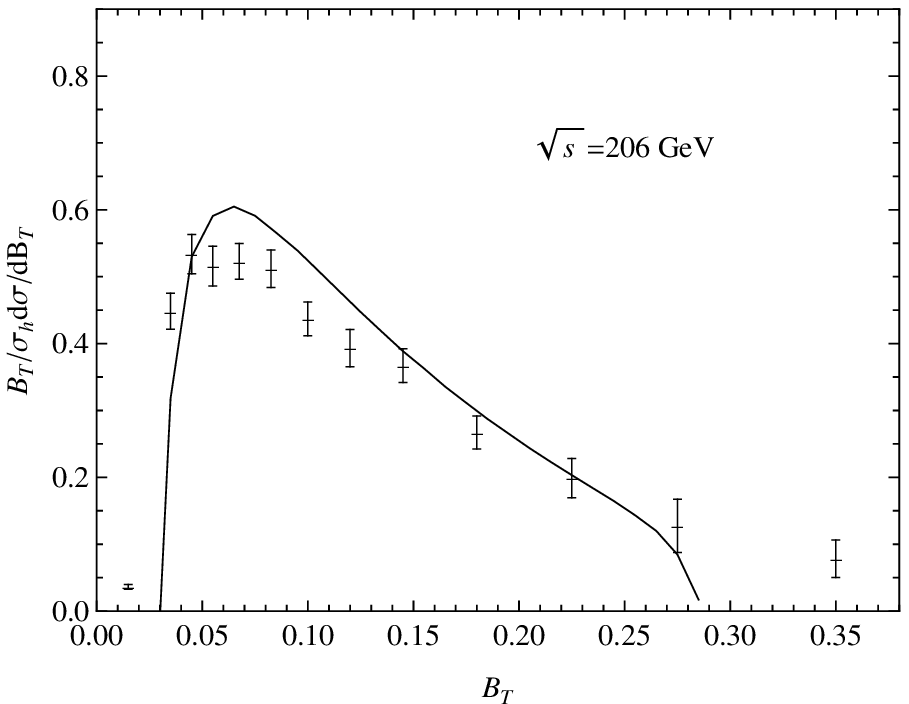}
\caption{The total jet broadening ($B_T$) distributions using PMC scale setting for $\sqrt{s}=91.2$, $133$, $161$, $172$, $183$, $189$, $200$, $206$ GeV. The experimental data are taken from the ALEPH Collaboration~\cite{Heister:2003aj}.}
\label{distriPMCcBT}
\end{center}
\end{figure*}

\begin{figure*}
\begin{center}
\includegraphics[width=0.45\textwidth]{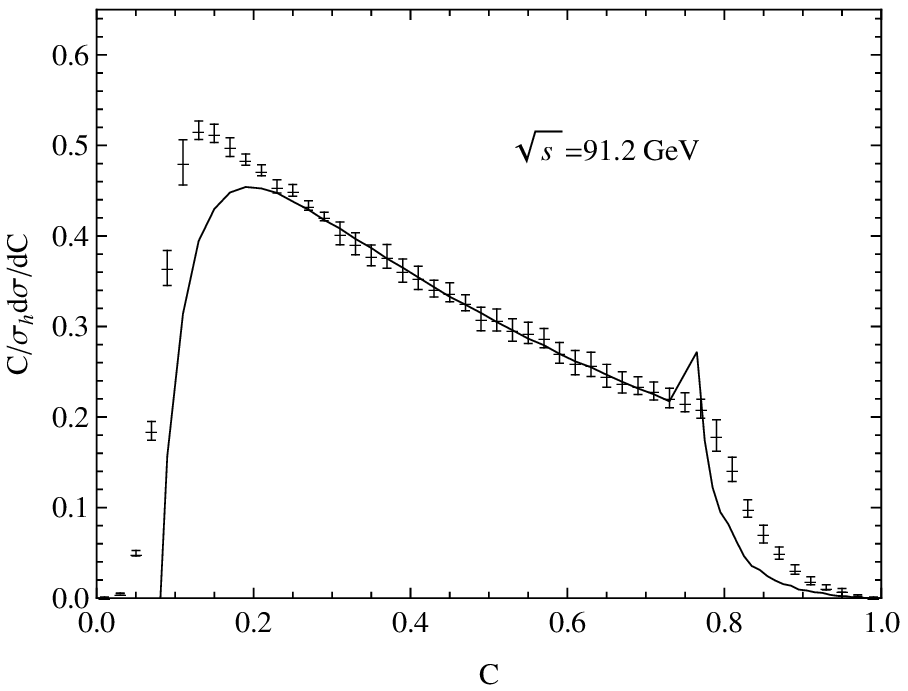}
\includegraphics[width=0.45\textwidth]{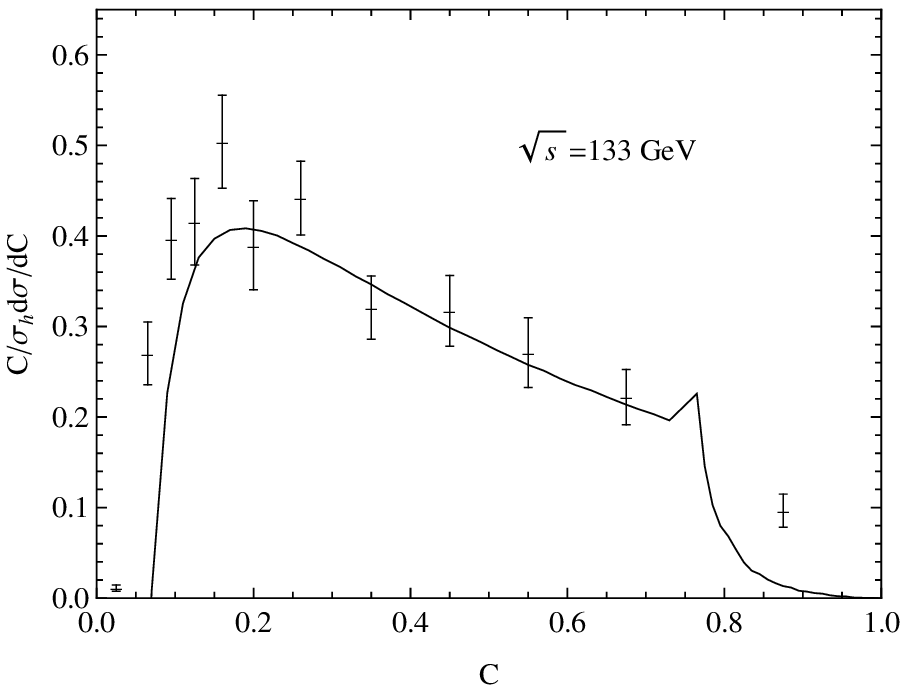}
\includegraphics[width=0.45\textwidth]{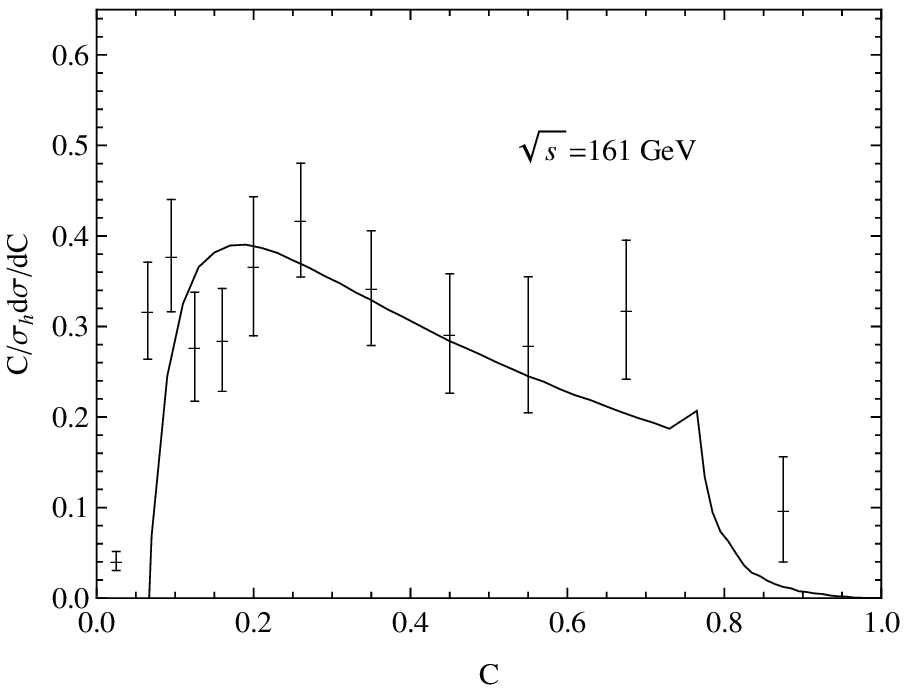}
\includegraphics[width=0.45\textwidth]{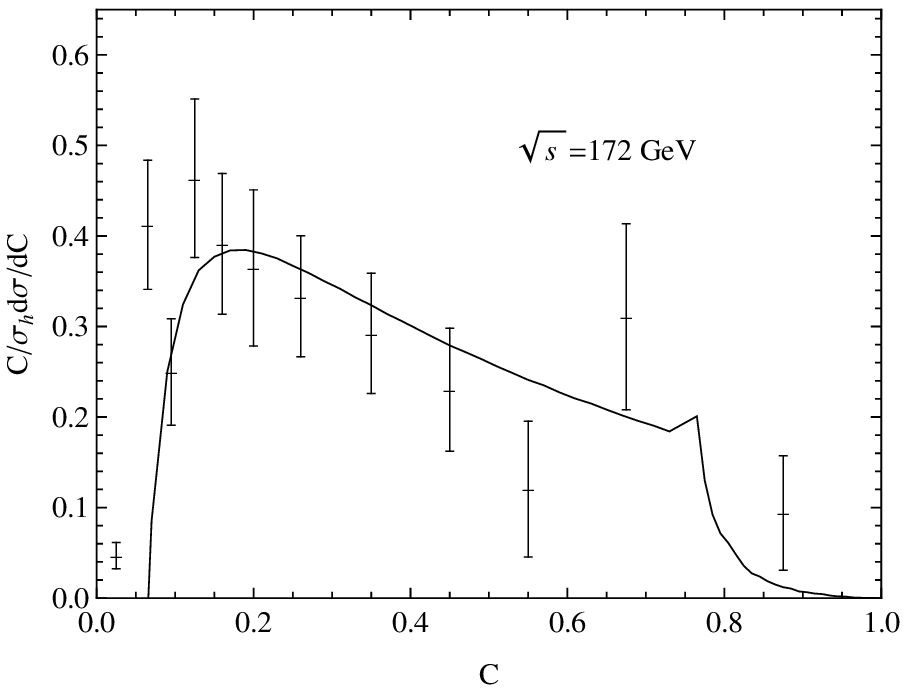}
\includegraphics[width=0.45\textwidth]{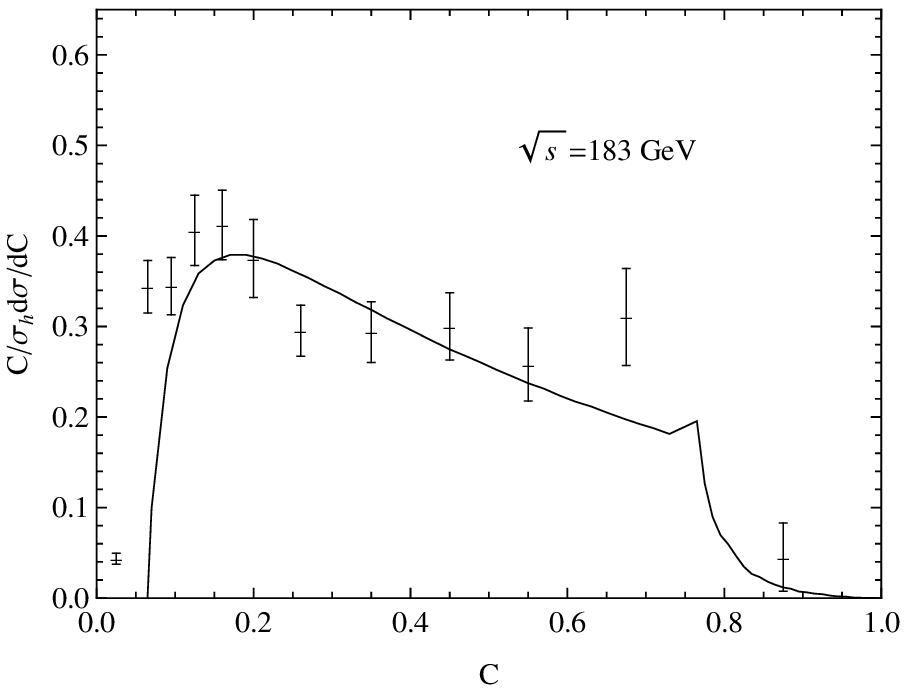}
\includegraphics[width=0.45\textwidth]{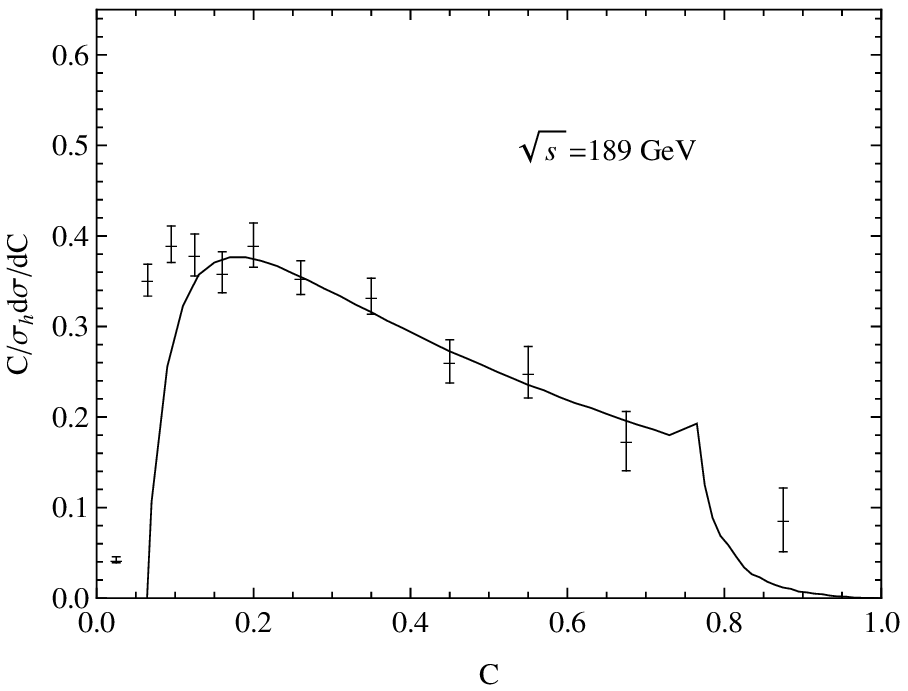}
\includegraphics[width=0.45\textwidth]{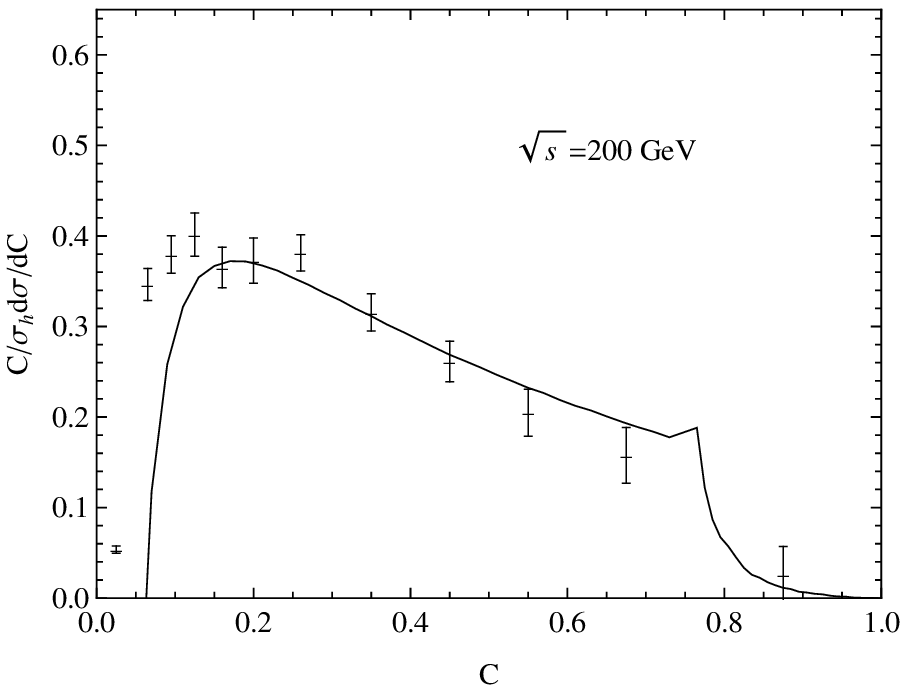}
\includegraphics[width=0.45\textwidth]{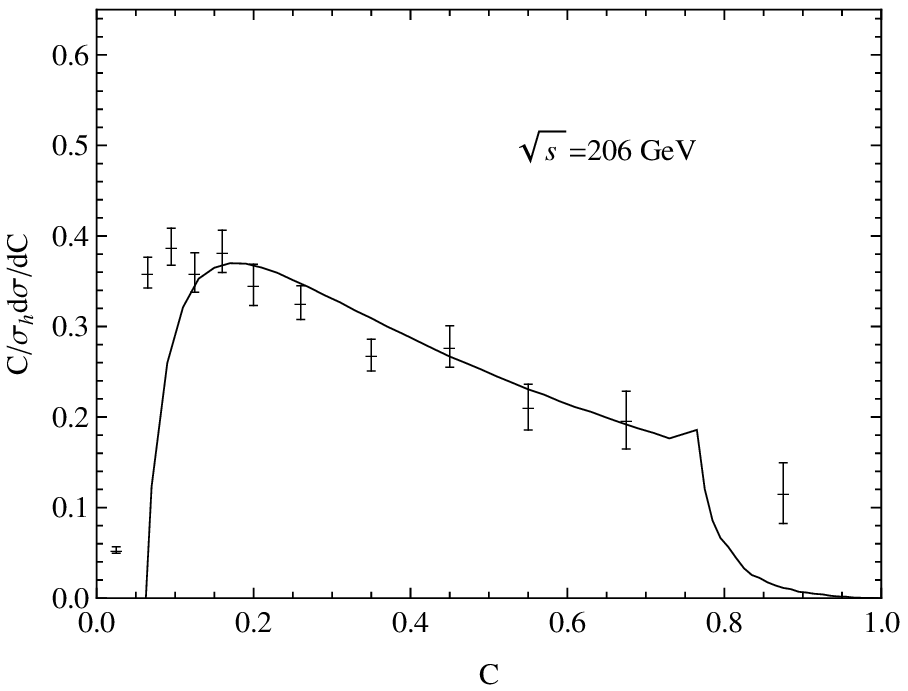}
\caption{The $C$-parameter ($C$) distributions using PMC scale setting for $\sqrt{s}=91.2$, $133$, $161$, $172$, $183$, $189$, $200$, $206$ GeV. The experimental data are taken from the ALEPH Collaboration~\cite{Heister:2003aj}.}
\label{distriPMCc}
\end{center}
\end{figure*}

In addition to the PMC scales, the behavior of the PMC conformal coefficients is very different from that of the conventional scale-setting method. Thus, the resulting PMC predictions are increased in wide kinematic regions compared to the conventional predictions. Since the PMC scales are independent of the choice of renormalization scale $\mu_r$ and the conformal coefficients are also renormalization scale-independent, the PMC predictions eliminate the renormalization scale uncertainty. By setting all input parameters to be their central values, the event shape distributions using PMC scale setting for the thrust ($1-T$), heavy jet mass ($\rho$), wide jet broadening ($B_W$), total jet broadening ($B_T$) and the C-parameter ($C$) are presented in Figs.(\ref{distriPMCT}) to (\ref{distriPMCc}). Each event shape distribution includes the results for $\sqrt{s}=91.2$, $133$, $161$, $172$, $183$, $189$, $200$, $206$ GeV, which have been measured at LEP experiment~\cite{Heister:2003aj}. Event shape distributions have been measured with a high precision, especially at $\sqrt{s}=91.2$ GeV. However, for the measurements at $\sqrt{s}=161$ and $172$ GeV, the experimental data have relatively large uncertainties.

These figures show that the PMC predictions are in agreement with the experimental data in wide intermediate kinematic regions, especially at $\sqrt{s}=91.2$ GeV. There is only a minor deviation in the intermediate region for the wide jet broadening ($B_W$) at $\sqrt{s}=91.2$ GeV, which is shown by Fig.(\ref{distriPMCBW}). For all the event shape observables, there are some deviations near the two-jet and multijet regions, which is expected since there are large logarithms that spoil the perturbative regime of the QCD. The resummation of large logarithms is thus required for the PMC results especially near the two-jet and multijet regions. In fact, the resummation of large logarithms has been extensively studied in the literature.

Due to kinematical constraints, the domain of event shape distributions at LO is restricted to certain ranges, e.g., $0\leq(1-T)\leq1/3$ for the thrust and $0\leq C\leq0.75$ for the C-parameter. The domain of the LO PMC scales using Eq.(\ref{evenPMCscale}) is also restricted in these ranges. In the range of $A(y) = 0$, e.g., $1/3\leq(1-T)\leq1/2$ for the thrust and $0.75\leq C\leq1$ for the C-parameter, a self-consistent PMC analysis can be obtained by using at least NNLO calculations. In our present analysis, we do not calculate the distributions in the range of $A(y) = 0$ for the thrust ($T$), heavy jet mass ($\rho=M^2_H/s$), wide jet broadening ($B_W$), total jet broadening ($B_T$). For the C-parameter, the NNLO calculations in the range of $A(y) = 0$ (i.e., $0.75\leq C\leq1$) are relatively stable, we can thus determine its NLO PMC scale in the range of $0.75\leq C\leq1$. We observed that similar to the case of the C-parameter distribution in the range of $0\leq C\leq0.75$, the PMC scale is not a single value and changes dynamically with the value of the C-parameter. Compared to the conventional method of fixing $\mu_r=\sqrt{s}$, the PMC scale is also small in the range of $0.75\leq C\leq1$. The resulting PMC predictions are also increased, which leads PMC results to be closer to the experimental data in $0.75\leq C\leq1$ range, as shown by Fig.(\ref{distriPMCc}).

After applying the PMC, although the logarithmic term $\ln(\mu^2_r/s)$ disappears, there are other types of logarithmic terms that appear in the conformal series. For example, at leading order the event shape observable $y$ has the logarithmic term $-\ln(y)/y$, and its behavior is $-\ln(y)/y\rightarrow +\infty$ for $y\rightarrow 0$ in the two-jet region. It is noted that the behavior of non-conformal terms and conformal terms is very different near the two-jet region. By taking the thrust ($1-T$) and the C-parameter ($C$) as the examples, we present the absolute value of the non-conformal terms $\bar{B}(y,\mu_r)_{\beta_0}\cdot\beta_0$ and conformal terms $\bar{B}(y,\mu_r)_{\rm con}$ in Figs.(\ref{connonconT}) and (\ref{connonconC}). We can see from Figs.(\ref{connonconT}) and (\ref{connonconC}) that compared to the non-conformal terms $\bar{B}(y,\mu_r)_{\beta_0}\cdot\beta_0$, the conformal terms $\bar{B}(y,\mu_r)_{\rm con}$ are relatively stable as the $(1-T)$ and the $C$ decrease in the intermediate region. In the wider region of the two-jet region, the magnitude of the non-conformal terms $\bar{B}(y,\mu_r)_{\beta_0}\cdot\beta_0$ is much larger than the that of the conformal terms $\bar{B}(y,\mu_r)_{\rm con}$. Although the conformal terms are relatively stable, they are infinite for $(1-T)\rightarrow 0$ and $C\rightarrow 0$ due to the presence of large logarithms. The behavior of the non-conformal terms is $\bar{B}(y,\mu_r)_{\beta_0}\cdot\beta_0\rightarrow +\infty$ for $(1-T)\rightarrow 0$ and $C\rightarrow 0$, and the conformal terms is $\bar{B}(y,\mu_r)_{\rm con}\rightarrow -\infty$ for $(1-T)\rightarrow 0$ and $C\rightarrow 0$.

\subsection{Event shape distributions below $Z^0$ peak }

\begin{figure*}
\begin{center}
\includegraphics[width=0.45\textwidth]{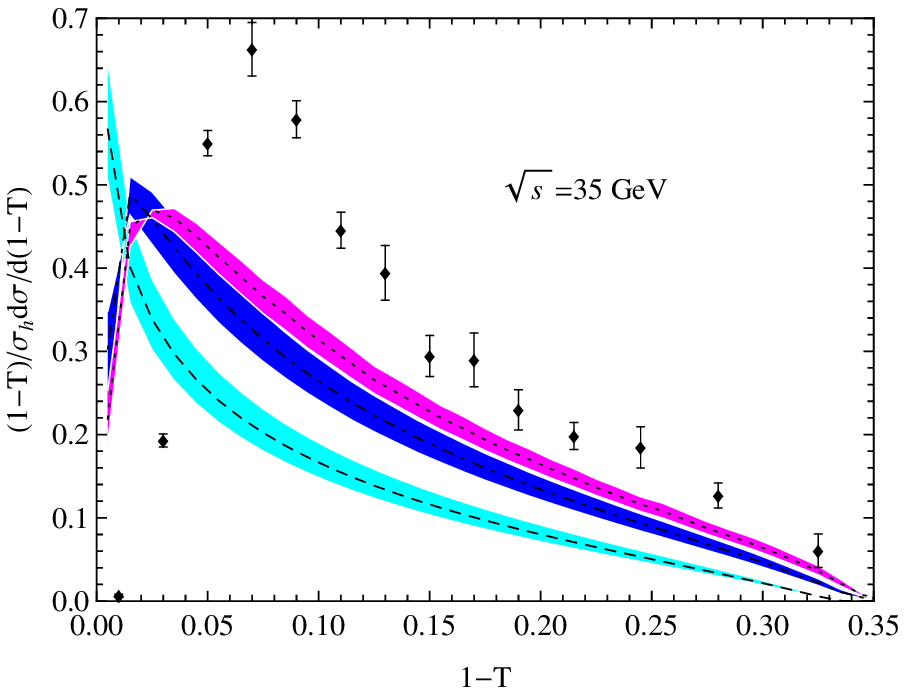}
\includegraphics[width=0.45\textwidth]{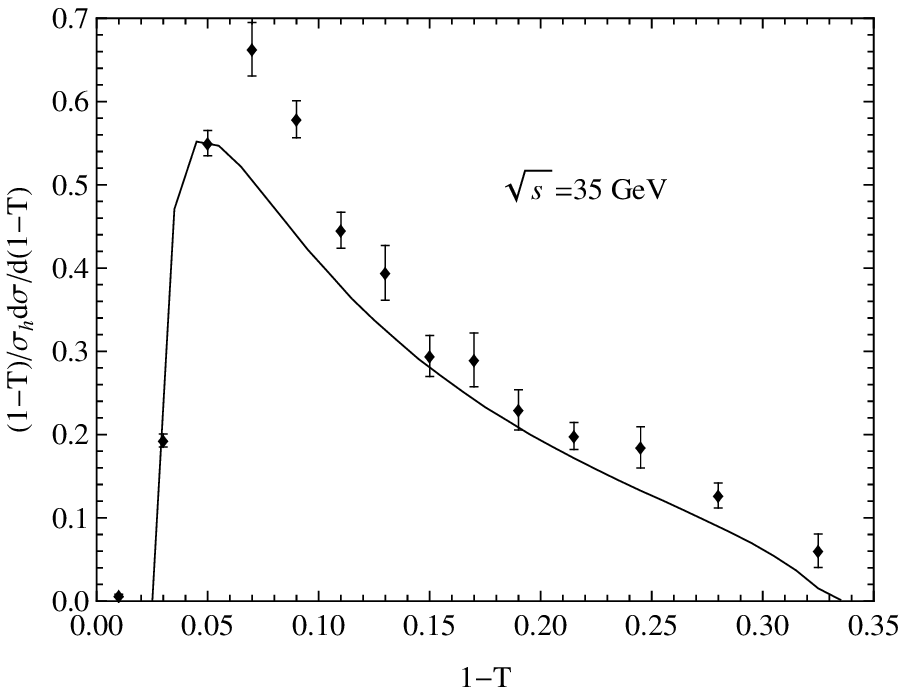}
\includegraphics[width=0.45\textwidth]{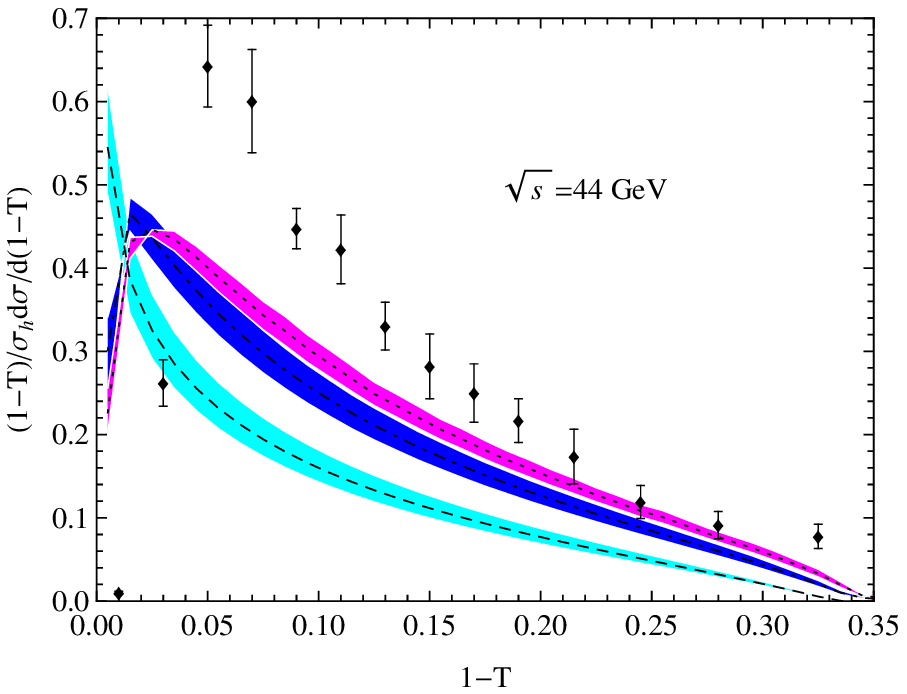}
\includegraphics[width=0.45\textwidth]{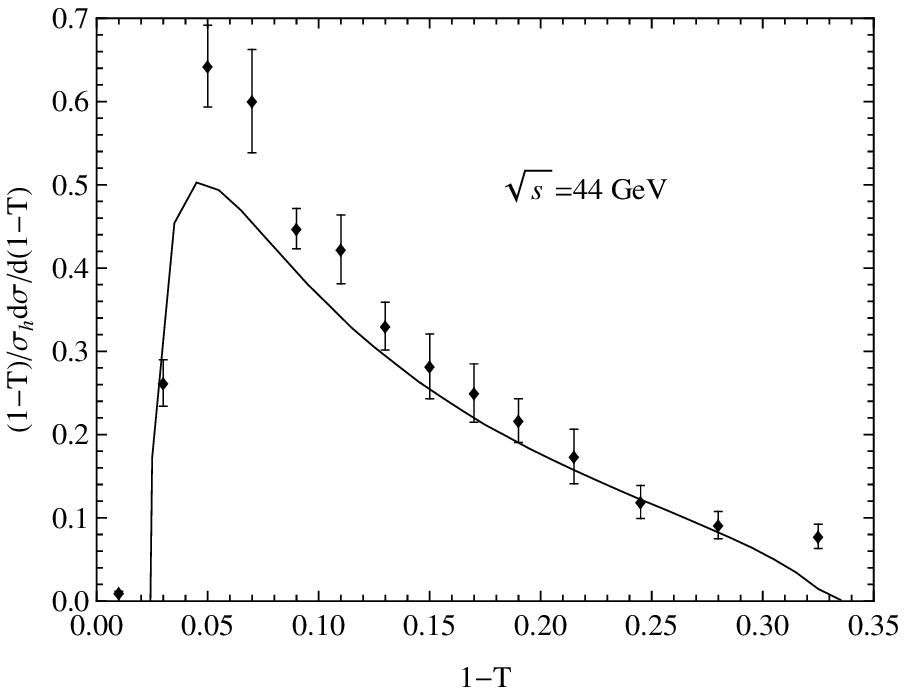}
\includegraphics[width=0.45\textwidth]{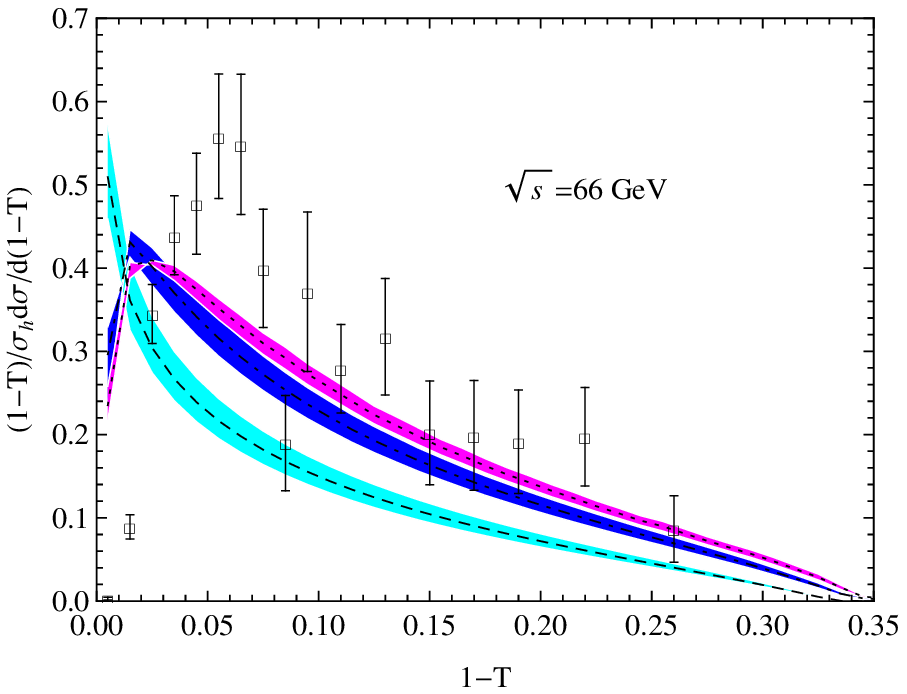}
\includegraphics[width=0.45\textwidth]{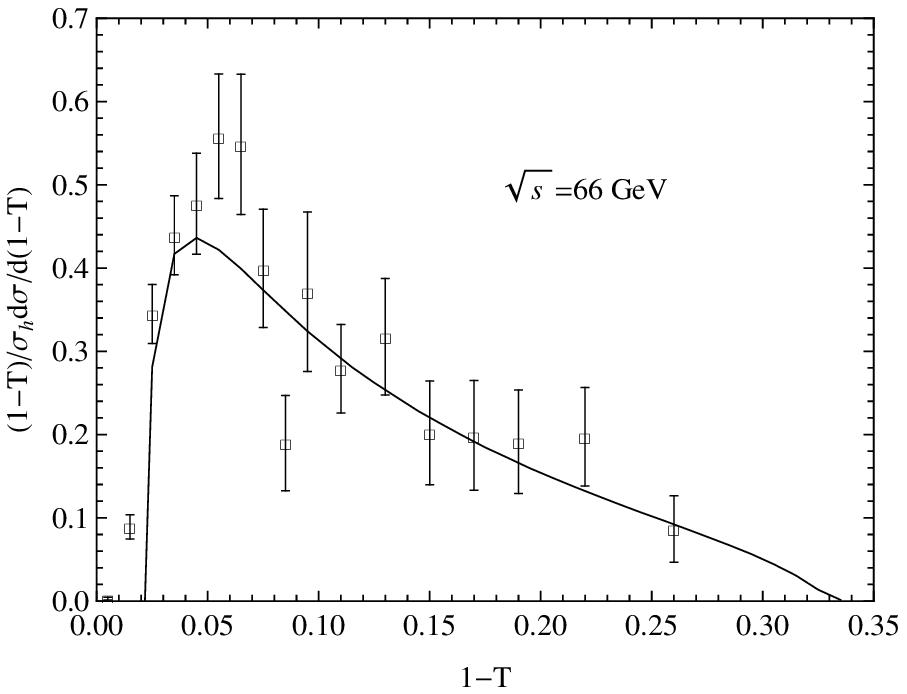}
\includegraphics[width=0.45\textwidth]{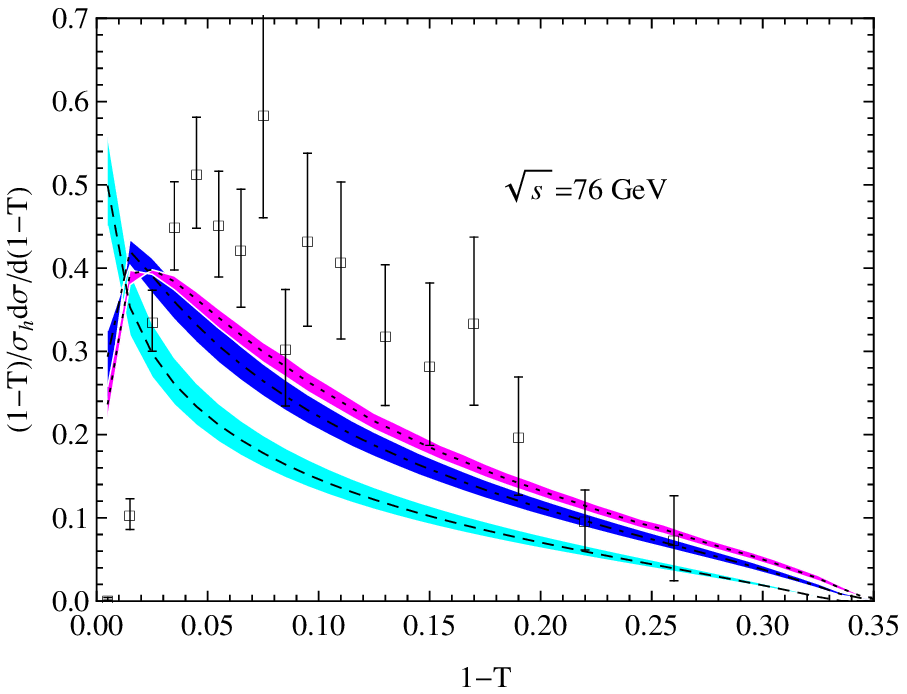}
\includegraphics[width=0.45\textwidth]{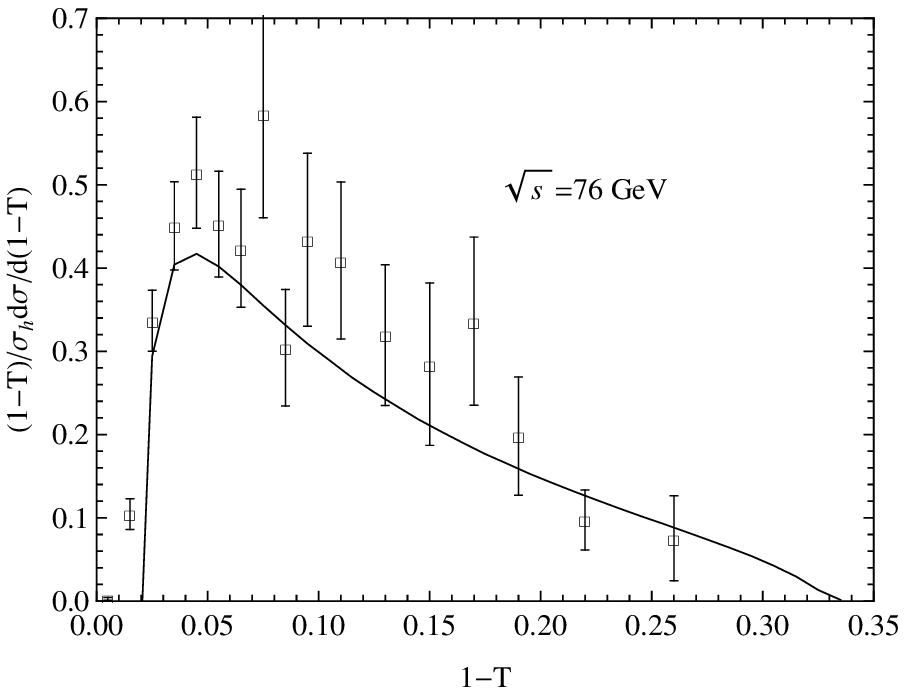}
\caption{The thrust ($1-T$) distributions using conventional (left column) and PMC (right column) scale-settings for $\sqrt{s}=35$, $44$, $66$, $76$ GeV. The experimental data for $\sqrt{s}=35$, $44$ GeV are taken from the JADE~\cite{MovillaFernandez:1997fr,Biebel:1999zt}, and $\sqrt{s}=66$, $76$ GeV data are taken from the DELPHI~\cite{Abdallah:2003xz} Collaboration.}
\label{distriTbelZ}
\end{center}
\end{figure*}

\begin{figure*}
\begin{center}
\includegraphics[width=0.45\textwidth]{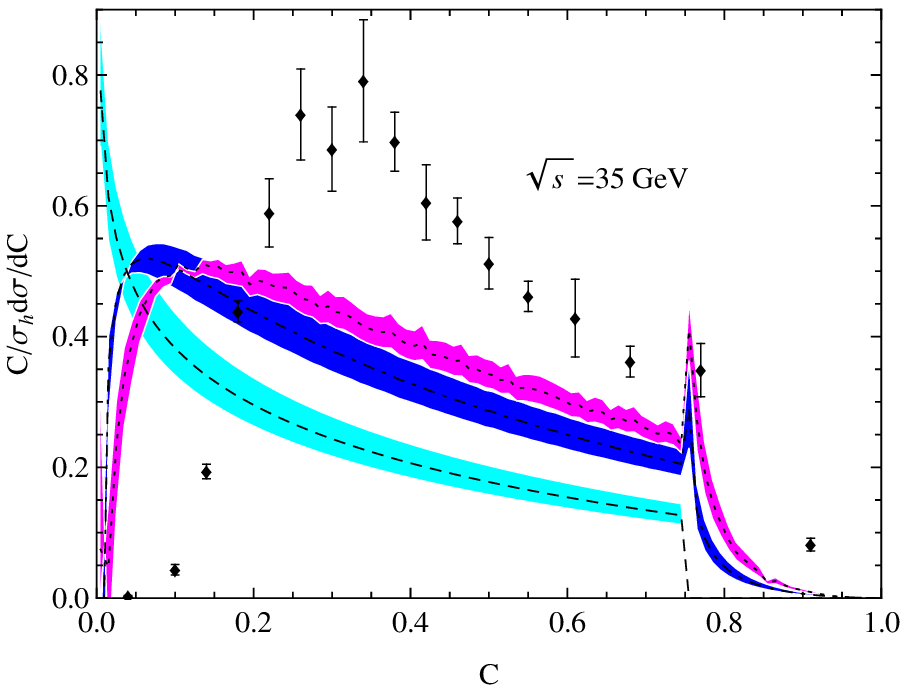}
\includegraphics[width=0.45\textwidth]{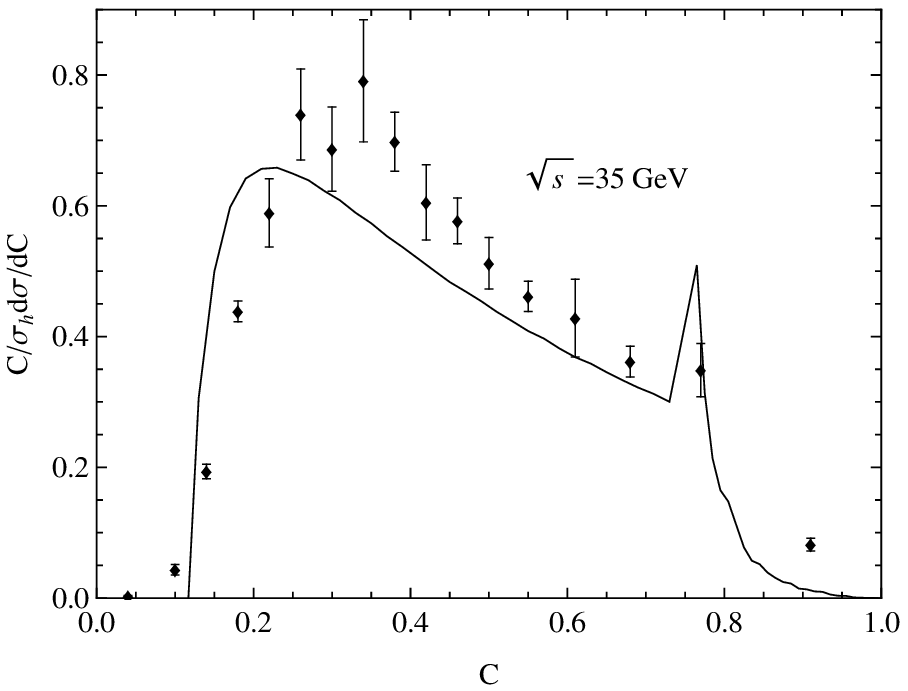}
\includegraphics[width=0.45\textwidth]{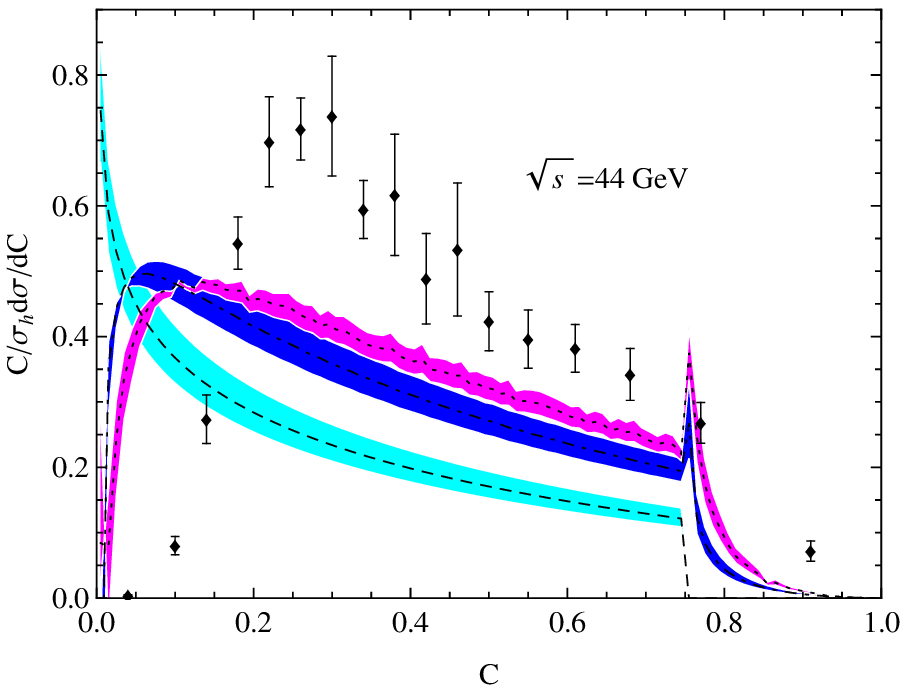}
\includegraphics[width=0.45\textwidth]{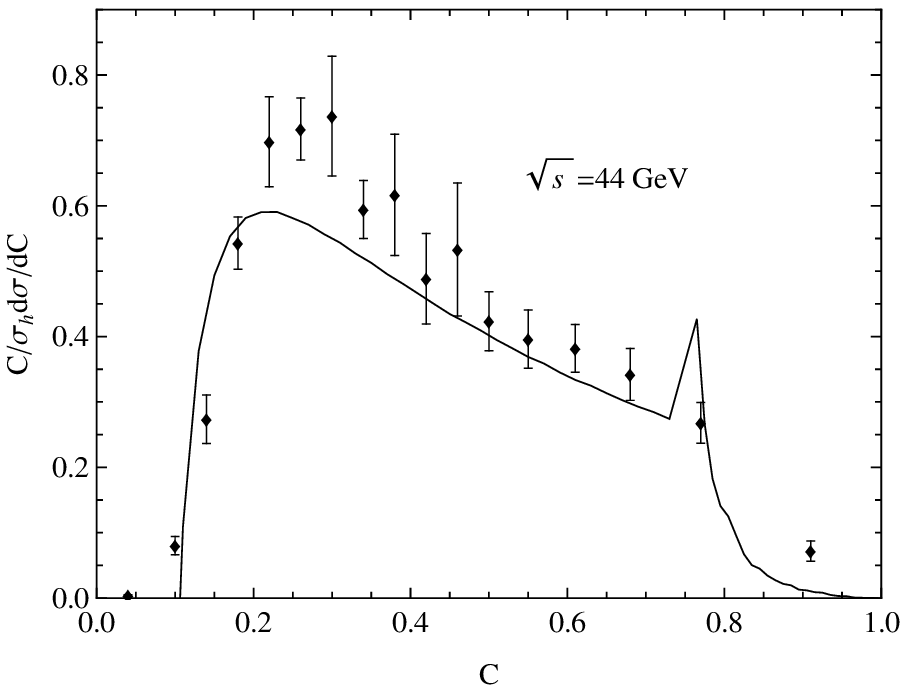}
\includegraphics[width=0.45\textwidth]{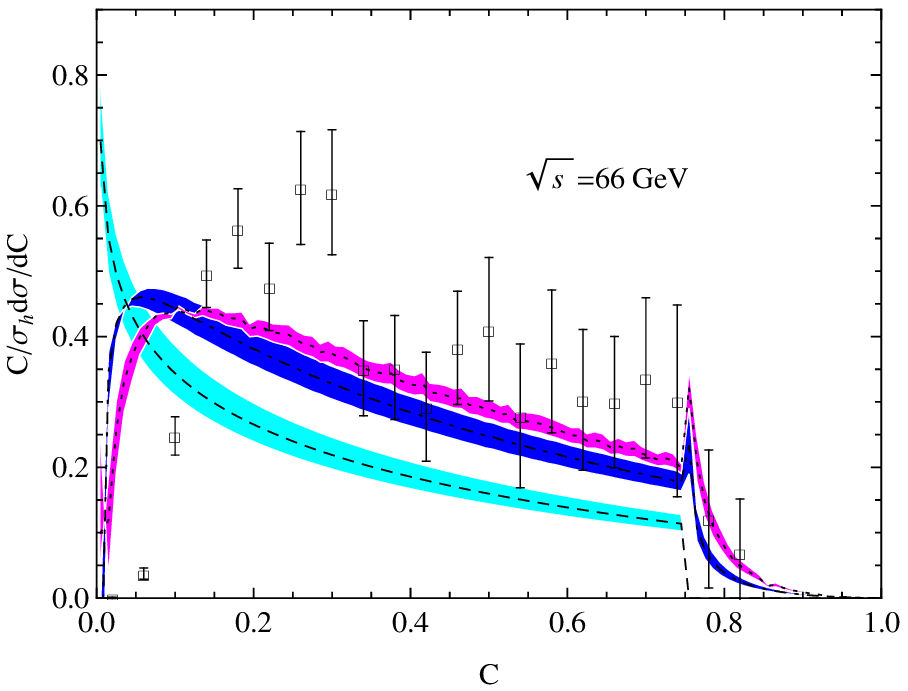}
\includegraphics[width=0.45\textwidth]{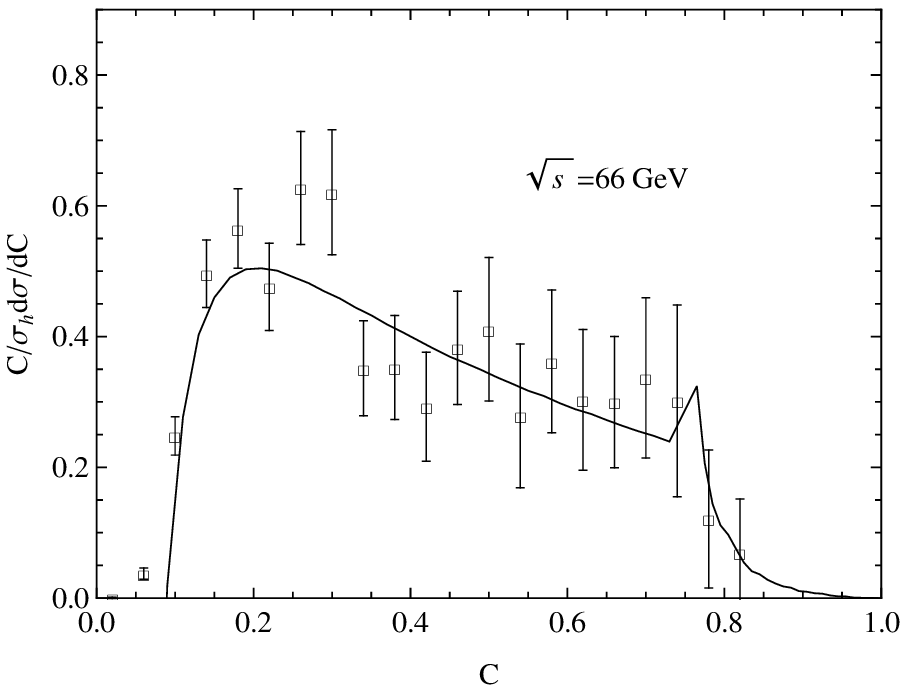}
\includegraphics[width=0.45\textwidth]{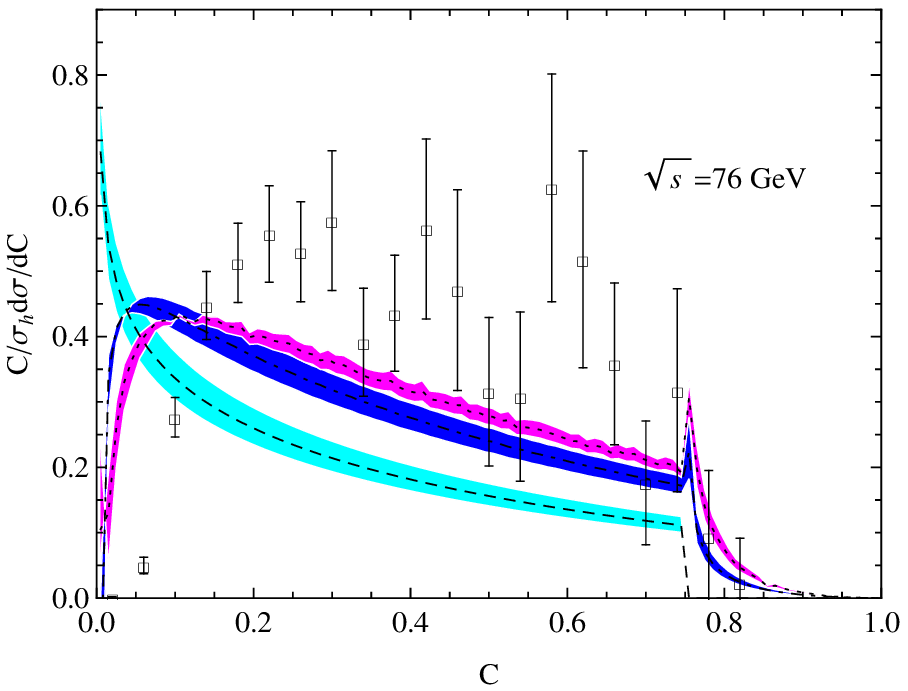}
\includegraphics[width=0.45\textwidth]{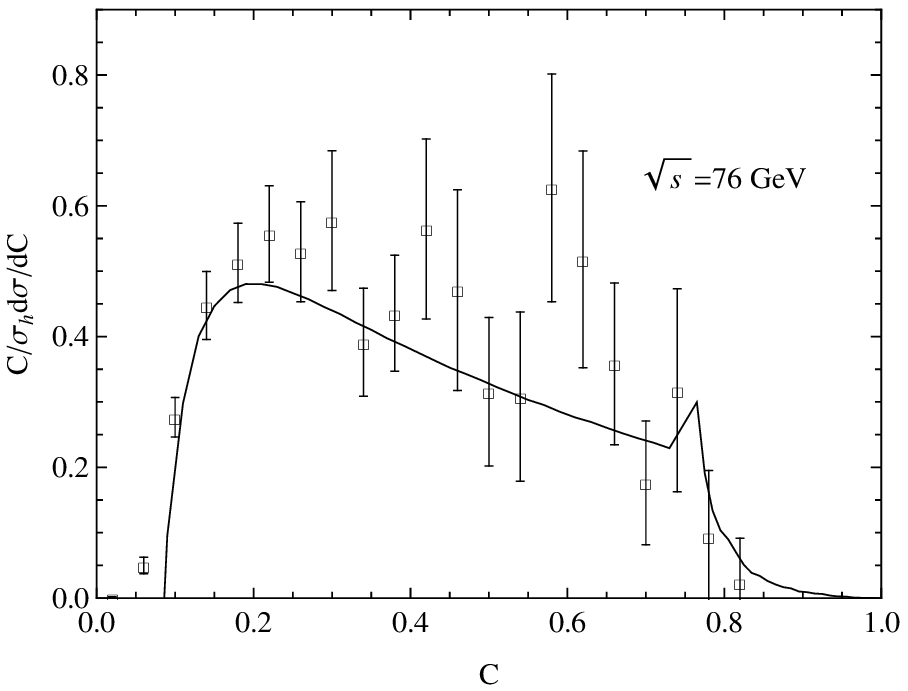}
\caption{The $C$-parameter ($C$) distributions using conventional (left column) and PMC (right column) scale-settings for $\sqrt{s}=35$, $44$, $66$, $76$ GeV. The experimental data for $\sqrt{s}=35$, $44$ GeV are taken from the JADE~\cite{MovillaFernandez:1997fr,Biebel:1999zt}, and $\sqrt{s}=66$, $76$ GeV data are taken from the DELPHI~\cite{Abdallah:2003xz} Collaboration. }
\label{distriCbelZ}
\end{center}
\end{figure*}

Event shape observables have been measured over a wide range of $\sqrt{s}$. As mentioned above, the PMC predictions are in agreement with the experimental data measured above $Z^0$ peak. The lower energy data below $Z^0$ peak for event shape observables are widely adopted to determine the QCD running coupling $\alpha_s$. It is of interest to calculate event shape distributions below $Z^0$ peak using PMC scale setting. It is noted that both the thrust ($T$) and the C-parameter ($C$) are measured at JADE for $\sqrt{s}=35$, $44$ GeV~\cite{MovillaFernandez:1997fr,Biebel:1999zt} and DELPHI~\cite{Abdallah:2003xz}. By taking the thrust ($T$) and the C-parameter ($C$) as examples, we calculate the corresponding distributions and present the results using conventional and PMC scale-settings in Figs.(\ref{distriTbelZ}) and (\ref{distriCbelZ}). The experimental data have relatively large uncertainties for the measurements at DELPHI Collaboration~\cite{Abdallah:2003xz}; we thus only present the distributions for $\sqrt{s}=66$ and $76$ GeV, which shows that both the conventional results and the PMC predictions are in agreement with the DELPHI data.

For $\sqrt{s}=35$, $44$ GeV, figures (\ref{distriTbelZ}) and (\ref{distriCbelZ}) show that similar to the case of distributions above $Z^0$ peak, the conventional results are plagued by the large renormalization scale uncertainty and underestimate the experimental data. The perturbative series shows a slow convergence, and the estimation of the magnitude of unknown higher-order QCD contributions by varying $\mu_r\in[\sqrt{s}/2,2\sqrt{s}]$ is also unreliable. After applying PMC scale setting, the resulting PMC predictions are greatly increased in wide kinematic regions, which leads PMC results to be closer to the experimental data. However, compared to the distributions above $Z^0$ peak shown in Figs.(\ref{distriPMCT}) and (\ref{distriPMCc}), the agreement between the theoretical predictions and the experimental data is better for the higher energy data. This is due to the non-perturbative effect is more important for the lower energy data.

\section{The determination of the QCD running coupling $\alpha_s$ in perturbative domain }
\label{sec:4}

Event shape observables provide an ideal platform for the determination of the QCD coupling $\alpha_s$. Given that the strong interactions occur at the lowest order $\mathcal{O}(\alpha_s)$, the comparison between theoretical predictions and experimental data is straightforward. Currently, the main obstacle for achieving a precise value of $\alpha_s$ is the ambiguity of theoretical predictions based on the conventional scale-setting method. One of the main source of error for extracted values of $\alpha_s$ from event shape observables is the renormalization scale uncertainty.

As pointed out by the Particle Data Group (PDG)~\cite{PDG:2020}, one often introduces corrections of non-perturbative hadronization effects using Monte Carlo generators, or one extracts the coupling $\alpha_s$ using non-perturbative analytic modelling such as power corrections, or dispersive models. However, the systematics of these non-perturbative methods have not yet been fully verified; the extracted values of $\alpha_s$ from an event shape have very different results according to the particular model used. For example, the extracted $\alpha_s$ is $\alpha_s(M^2_Z)=0.1224\pm0.0039$~\cite{Dissertori:2009ik} by using Monte Carlo generators to estimate the power corrections (the central values are $\alpha_s(M^2_Z)=0.1266$ for thrust and $\alpha_s(M^2_Z)=0.1252$ for C-parameter); however, by using analytic models, the results are $\alpha_s(M^2_Z)=0.1135\pm0.0011$~\cite{Abbate:2010xh} from the thrust and $\alpha_s(M^2_Z)=0.1123\pm0.0015$~\cite{Hoang:2015hka} from the C-parameter, which are rather smaller than the PDG world average.

\begin{figure}[htb]
\centering
\includegraphics[width=0.50\textwidth]{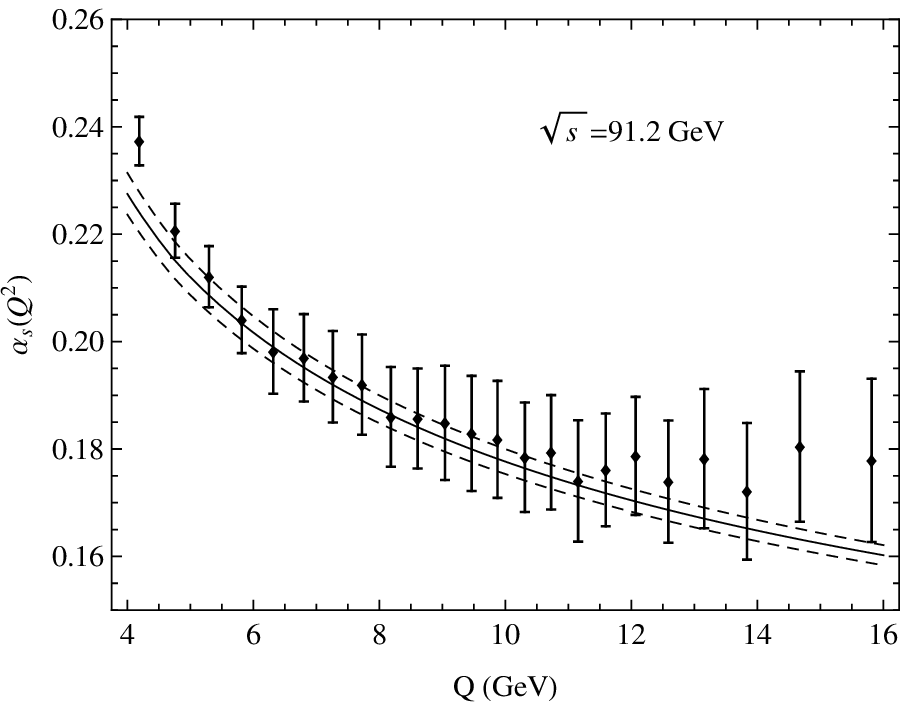}
\caption{The extracted running coupling $\alpha_s(Q^2)$ from the thrust ($1-T$) distribution by comparing the PMC predictions with the ALEPH data~\cite{Heister:2003aj} measured at a single energy of $\sqrt{s}=91.2$ GeV. As a comparison, the solid line is the world average evaluated from $\alpha_s(M^2_Z)=0.1179$~\cite{PDG:2020}, and two dashed lines represent its uncertainty. }
\label{alphasPMCT}
\end{figure}

\begin{figure}[htb]
\centering
\includegraphics[width=0.50\textwidth]{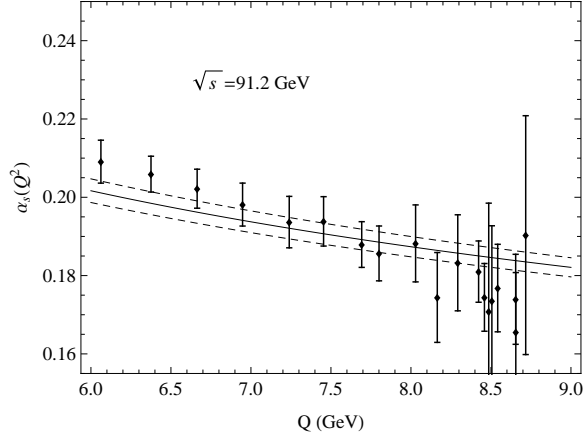}
\caption{Similar to Fig. (\ref{alphasPMCT}), but $\alpha_s(Q^2)$ extracted from the heavy jet mass ($\rho$) distribution.}
\label{alphasPMCR}
\end{figure}

\begin{figure}[htb]
\centering
\includegraphics[width=0.50\textwidth]{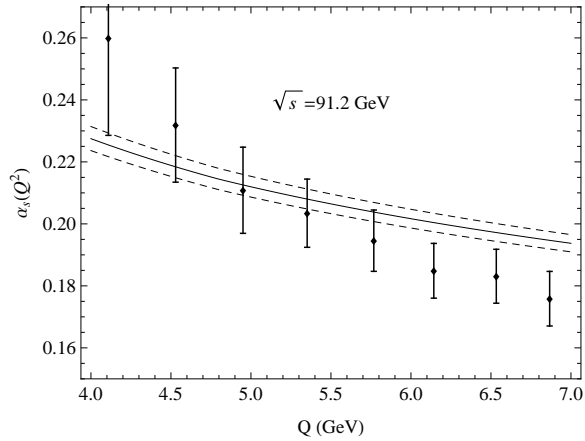}
\caption{Similar to Fig. (\ref{alphasPMCT}), but $\alpha_s(Q^2)$ extracted from the wide jet broadening ($B_W$) distribution. }
\label{alphasPMCBW}
\end{figure}

\begin{figure}[htb]
\centering
\includegraphics[width=0.50\textwidth]{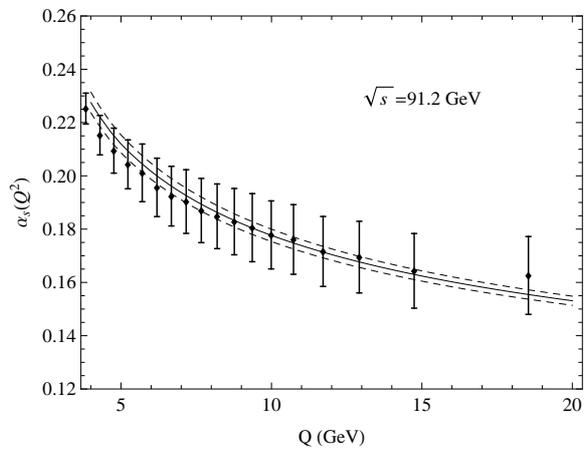}
\caption{Similar to Fig. (\ref{alphasPMCT}), but $\alpha_s(Q^2)$ extracted from the total jet broadening ($B_T$) distribution.}
\label{alphasPMCBT}
\end{figure}

\begin{figure}[htb]
\centering
\includegraphics[width=0.50\textwidth]{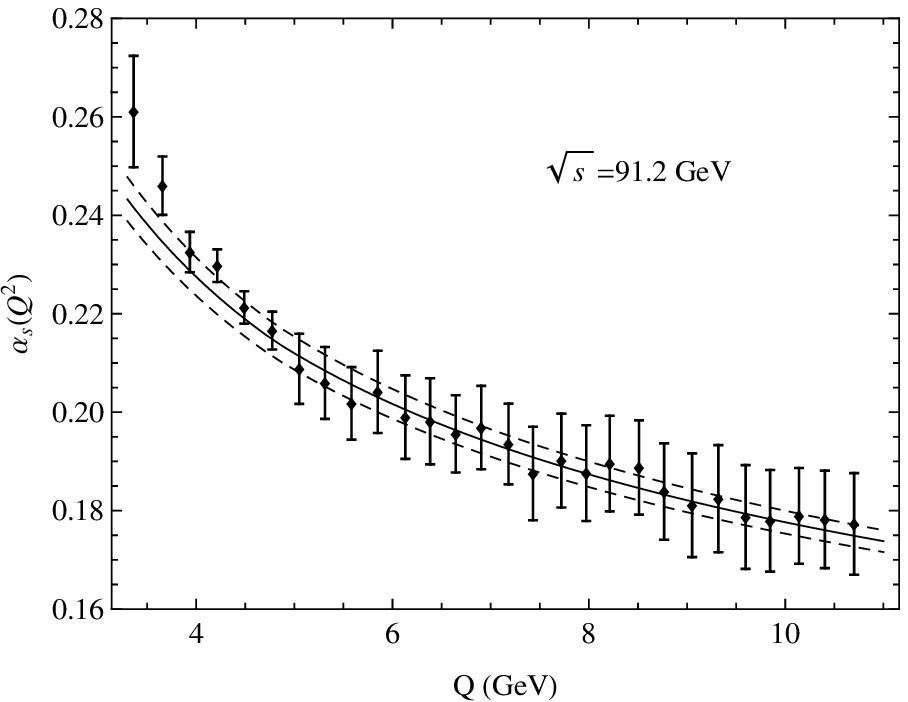}
\caption{Similar to Fig. (\ref{alphasPMCT}), but $\alpha_s(Q^2)$ extracted from the C-parameter ($C$) distribution. }
\label{alphasPMCC}
\end{figure}

In the case of conventional scale setting, since the scale in $\alpha_s$ is simply set to $\mu_r=\sqrt{s}$, one only extracts one value of $\alpha_s$ at scale $\sqrt{s}$. In contrast, the PMC scales are not a single value but they change with event shape values. We can thus extract
$\alpha_s(Q^2)$ over a wide range of $Q^2$ by the comparison of PMC predictions with experimental data measured at a single energy of $\sqrt{s}$. By adopting the precise experimental data at $\sqrt{s}=91.2$ GeV and using the method similar to that of Ref.\cite{Becher:2008cf}, the extracted running coupling $\alpha_s(Q^2)$ in perturbative domain are presented in Figs.(\ref{alphasPMCT}) to (\ref{alphasPMCC}), where the error bars are the squared averages of the experimental and theoretical errors.

Since the renormalization scale uncertainty is eliminated by the PMC method, the extracted $\alpha_s(Q^2)$ are not plagued by the uncertainty from the choice of scale $\mu_r$. It is useful to estimate the magnitude of unknown higher-order terms. As mentioned above, the conventional estimate obtained by simply varying the scale $\mu_r\in[\sqrt{s}/2, 2\sqrt{s}]$ is unreliable. A reasonable estimate for unknown higher-order terms can be characterized by the convergence of the pQCD series as well as the magnitude of the last known higher-order term. In the case of conventional scale setting, the relative importance of pQCD corrections in the intermediate region are about LO:NLO$\sim1:0.5$ for the thrust, $1:0.3$ for the heavy jet mass, $1:0.3$ for the wide jet broadening, $1:0.6$ for the total jet broadening and $1:0.5$ for the C-parameter. After using the PMC, due to the elimination of divergent renormalon terms, the convergence of the pQCD series is greatly improved, which are about LO:NLO$\sim1:0.2$, $1:0.1$, $1:0.1$, $1:0.2$ and $1:0.2$ for the thrust, heavy jet mass, wide jet broadening, total jet broadening and the C-parameter in the intermediate region, respectively. In our present analysis, if one assumes that the relative importance of the unknown ($n+1$)th-order term $y_{n+1}$ is the same as that of the last-known $n$th-order term $y_{n}$, the unknown higher-order terms are estimated by using $\pm0.2y_n$ for the thrust, total jet broadening and C-parameter, and $\pm0.1y_n$ for the heavy jet mass and wide jet broadening. This estimate of the unknown higher-order terms is natural for a convergent pQCD series.

\begin{figure}[htb]
\centering
\includegraphics[width=0.50\textwidth]{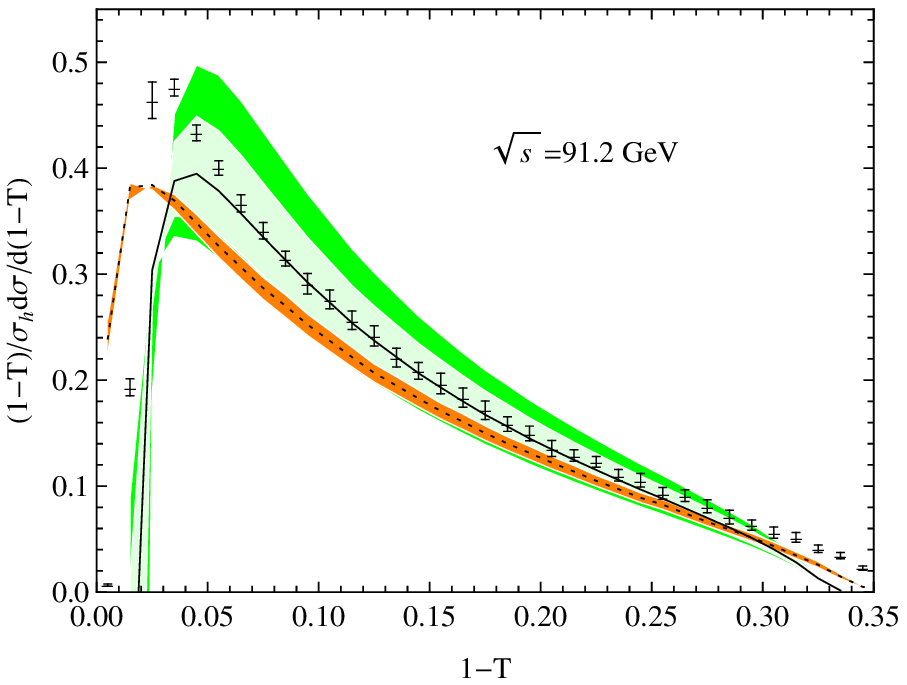}
\caption{The PMC theoretical uncertainties for the thrust ($1-T$) distribution at $\sqrt{s}=91.2$ GeV obtained by simply varying the PMC scale. The green and light green bands represent the variation of the PMC scale in the ranges $[Q_\star/2, 2Q_\star]$ and $[Q_\star/1.5, 1.5Q_\star]$, respectively. The NNLO conventional theoretical uncertainty (orange band) obtained by varying $\mu_r\in[\sqrt{s}/2, 2\sqrt{s}]$ is also presented as a comparison. }
\label{PMCscalevaryT}
\end{figure}

Due to the PMC scales are determined unambiguously by absorbing the non-conformal $\beta$ terms, the variation of the PMC scales will explicitly break the RGI, leading to unreliable PMC predictions~\cite{Wu:2014iba}. The conventional estimate of unknown higher-order terms by simply varying the scale is not applicable to the PMC method. We take the thrust ($T$) distribution at $\sqrt{s}=91.2$ GeV as a example, and then present the PMC theoretical uncertainties obtained by simply varying the PMC scale in the ranges $[Q_\star/2, 2Q_\star]$ and $[Q_\star/1.5, 1.5Q_\star]$ in Fig.(\ref{PMCscalevaryT}). It shows that simply varying the PMC scale leads to the large theoretical uncertainty. Although the NNLO conventional theoretical uncertainty obtained by $\mu_r\in[\sqrt{s}/2, 2\sqrt{s}]$ is very large, the PMC theoretical uncertainty obtained by simply varying the PMC scale is far beyond the conventional results. Thus, simply varying the PMC scale would lead to unreliable PMC predictions.

In the case of $\sqrt{s}=91.2$ GeV, Figures (\ref{alphasPMCT}) to (\ref{alphasPMCC}) show that the extracted $\alpha_s(Q^2)$ in the ranges $4<Q<16$ GeV from the thrust, $6<Q<9$ GeV from the heavy jet mass, $4<Q<7$ GeV from the wide jet broadening, $4<Q<19$ GeV from the total jet broadening and $3<Q<11$ GeV from the C-parameter are in excellent agreement with the world average evaluated from $\alpha_s(M^2_Z)=0.1179\pm 0.0010$~\cite{PDG:2020}. Since the PMC scales are increased with the event shape values and the PMC predictions are in agreement with the experimental data in wide kinematic regions, we can determine the running of $\alpha_s(Q^2)$ over a wide range of $Q^2$, especially from the thrust, total jet broadening and the C-parameter. We observe that the PMC scale for the heavy jet mass is increased with event shape value $\rho$, and stable at around $8.5$ GeV for $\sqrt{s}=91.2$ GeV in intermediate regions; thus more values of $\alpha_s(Q^2)$ around $8.5$ GeV are extracted.

\begin{figure*}
\begin{center}
\includegraphics[width=0.45\textwidth]{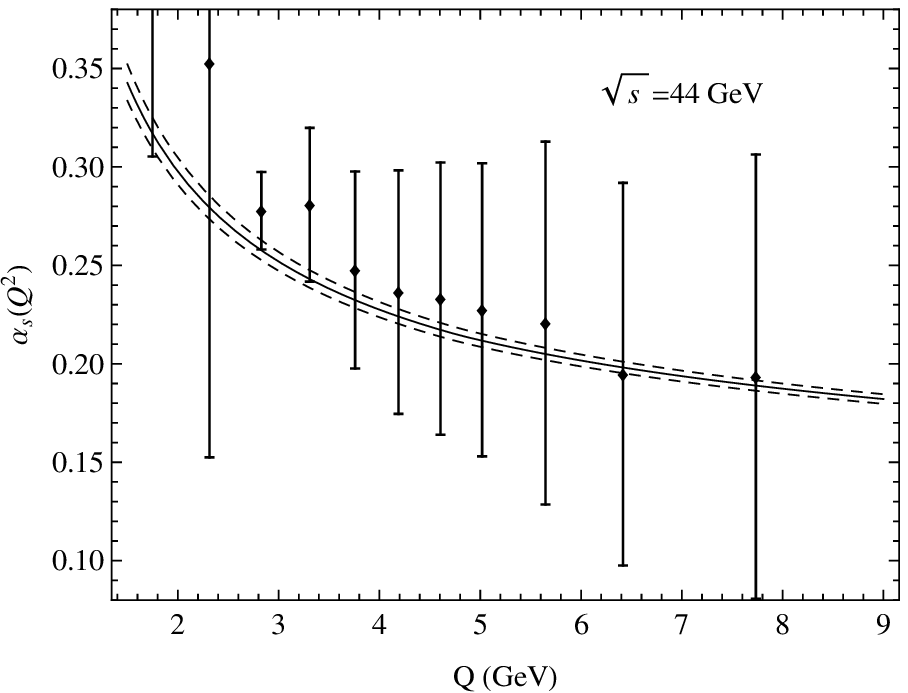}
\includegraphics[width=0.45\textwidth]{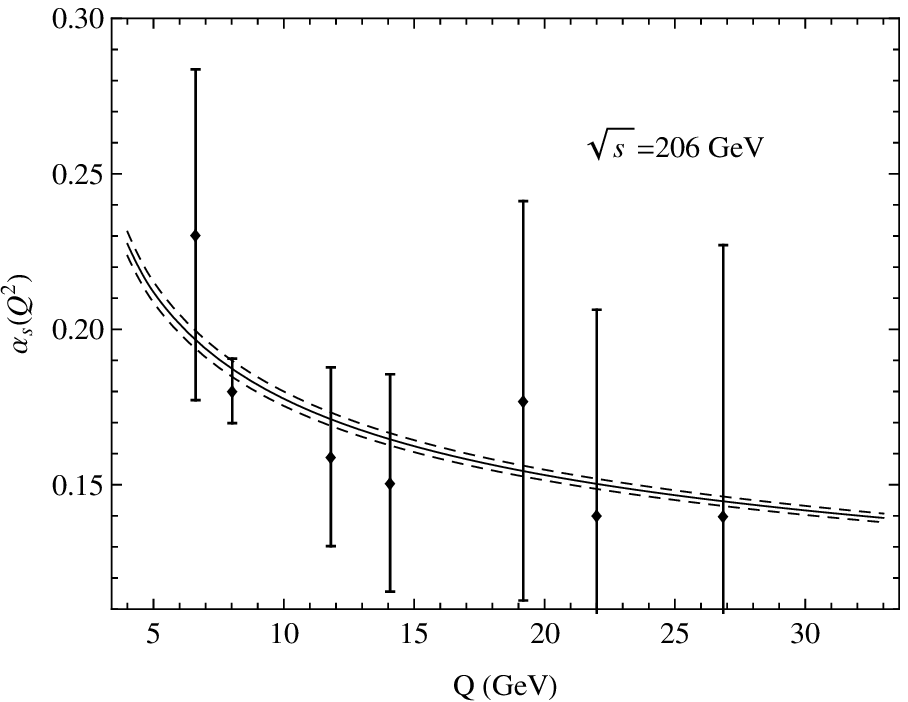}
\caption{The extracted running coupling $\alpha_s(Q^2)$ from the thrust ($1-T$) distributions by comparing the PMC predictions with the experimental data measured at $\sqrt{s}= 44$ GeV~\cite{MovillaFernandez:1997fr} and $206$ GeV~\cite{Heister:2003aj}. As a comparison, the solid line is the world average evaluated from $\alpha_s(M^2_Z)=0.1179$~\cite{PDG:2020}, and two dashed lines represent its uncertainty. }
\label{alphasPMCC44206}
\end{center}
\end{figure*}

In addition to the extracted running coupling $\alpha_s(Q^2)$ in the case of $\sqrt{s}=91.2$ GeV, the PMC scales are increased with the center-of-mass energy $\sqrt{s}$, we can extract $\alpha_s(Q^2)$ over a wider range by comparing the PMC predictions with experimental data measured at higher center-of-mass energy $\sqrt{s}$. For example, we determine the running coupling $\alpha_s(Q^2)$ over a wider range of $4<Q<22$ GeV by comparing the $C$-parameter distribution with the experimental data measured at a single energy of $\sqrt{s}=206$ GeV. By comparing the thrust distribution with the measurement at $\sqrt{s}=206$ GeV, we determine the running coupling $\alpha_s(Q^2)$ over a wider range of $5<Q<27$ GeV. For the lower energy data, we determine the running coupling $\alpha_s(Q^2)$ over a narrow range of $2<Q<8$ GeV by comparing the thrust distribution with the measurement at $\sqrt{s}=44$ GeV. Taking the thrust as an example, we present the extracted running coupling $\alpha_s(Q^2)$ obtained at $\sqrt{s}= 44$ GeV~\cite{MovillaFernandez:1997fr} and $206$ GeV~\cite{Heister:2003aj} in Fig.(\ref{alphasPMCC44206}). Due to the experimental uncertainty dominates the error of the coupling constant, the extracted $\alpha_s(Q^2)$ have relatively large uncertainty and few experimental measurement points are given compared to the case of $\sqrt{s}=91.2$ GeV.

Remarkably, the PMC scale-setting method provides a novel approach for measuring the running of $\alpha_s(Q^2)$ over a wide range of $Q^2$ in perturbative domain from event shape observables measured at a single energy scale of $\sqrt{s}$. Note that the running of the QED coupling constant $\alpha(Q^2)$ can be obtained from events at a single energy of $\sqrt{s}$~\cite{Abbiendi:2005rx}. As mentioned above, there are some deviations near the two-jet and multijet regions, and thus we can not determine reliable running coupling $\alpha_s(Q^2)$ in these kinematic regions. In order to improve the PMC predictions, the resummation of large logarithms and higher pQCD calculations should be included for the PMC analysis near the two-jet and multijet regions.

\section{Summary}
\label{sec:5}

Event shape observables in electron-positron annihilation provide an ideal platform for the determination of the QCD coupling $\alpha_s$. Conventionally, one often sets the renormalization scale equal to the center-of-mass energy $\sqrt{s}$, and theoretical uncertainties are estimated by varying the renormalization scale over an arbitrary range for the pQCD predictions. Such conventional procedure introduces an inherent scheme-and-scale dependence, which becomes one of the most important systematic errors for pQCD predictions. Currently, the ambiguity of theoretical predictions based on conventional scale setting is the main obstacle for achieving a precise value of $\alpha_s$. The event shape distributions in general do not match the experimental data, and only one value of $\alpha_s$ at the scale $\sqrt{s}$ can be extracted and the main source of error is from the renormalization scale ambiguity.

We analyze event shape observables by using the PMC which is a rigorous scale-setting method to eliminate the renormalization scheme and scale ambiguity. The PMC scales are determined by absorbing all of the nonconformal $\beta$ contributions into the running coupling via the RGE. Very different from the conventional method of simply setting the scale $\mu_r=\sqrt{s}$, the resulting PMC scales are not a single value but depend dynamically on the virtuality of the underlying quark and gluon subprocess and thus the specific kinematics of each event. The PMC scales thus yield the correct physical behavior of the scale and the scale-independent PMC predictions for event shape distributions agree with precise experimental data. Remarkably, the PMC method provides a novel method for the precise determination of the running of $\alpha_s(Q^2)$ over a wide range of $Q^2$ in perturbative domain from event shape observables measured at a single energy of $\sqrt{s}$. Our analyses show the importance of a correct renormalization scale setting, and open a new way for the calculation of event shape observables in electron-electron, electron-proton or proton-proton collisions.

{\bf Acknowledgements}:

We thank Stanley J. Brodsky for his valuable comments and for useful discussions. This work was supported in part by the Natural Science Foundation of China under Grants No.12265011, No.11905056, No.12175025 and No.12147102; by the Project of Guizhou Provincial Department under Grant No.KY[2021]003.


\begin{thebibliography}{99}

\bibitem{Heister:2003aj}
  A.~Heister {\it et al.} [ALEPH Collaboration],
  Studies of QCD at $e^+ e^-$ centre-of-mass energies between 91 GeV and 209 GeV,
  Eur.\ Phys.\ J.\ C {\bf 35}, 457 (2004).

\bibitem{Abdallah:2003xz}
  J.~Abdallah {\it et al.} [DELPHI Collaboration],
  A Study of the energy evolution of event shape distributions and their means with the DELPHI detector at LEP,
  Eur.\ Phys.\ J.\ C {\bf 29}, 285 (2003).

\bibitem{Abbiendi:2004qz}
  G.~Abbiendi {\it et al.} [OPAL Collaboration],
  Measurement of event shape distributions and moments in $e^{+} e^{-}\rightarrow$ hadrons at 91-209 GeV and a determination of $\alpha_s$,
  Eur.\ Phys.\ J.\ C {\bf 40}, 287 (2005).

\bibitem{Achard:2004sv}
  P.~Achard {\it et al.} [L3 Collaboration],
  Studies of hadronic event structure in $e^{+} e^{-}$ annihilation from 30 to 209 GeV with the L3 detector,
  Phys.\ Rept.\  {\bf 399}, 71 (2004).

\bibitem{Abe:1994mf}
  K.~Abe {\it et al.} [SLD Collaboration],
  Measurement of $\alpha_s(M_Z^2)$ from hadronic event observables at the Z$^0$ resonance,
  Phys.\ Rev.\ D {\bf 51}, 962 (1995).

\bibitem{Ellis:1980wv}
  R.~K.~Ellis, D.~A.~Ross and A.~E.~Terrano,
  The Perturbative Calculation of Jet Structure in $e^+ e^-$ Annihilation,
  Nucl.\ Phys.\ B {\bf 178}, 421 (1981).

\bibitem{Kunszt:1980vt}
  Z.~Kunszt,
  Comment on the $\mathcal{O}(\alpha^2_S)$ Corrections to Jet Production in $e^+ e^-$ Annihilation,
  Phys.\ Lett.\ B  {\bf 99}, 429 (1981).

\bibitem{Vermaseren:1980qz}
  J.~A.~M.~Vermaseren, K.~J.~F.~Gaemers and S.~J.~Oldham,
  Perturbative QCD Calculation of Jet Cross-Sections in $e^+ e^-$ Annihilation,
  Nucl.\ Phys.\ B {\bf 187}, 301 (1981).

\bibitem{Fabricius:1981sx}
  K.~Fabricius, I.~Schmitt, G.~Kramer and G.~Schierholz,
  Higher Order Perturbative QCD Calculation of Jet Cross-Sections in $e^+ e^-$ Annihilation,
  Z.\ Phys.\ C {\bf 11}, 315 (1981).

\bibitem{Giele:1991vf}
  W.~T.~Giele and E.~W.~N.~Glover,
  Higher order corrections to jet cross-sections in $e^+ e^-$ annihilation,
  Phys.\ Rev.\ D {\bf 46}, 1980 (1992).

\bibitem{Catani:1996jh}
  S.~Catani and M.~H.~Seymour,
  The Dipole formalism for the calculation of QCD jet cross-sections at next-to-leading order,
  Phys.\ Lett.\ B {\bf 378}, 287 (1996).

\bibitem{Gehrmann-DeRidder:2007nzq}
  A.~Gehrmann-De Ridder, T.~Gehrmann, E.~W.~N.~Glover and G.~Heinrich,
  Second-order QCD corrections to the thrust distribution,
  Phys.\ Rev.\ Lett.\  {\bf 99}, 132002 (2007).

\bibitem{GehrmannDeRidder:2007hr}
  A.~Gehrmann-De Ridder, T.~Gehrmann, E.~W.~N.~Glover and G.~Heinrich,
  NNLO corrections to event shapes in $e^+ e^-$ annihilation,
  JHEP {\bf 0712}, 094 (2007).

\bibitem{Ridder:2014wza}
  A.~Gehrmann-De Ridder, T.~Gehrmann, E.~W.~N.~Glover and G.~Heinrich,
  EERAD3: Event shapes and jet rates in electron-positron annihilation at order $\alpha_s^3$,
  Comput.\ Phys.\ Commun.\  {\bf 185}, 3331 (2014).

\bibitem{Weinzierl:2008iv}
  S.~Weinzierl,
  NNLO corrections to 3-jet observables in electron-positron annihilation,
  Phys.\ Rev.\ Lett.\  {\bf 101}, 162001 (2008).

\bibitem{Weinzierl:2009ms}
  S.~Weinzierl,
  Event shapes and jet rates in electron-positron annihilation at NNLO,
  JHEP {\bf 0906}, 041 (2009).

\bibitem{DelDuca:2016csb}
  V.~Del Duca, C.~Duhr, A.~Kardos, G.~Somogyi and Z.~Tr\'{o}cs\'{a}nyi,
  Three-Jet Production in Electron-Positron Collisions at Next-to-Next-to-Leading Order Accuracy,
  Phys.\ Rev.\ Lett.\  {\bf 117}, 152004 (2016).

\bibitem{DelDuca:2016ily}
  V.~Del Duca, C.~Duhr, A.~Kardos, G.~Somogyi, Z.~Sz\H{o}r, Z.~Tr\'{o}cs\'{a}nyi and Z.~Tulip\'{a}nt,
  Jet production in the CoLoRFulNNLO method: event shapes in electron-positron collisions,
  Phys.\ Rev.\ D {\bf 94}, 074019 (2016).

\bibitem{Catani:1992ua}
  S.~Catani, L.~Trentadue, G.~Turnock and B.~R.~Webber,
  Resummation of large logarithms in $e^{+} e^{-}$ event shape distributions,
  Nucl. Phys. B \textbf{407} (1993), 3-42.

\bibitem{Banfi:2004yd}
 A.~Banfi, G.~P.~Salam and G.~Zanderighi,
  Principles of general final-state resummation and automated implementation,
  JHEP \textbf{03}, 073 (2005).

\bibitem{Banfi:2014sua}
  A.~Banfi, H.~McAslan, P.~F.~Monni and G.~Zanderighi,
  A general method for the resummation of event-shape distributions in $e^{+} e^{-}$ annihilation,
  JHEP \textbf{05}, 102 (2015).

\bibitem{Chien:2010kc}
  Y.~T.~Chien and M.~D.~Schwartz,
  Resummation of heavy jet mass and comparison to LEP data,
  JHEP \textbf{08}, 058 (2010).

\bibitem{Becher:2012qc}
  T.~Becher and G.~Bell,
  NNLL Resummation for Jet Broadening,
  JHEP \textbf{11}, 126 (2012).

\bibitem{Abbate:2010xh}
  R.~Abbate, M.~Fickinger, A.~H.~Hoang, V.~Mateu and I.~W.~Stewart,
  Thrust at $N^{3}LL$ with Power Corrections and a Precision Global Fit for $\alpha_{s}(m_Z)$,
  Phys. Rev. D \textbf{83}, 074021 (2011).

\bibitem{Hoang:2014wka}
  A.~H.~Hoang, D.~W.~Kolodrubetz, V.~Mateu and I.~W.~Stewart,
  $C$-parameter distribution at N$^3$LL' including power corrections,
  Phys. Rev. D \textbf{91}, 094017 (2015).

\bibitem{Chiu:2011qc}
  J.~y.~Chiu, A.~Jain, D.~Neill and I.~Z.~Rothstein,
  The Rapidity Renormalization Group,
  Phys. Rev. Lett. \textbf{108}, 151601 (2012).

\bibitem{Chiu:2012ir}
  J.~Y.~Chiu, A.~Jain, D.~Neill and I.~Z.~Rothstein,
  A Formalism for the Systematic Treatment of Rapidity Logarithms in Quantum Field Theory,
  JHEP \textbf{05}, 084 (2012).

\bibitem{PDG:2020}
 P.~A.~Zyla {\it et al.} [Particle Data Group], Prog. Theor. Exp. Phys. 2020 (2020), 083C01.

\bibitem{Brodsky:2011ta}
  S.~J.~Brodsky and X.~G.~Wu,
  Scale Setting Using the Extended Renormalization Group and the Principle of Maximum Conformality: the QCD Coupling Constant at Four Loops,
  Phys.\ Rev.\ D {\bf 85}, 034038 (2012).

\bibitem{Brodsky:2012rj}
  S.~J.~Brodsky and X.~G.~Wu,
  Eliminating the Renormalization Scale Ambiguity for Top-Pair Production Using the Principle of Maximum Conformality,
  Phys.\ Rev.\ Lett.\  {\bf 109}, 042002 (2012).

\bibitem{Brodsky:2011ig}
  S.~J.~Brodsky and L.~Di Giustino,
  Setting the Renormalization Scale in QCD: The Principle of Maximum Conformality,
  Phys.\ Rev.\ D {\bf 86}, 085026 (2012).

\bibitem{Mojaza:2012mf}
  M.~Mojaza, S.~J.~Brodsky and X.~G.~Wu,
  Systematic All-Orders Method to Eliminate Renormalization-Scale and Scheme Ambiguities in Perturbative QCD,
  Phys.\ Rev.\ Lett.\  {\bf 110}, 192001 (2013).

\bibitem{Brodsky:2013vpa}
  S.~J.~Brodsky, M.~Mojaza and X.~G.~Wu,
  Systematic Scale-Setting to All Orders: The Principle of Maximum Conformality and Commensurate Scale Relations,
  Phys.\ Rev.\ D {\bf 89}, 014027 (2014).

\bibitem{Wang:2019ljl}
  S.~Q.~Wang, S.~J.~Brodsky, X.~G.~Wu and L.~Di Giustino,
  Thrust Distribution in Electron-Positron Annihilation using the Principle of Maximum Conformality,
  Phys. Rev. D \textbf{99}, 114020 (2019).

\bibitem{Wang:2019isi}
  S.~Q.~Wang, S.~J.~Brodsky, X.~G.~Wu, J.~M.~Shen and L.~Di Giustino,
  Novel method for the precise determination of the QCD running coupling from event shape distributions in electron-positron annihilation,
  Phys. Rev. D \textbf{100}, 094010 (2019).

\bibitem{Brodsky:1982gc}
  S.~J.~Brodsky, G.~P.~Lepage and P.~B.~Mackenzie,
  On the Elimination of Scale Ambiguities in Perturbative Quantum Chromodynamics,
  Phys.\ Rev.\ D {\bf 28}, 228 (1983).

\bibitem{GellMann:1954fq}
  M.~Gell-Mann and F.~E.~Low,
  Quantum electrodynamics at small distances,
  Phys.\ Rev.\  {\bf 95}, 1300 (1954).

\bibitem{Brodsky:2012ms}
  S.~J.~Brodsky and X.~G.~Wu,
  Self-Consistency Requirements of the Renormalization Group for Setting the Renormalization Scale,
  Phys.\ Rev.\ D {\bf 86}, 054018 (2012).

\bibitem{Wu:2014iba}
  X.~G.~Wu, Y.~Ma, S.~Q.~Wang, H.~B.~Fu, H.~H.~Ma, S.~J.~Brodsky and M.~Mojaza,
  Renormalization Group Invariance and Optimal QCD Renormalization Scale-Setting,
  Rept.\ Prog.\ Phys.\  {\bf 78}, 126201 (2015).

\bibitem{Wu:2019mky}
  X.~G.~Wu, J.~M.~Shen, B.~L.~Du, X.~D.~Huang, S.~Q.~Wang and S.~J.~Brodsky,
  The QCD Renormalization Group Equation and the Elimination of Fixed-Order Scheme-and-Scale Ambiguities Using the Principle of Maximum Conformality,
  Prog. Part. Nucl. Phys. {\bf 108}, 103706 (2019).

\bibitem{Zheng:2013uja}
  X.~C.~Zheng, X.~G.~Wu, S.~Q.~Wang, J.~M.~Shen and Q.~L.~Zhang,
  Reanalysis of the BFKL Pomeron at the next-to-leading logarithmic accuracy,
  JHEP \textbf{10}, 117 (2013).

\bibitem{Shen:2017pdu}
  J.~M.~Shen, X.~G.~Wu, B.~L.~Du and S.~J.~Brodsky,
  Novel All-Orders Single-Scale Approach to QCD Renormalization Scale-Setting,
  Phys.\ Rev.\ D {\bf 95}, 094006 (2017).

\bibitem{Webber:1994cp}
  B.~R.~Webber,
  Estimation of power corrections to hadronic event shapes,
  Phys. Lett. B \textbf{339}, 148-150 (1994).

\bibitem{Beneke:1995pq}
  M.~Beneke and V.~M.~Braun,
  Power corrections and renormalons in Drell-Yan production,
  Nucl. Phys. B \textbf{454}, 253-290 (1995).

\bibitem{Gardi:2001ny}
E.~Gardi and J.~Rathsman,
  Renormalon resummation and exponentiation of soft and collinear gluon radiation in the thrust distribution,
  Nucl. Phys. B \textbf{609}, 123-182 (2001).

\bibitem{Hoang:2007vb}
  A.~H.~Hoang and I.~W.~Stewart,
  Designing gapped soft functions for jet production,
  Phys. Lett. B \textbf{660}, 483-493 (2008).

\bibitem{Gracia:2021nut}
  N.~G.~Gracia and V.~Mateu,
  Toward massless and massive event shapes in the large-\ensuremath{\beta}$_{0}$ limit,
  JHEP \textbf{07}, 229 (2021).

\bibitem{Chetyrkin:2000yt}
  K.~G.~Chetyrkin, J.~H.~Kuhn and M.~Steinhauser,
  RunDec: A Mathematica package for running and decoupling of the strong coupling and quark masses,
  Comput.\ Phys.\ Commun.\  {\bf 133}, 43 (2000).

\bibitem{Beneke:1994qe}
  M.~Beneke and V.~M.~Braun,
  Naive nonAbelianization and resummation of fermion bubble chains,
  Phys. Lett. B \textbf{348}, 513-520 (1995).

\bibitem{Beneke:1998ui}
  M.~Beneke,
  Renormalons,
  Phys.\ Rept.\  {\bf 317}, 1 (1999).

\bibitem{Wu:2013ei}
 X.~G.~Wu, S.~J.~Brodsky and M.~Mojaza,
  The Renormalization Scale-Setting Problem in QCD,
  Prog. Part. Nucl. Phys. \textbf{72}, 44-98 (2013).

\bibitem{Brodsky:2014yha}
  S.~J.~Brodsky, G.~F.~de Teramond, H.~G.~Dosch and J.~Erlich,
  Light-Front Holographic QCD and Emerging Confinement,
  Phys.\ Rept.\  {\bf 584}, 1 (2015).

\bibitem{Kramer:1990zt}
  G.~Kramer and B.~Lampe,
  Jet production rates at LEP and the scale of $\alpha_s$,
  Z.\ Phys.\ A {\bf 339}, 189 (1991).

\bibitem{Gehrmann:2014uva}
  T.~Gehrmann, N.~H\"{a}fliger and P.~F.~Monni,
  BLM Scale Fixing in Event Shape Distributions,
  Eur.\ Phys.\ J.\ C {\bf 74}, 2896 (2014).

\bibitem{Pietrulewicz:2014qza}
  P.~Pietrulewicz, S.~Gritschacher, A.~H.~Hoang, I.~Jemos and V.~Mateu,
  Variable Flavor Number Scheme for Final State Jets in Thrust,
  Phys. Rev. D \textbf{90}, 114001 (2014).

\bibitem{MovillaFernandez:1997fr}
  P.~A.~Movilla Fernandez \textit{et al.} [JADE],
  A Study of event shapes and determinations of $\alpha_s$ using data of $e^+ e^-$ annihilations at $\sqrt{s}=22$ GeV to $44$ GeV,
  Eur. Phys. J. C \textbf{1}, 461-478 (1998).

\bibitem{Biebel:1999zt}
  O.~Biebel \textit{et al.} [JADE],
  C parameter and jet broadening at PETRA energies,
  Phys. Lett. B \textbf{459}, 326-334 (1999).

\bibitem{Dissertori:2009ik}
  G.~Dissertori, A.~Gehrmann-De Ridder, T.~Gehrmann, E.~W.~N.~Glover, G.~Heinrich, G.~Luisoni and H.~Stenzel,
  Determination of the strong coupling constant using matched NNLO+NLLA predictions for hadronic event shapes in e+e- annihilations,
  JHEP \textbf{08}, 036 (2009).

\bibitem{Hoang:2015hka}
  A.~H.~Hoang, D.~W.~Kolodrubetz, V.~Mateu and I.~W.~Stewart,
  Precise determination of $\alpha_s$ from the $C$-parameter distribution,
  Phys.\ Rev.\ D {\bf 91}, 094018 (2015).

\bibitem{Becher:2008cf}
  T.~Becher and M.~D.~Schwartz,
  A precise determination of $\alpha_s$ from LEP thrust data using effective field theory,
  JHEP {\bf 0807}, 034 (2008).

\bibitem{Abbiendi:2005rx}
  G.~Abbiendi {\it et al.} [OPAL Collaboration],
  Measurement of the running of the QED coupling in small-angle Bhabha scattering at LEP,
  Eur.\ Phys.\ J.\ C {\bf 45}, 1 (2006).

\end{thebibliography}
\end{document}